\documentclass[preprint]{aastex}

\usepackage{isotope}
\usepackage{capt-of}
\usepackage{multirow}
\usepackage{longtable}
\usepackage{lscape}
\usepackage{morefloats}
\usepackage{color}
\usepackage{wasysym}
\usepackage{hyperref}
\usepackage[section] {placeins}
\newcommand{\fig}[1]{Fig.\,\ref{#1}}

\newcommand{\tab}[1]{Tab.\,\ref{#1}}

\newcommand{\ms}{main sequence}

\newcommand{\spr}{\mbox{$s$-process}}
\newcommand{\sprn}{\mbox{$s$ process}}

\newcommand{\hevi}{\ensuremath{^{4}\mem{He}}}

\newcommand{\cdr}{\isotope[13]{C}}
\newcommand{\czw}{\isotope[12]{C}}

\newcommand{\cvi}{\ensuremath{^{14}\mem{C}}}

\newcommand{\nvi}{\ensuremath{^{14}\mem{N}}}

\newcommand{\ofu}{\ensuremath{^{15}\mem{O}}}
\newcommand{\ose}{\isotope[16]{O}}

\newcommand{\oac}{\ensuremath{^{18}\mem{O}}}

\newcommand{\fac}{\ensuremath{^{18}\mem{F}}}
\newcommand{\fne}{\ensuremath{^{19}\mem{F}}}

\newcommand{\nezw}{\ensuremath{^{22}\mem{Ne}}}

\newcommand{\mgfu}{\ensuremath{^{25}\mem{Mg}}}

\newcommand{\fese}{\ensuremath{^{56}\mem{Fe}}}

\newcommand{\zrse}{\ensuremath{^{96}\mem{Zr}}}

\newcommand{\zrfu}{\ensuremath{^{95}\mem{Zr}}}
\newcommand{\srac}{\ensuremath{^{88}\mem{Sr}}}

\newcommand{\baac}{\ensuremath{^{138}\mem{Ba}}}

\newcommand{\blac}{\ensuremath{^{208}\mem{Pb}}}

\newcommand{\msun}{\ensuremath{\, M_\odot}}

\newcommand{\mppnp}{\textsf{mppnp}}

\newcommand{\MESA}{\texttt{MESA}}

\newcommand{\beq}{\begin{equation}}
\newcommand{\beqa}{\begin{eqnarray}}
\newcommand{\eeq}{\end{equation}}
\newcommand{\eeqa}{\end{eqnarray}}
\newcommand{\bedis}{\begin{displaymath}}
\newcommand{\edis}{\end{displaymath}}

\newcommand{\mem}[1]{\ensuremath{\mathrm{ #1}}}

\newcommand{\ov}{{\it overshoot}}

\graphicspath{{figures/}}

\slugcomment{DRAFT: \today}

\shorttitle{AGB star evolution and nucleosynthesis with a new CBM model}
\shortauthors{}

\begin{document}

\title{Application of a theory and simulation based convective boundary mixing model for AGB star evolution and nucleosynthesis}
\author {U.~Battino\altaffilmark{1,11},
M. Pignatari\altaffilmark{2,3,11},
C.~Ritter\altaffilmark{4,5,11},
F.~Herwig\altaffilmark{4,5,11},
P.~Denisenkov\altaffilmark{4,5,9,11},
J.W.~Den Hartogh\altaffilmark{6,11},
R.~Trappitsch\altaffilmark{8,11},
R.~Hirschi\altaffilmark{6,7,11},
B.~Freytag\altaffilmark{10},
F.~Thielemann\altaffilmark{1},
B.~Paxton\altaffilmark{12}
}

\altaffiltext{1}{Department of Physics, University of Basel, Klingelbergstrasse 82, CH-4056 Basel, Switzerland}
\altaffiltext{2}{Konkoly Observatory, Research Center for Astronomy and Earth Sciences, Hungarian Academy of Sciences, Konkoly Thege Miklós út 15-17, H-1121 Budapest, Hungary}
\altaffiltext{3}{E.A. Milne Centre for Astrophysics, Dept of Physics \& Mathematics, University of Hull, HU6 7RX, United Kingdom}
\altaffiltext{4}{Department of Physics \& Astronomy, University of Victoria, Victoria, BC, V8P5C2 Canada.}
\altaffiltext{5}{JINA Center for the Evolution of the Elements, USA.}
\altaffiltext{6}{Keele University, Keele, Staffordshire ST5 5BG, United Kingdom.}
\altaffiltext{7}{Institute for the Physics and Mathematics of the Universe (WPI), University of Tokyo, 5-1-5 Kashiwanoha, Kashiwa 277-8583, Japan}
\altaffiltext{8}{Department of the Geophysical Sciences and Chicago Center for Cosmochemistry, Chicago, IL 60637, USA.}
\altaffiltext{9}{TRIUMF, 4004 Wesbrook Mall, Vancouver, BC V6T~2A3, Canada}
\altaffiltext{10}{Department of Physics and Astronomy at Uppsala University,Regementsvagen 1, Box 516, SE-75120 Uppsala, Sweden}
\altaffiltext{11}{The NuGrid collaboration, \url{http://www.nugridstars.org}}
\altaffiltext{12}{Kavli Institute for Theoretical Physics and Department of Physics, Kohn Hall, University of California, Santa Barbara, CA 93106, USA}


\begin{abstract}
The \spr\ nucleosynthesis in Asymptotic Giant Branch (AGB) stars 
depends on the modeling of convective boundaries. 
We present models and \spr\ simulations that adopt a treatment of convective boundaries based on the results of hydrodynamic simulations and on the theory of mixing due to gravity waves in the vicinity of convective boundaries. 
Hydrodynamics simulations suggest the presence of convective
boundary mixing (CBM) at the bottom of the thermal pulse-driven convective
zone. Similarly, convection-induced mixing processes are proposed
for the mixing below the convective envelope during third dredge-up
where the \cdr\ pocket for the \sprn\ in AGB stars forms.  In this work
we apply a CBM 
model motivated by simulations and theory to models with initial
mass $M=2$ and $M=3M_\odot$, and with initial metal content $Z=0.01$
and $Z=0.02$. As reported previously, 
the He-intershell abundance of \czw\ and \ose\ are increased by
CBM 
at the bottom of pulse-driven convection zone. This mixing is affecting the \nezw($\alpha$,n)\mgfu\ activation and the \spr\ efficiency in the
\cdr-pocket.
In our model CBM 
at the bottom of the convective envelope during the third dredge-up represents gravity wave mixing. We take further into account that hydrodynamic simulations indicate a declining mixing efficiency already about a pressure scale height from the convective boundaries, compared to mixing-length theory. 
We obtain the formation of the \cdr-pocket with a mass of $\approx~10^{-4}\msun$. 
  The final
\spr\ abundances are characterized by $0.36<\mathrm{[s/Fe]}<0.78$ and the
 heavy-to-light \spr\ ratio is $-0.23<\mathrm{[hs/ls]}<0.45$.  
Finally, we compare our results with stellar observations, pre-solar grain measurements and previous work. 
\end{abstract}

\clearpage


\section{Introduction}
\label{intro}

The Asymptotic giant branch (AGB) phase is the final evolutionary 
stage of low- and intermediate-mass stars, during which all their envelope is lost by
stellar wind forming a Planetary Nebula \citep{renzini:83,kwok:90}.  
During this phase, the energy output is dominated by the H-burning shell and the He-burning shell, 
activated alternatively on top of a degenerate core, mainly made of C and O
\citep{schwarzschild:65}.

AGB stars have a fundamental role in the chemical evolution of the galaxy,
producing among light elements a relevant amount of C, N, F and Na
observed today in the solar system \citep[e.g.,][]{tosi:07,kobayashi:11}. 
Beyond Fe, about half of the heavy isotope abundances are made by the slow neutron capture process 
\citep[\spr,][]{burbidge:57,cameron:57}. In particular, AGB stars 
have been identified as the site of the main \spr\ component 
of the solar abundance distribution between the Sr neutron magic peak and Pb, 
and the strong \spr\ component, explaining 
half of the solar \blac\ \citep[see][and references therein]{kaeppeler:11}.
Most of the neutrons for the \spr\ come from the \cdr($\alpha$,n)\ose\ 
neutron source, activated in the radiative \cdr-pocket in the He intershell 
stellar region \citep[][]{straniero:95}. The physics mechanisms driving the formation of the \cdr-pocket are still matter of debate \citep[see][and references therein]{herwig:05}, and will also be discussed in this work.

Neutrons are also made by the \nezw($\alpha$,n)\mgfu\ reaction, 
activated at the bottom of the He intershell during the Thermal Pulses (TPs). 
Whereas the contribution to the total amount of 
neutrons is smaller compared to the \cdr\ neutron source, the activation 
of the \nezw($\alpha$,n)\mgfu\ generates higher neutron densities above
10$^{10}$ neutrons cm$^{-3}$, leaving its fingerprints in the final \spr\ AGB
stellar yields \citep[e.g.,][]{gallino:98,cristallo:11,karakas:14}. 

The production of the \spr\ elements has been directly 
observed for a large sample of intrinsic or extrinsic AGB stars 
at different metallicities 
\citep[e.g.,][and references therein]{busso:01,abia:02,sneden:08,zamora:09}, in grains of
presolar origin condensed in the winds of old AGB stars and 
found in pristine carbonaceous meteorites 
\citep[e.g.,][]{lugaro:03a,avila:12,zinner:14,liu:14a,liu:14b}, in post-AGB stars \citep[e.g.,][]{reddy:02,reyniers:04,reyniers:07,vanAarle:13,desmedt:14} 
and in ionized material of planetary nebulae around their central remnant star 
after the AGB phase \citep[e.g.,][]{sterling:02,sharpee:07,sterling:09,otsuka:13}.
The abundances of the He intershell have been directly observed in 
post-AGB H-deficient stars
\citep[e.g.,][]{werner:06,werner:14} and in planetary nebulae \citep[e.g.,][]{pequignot:00,rodriguez:11,delgado-inglada:15}, still carrying the abundance signatures original of 
their previous AGB phase, in particular for light elements like He, C and O.

The possibility to compare stellar-models predictions with such a large
variety of independent observations together with the
needs for galactic chemical evolution calculations \citep[e.g.,][]{travaglio:04}, 
has motivated the production of different sets of 
AGB stellar yields \citep[e.g.,][]{bisterzo:11,cristallo:11,lugaro:12,karakas:14}.
The \spr\ nucleosynthesis is extremely sensitive to
thermodynamic conditions, abundances and convective boundary mixing mechanisms 
in the parent AGB stars, providing
fundamental constraints for the macro- and micro-physics inputs
used to produce theoretical stellar AGB models \citep[e.g.,][]{herwig:05}. 
Mixing at two convection boundaries, the bottom of the convective envelope during the third dredge-up, and at the bottom of the pulse-driven convection zone (PDCZ) have been identified as particularly relevant for the nucleosynthesis and evolution of the elements. 
The latter affects 
the abundances of the most abundant species (e.g., \hevi, \isotope[12]{C} and \isotope[16]{O}), 
and therefore the evolution and the nucleosynthesis 
in the He intershell during the AGB phase \citep[e.g.,][]{herwig:97,lugaro:03a}.
CBM below the envelope during 
the third dredge-up (TDU) facilitates the formation of the \cdr-pocket  \citep[][]{straniero:95}. Neither of these inherently multi-dimensional fluid dynamics processes
can be simulated ab-initio in hydrostatic one-dimensional stellar evolution models.

CBM at the bottom of the convective envelope has been represented as semiconvection \citep[][]{iben:82a} 
, overshooting \citep[][]{herwig:97}, or exponential decay of convective velocities \citep[][]{cristallo:00a}.
To address this challenge  \cite[][hereafter De03]{denissenkov:03} investigated mixing induced by 
internal gravity waves (IGWs) and found a \cdr-pocket with approximately the size of $\sim 10^{-4}M_\odot$
(see their Fig.~5). 
Other mechanisms that have been proposed considered mixing driven by magnetic buoyancy
\citep[][]{busso:07,nucci:14,trippella:16}. In the first work the efficiency of mixing was overestimated by
several orders of magnitude \citep{denissenkov:09}. 
  In the second work, the authors 
found the velocity and magnetic field distributions 
that satisfy the MHD equations 
under 
restricted assumptions, but it still needs to be explored what physical process, including magnetic buoyancy, 
could lead to such distributions.


Limitations in distinguishing between these scenarios include also 
the uncertainty of their implementation in hydrostatic models, leading to 
different nucleosynthesis results to compare with observations. 
For instance, starting from indications of hydrodynamics simulations by \cite{freytag:96}, 
In 1D models, \citet{herwig:97} applied their parameterized description of the velocities of the convective elements to the inclusion of overshoot in stellar evolution calculations up to the AGB.
\citet{cristallo:00a} implemented a CBM formalism based on the same work by \cite{freytag:96}, but did it differently and got different results,
with higher \spr\ production of heavy elements by at least one order of magnitude.

\cite{herwig:07} (hereafter He07) studied the CBM at the bottom of the PDCZ 
via 2-dimensional hydrodynamical simulations, showing that their results can be reproduced by 
a first initial decay of the mixing efficiency, followed by a second shallower decay term. 
Even if He07 simulations do not define this clearly, we believe that the first term is due to Kelvin-Helmholtz instabilities. 
\citet{casanova:11} interpreted CBM taking place in their 3D simulations of nova explosions as Kelvin-Helmholtz instabilities as the source of inhomogeneous mixing. 
Since the physical mechanism driving a nova-outburst is similar to the one driving a Helium-flash in AGB stars, we expect Kelvin-Helmholtz instabilities to dominate the CBM at the bottom of convective PDCZ as well. 
Concerning the second mixing term obtained by He07, we interpreted it as due to IGW, which were seen them plentifully in the
hydrodynamic simulations by He07.

  IGWs have mostly been considered as an efficient mechanism for angular momentum redistribution in rotating low- and
intermediate-mass stars,
particularly, in the Sun \citep[e.g.,][]{press:81,ringot:98,talon:05,talon:08,fuller:14}. Chemical mixing is produced by
IGWs indirectly, when they modify a velocity field in a stellar radiative zone, which may either bring the rate of rotational mixing
in agreement with observations \citep{charbonnel:05} or lead to a velocity distribution that becomes unstable on a small 
length scale when radiative damping is taken
into account \citep{garcia-lopez:91,montalban:00}. De03 implemented the last two IGW mixing mechanisms at the bottom of
the convective envelope of a $3\,M_\odot$ TP-AGB star and showed that both of them could result in the formation of
a $^{13}$C pocket wide enough for the s process.

In this work, we apply the CBM model parameters by He07 as well as a CBM model representing IGW mixing proposed by De03 at the bottom of the convective envelope for the formation of the \cdr\ pocket. 
The resulting abundance predictions are confronted with \spr\ observables in stars and pre-solar grains.

In Section \ref{sec:tools} we briefly describe the \MESA\ stellar code and \mppnp\ post-processing nucleosynthesis tool. 
In Section \ref{mesa-mppnp:calc} particular
attention is given to  \cdr-pocket formation and intershell abundances evolution.
In Section \ref{sec:postpro} we describe the post-processing method applied to compute \spr\ nucleosynthesis 
using the NuGrid \mppnp\ code, also comparing our results with observations and other stellar models. Our conclusions are given in Section \ref{sec:conclusions}. 
Finally, in the Appendix 
more details are given about the simulations setup of our \MESA\ stellar models, also comparing with different options and \MESA\ revisions.

\clearpage

\section{Physical ingredients}
\label{sec:tools}

\subsection{Stellar Evolution code -- MESA}
\label{sec:mesa}

In this work we present eleven AGB stellar models with initial mass 2 and 3\msun\ and initial metallicity Z = 0.01 and Z = 0.02.
Their main features are given in \tab{tab:model_name}\, \ref{tab:model_prop} and \ref{tab:m3z2m2_TPprop}, and they will be discussed in detail in \S \ref{mesa-mppnp:calc}.
These models were computed using the stellar code \MESA\ \citep[MESA revision 4219,][]{mesa}.

The solar distribution used as a reference is given by \cite{grevesse:93}.
The CO-enhanced opacities are used throughout the calculations, using OPAL tables \citep{Iglesias:96}.
For lower temperatures, we adopt the corresponding opacities from \cite{ferguson:05}.
We use the Reimers formula \citep{reimers:75} with $\eta_{R}=0.5$ for the mass loss up to the end of the RGB phase.
Along the AGB phase we use instead the
\citet{bloecker:95a} formula with $\eta_{B}=0.01$ for the O-rich phase, $\eta_{B}=0.04$ and $\eta_{B}=0.08$ for the 2 and 3 \msun\ models respectively after the TDU event that makes the surface C/O ratio larger than 1.15. 
This choice is motivated by observational constraints, as for example the maximum level of C enhancement seen in C-rich
stars and planetary nebulae \citep{herwig:05}, as well as by hydrodynamics simulations
investigating mass loss rates in C-rich giants \citep{mattsson:11}.
For the simulations the \MESA\ nuclear network \emph{agb.net} is used, 
including the pp chains, the CNO tri-cycle, the
triple-$\alpha$ and the $\alpha$-capture reactions
\czw($\alpha$, $\gamma$)\ose, \nvi($\alpha$,$\gamma$)\fac(e$^+$,$\nu$)\oac, \oac($\alpha$,$\gamma$)\nezw, \cdr($\alpha$, n)\ose\ and
\fne($\alpha$, p)\nezw. We use the NACRE \citep{angulo:99} reaction rate compilation for most reactions. 
For the \czw\ ($\alpha$,$\gamma$)\ose\ we adopt the rate by \citep{kunz:02}, \nvi(p, $\gamma$)\ofu\ is by
\citep{imbriani:04} and the triple-$\alpha$ by \citep{fynbo:05}.
Convective mixing follows the standard mixing length theory \citep{vitense:53,boehm-vitense:58}
taking into account also CBM treatment.

\MESA\ provides the exponential convective boundary mixing model of \citet{freytag:96} and \citet{herwig:00} 

\beq
D_\mathrm{CBM}(z) = D_{0}  \exp^{-2z/f_1H_{P0}}
\eeq

where 
$z$ is the distance in the radiative layer away from the Schwarzschild boundary.
The term  
$f_\mathrm{1} H_\mathrm{P0}$  is the scale height of the \ov\ regime.



$D_\mathrm{0}$ and $H_\mathrm{P0}$ are the diffusion coefficient and the 
pressure scale height at the convective boundary respectively.
This model describes the rapid decrease of the mixing efficiency at the convective boundary observed in hydrodynamic simulations of efficient, adiabatic convection in the deep stellar interior \citep[e.g.\ He07,][]{herwig:06a,woodward:15}.
He07 reported that mixing below the PDCZ according to their hydrodynamic simulations is best described by combining this initial decay of the mixing efficiency with a second, shallower exponential diffusion profile. \MESA\ lets this second decay to start as soon as the mixing coefficient drops under a value $D_\mathrm{2}$ given by
\beq
D_2 = D_0  \exp^{-2z_2/(f_2H_{P0})}.
\eeq
with an e-folding distance
$f_\mathrm{2} H_\mathrm{P0}$, that is adopted for distances z $>$ $z_\mathrm{2}$.

Therefore, for z $>$ $z_\mathrm{2}$: 
\beq\
D_\mathrm{CBM}(z) = D_2  \exp^{-2(z-z_2)/(f_2H_{P0})}
\eeq\

Hydrodynamic simulations show that the exponential decay starts before reaching the 
Schwarzschild boundary. In MESA the switch from convective mixing to overshooting happens at a distance $f_\mathrm{0} H_\mathrm{P0}$ from the estimated location 
of the Schwarzschild boundary, where $H_\mathrm{P0}$ is the pressure scale height at that location. in this paper, we always assume that $f_\mathrm{0}$ = $f_\mathrm{1}$.



\begin{figure}[htbp]
\begin{center}
\includegraphics[scale=0.7]{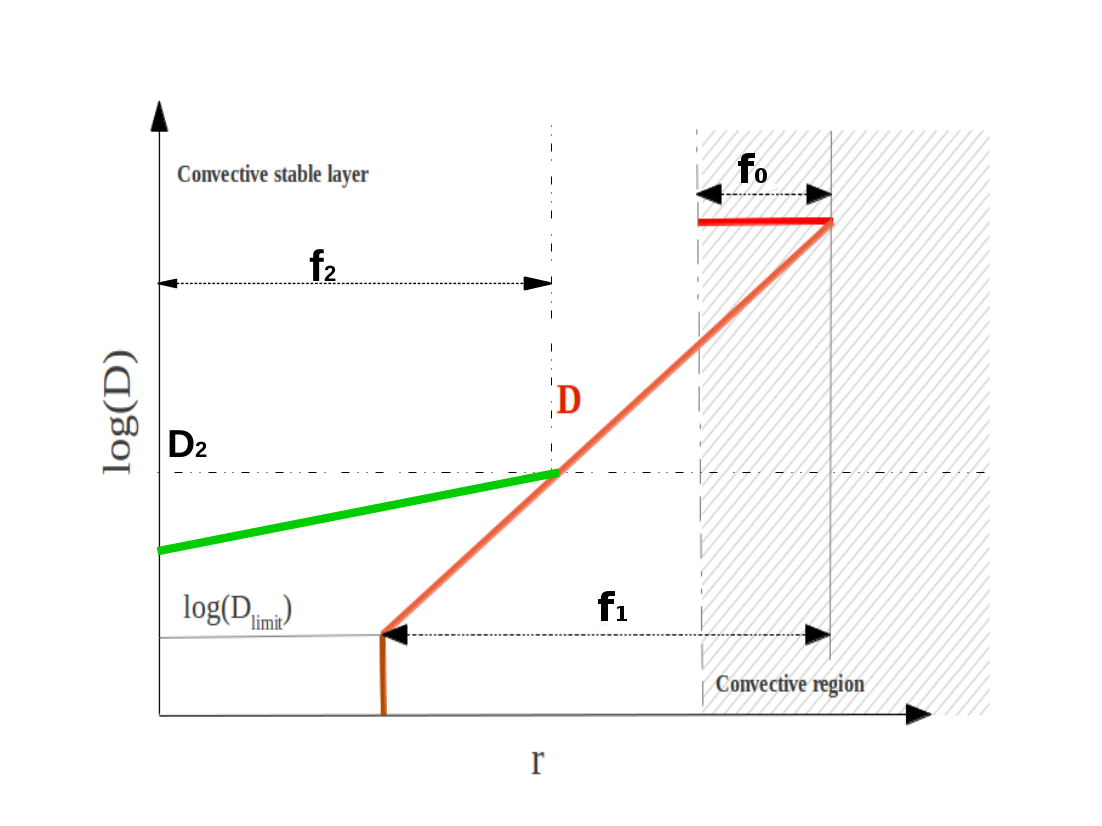}
\end{center}
\caption{Schematic description of the double-exponential CBM applied in this work. The red line is the standard overshooting mixing coefficient profile following the single-exponential decay. This profile is dominated by a single '$f_\mathrm{1}$' parameter which determines the slope of the mixing profile: the lower the 'f' value, the steeper the profile is. In order to take into account IGW, in this work we apply a second, slower, decreasing profile (green line) that becomes more relevant than the first one as soon as the mixing coefficient is equal or lower that a '$D_\mathrm{2}$' value, the slope of which is determined by the '$f_\mathrm{2}$' parameter. Check the text for the relation between D and and all the CBM parameters.}
\label{CBM:schematic}
\end{figure}

During the pre-AGB phase the default overshooting, with a single-exponential decay of the diffusion coefficient in the radiative layer as described in \citet{herwig:00}, is applied.  
  A single-exponential decay is also used to account the CBM at the top of the PDCZ, using a value $f$=0.014. This low value is constrained by the increase in entropy across
the hydrogen-burning shell and is expected to have an impact on nucleosynthesis only at much lower metallicity, around Z=0.0001 \citep{fujimoto:00,stancliffe:11,herwig:05}.
On the other hand, for the AGB phase in this work we adopt a three parameter CBM model with two exponential decay regions characterized $f_\mathrm{1}$ and $f_\mathrm{2}$ while $D_\mathrm{2}$ defines the boundary between the two regions. These three parameters are inputs to the CBM model in  \MESA\ in order to determine the mixing profile at the convective boundary.
A schematic description of this formalism is given in \fig{CBM:schematic}.
This CBM scheme is only applied during the AGB phase, since the mixing that our CBM model represents has been specifically studied in this phase. Model parameters for CBM at the bottom of the PDCZ and at the bottom of the convective envelope during the TDU are given in \tab{tab:model_name}\footnote{The inlist files and any additional information to reproduce our results are provided on \url{htp://www.mesastars.org} and in \url{data.nugridstars.org}}. 
The model parameters $f_\mathrm{1}$, $f_\mathrm{2}$ and $D_\mathrm{2}$ at each of these two convective boundaries are taken from He07 and from theoretical work by \citet{denissenkov:03}. For the PDCZ, He07 extracted the following values as upper limits: $f_\mathrm{1}$=0.01, $D_\mathrm{2}$=10$^{5}$cm$^{2}$s$^{-1}$, $f_\mathrm{2}$=0.14. 

\begin{figure}[htbp]
\begin{center}
\includegraphics{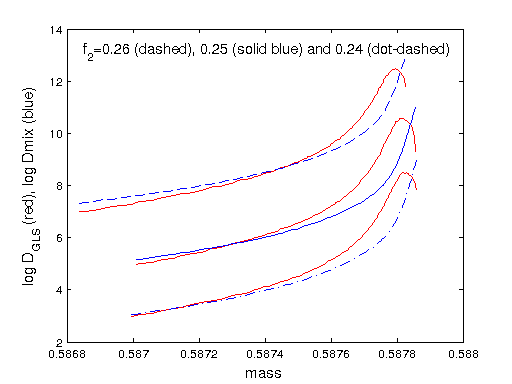}
\end{center}
\caption{Comparison between the diffusion coefficient profile calculated using the GLS prescription for the IGW mixing from De03
(the middle red curve) and the one derived for the CBM with the parameterization used in this work (the solid blue curve).
The dashed and dot-dashed blue curves with their adjacent red curves show comparisons for the cases of
$f_2 = 0.26$ and $f_2 = 0.24$. They are artificially shifted along the vertical axis by $\Delta\log D = 2$ 
up and down relative to the standard case of $f_2 = 0.25$. The bump on the $\log D_\mathrm{GLS}$ profile near 
the convective boundary is produced by a fast increase of the buoyancy frequency $N$ accompanied by a rapid decrease of
the thermal diffusivity $K$ with depth and by the fact that $D_\mathrm{GLS}\propto NK$ (eq. 15 in De03). } 
\label{igw:comp}
\end{figure}

Concerning the bottom of convective envelope during TDU, we chose $f_\mathrm{2}$ to match the mixing profile of IGWs derived by De03, and $D_\mathrm{2}$ to match the maximum of the IGW profile modelling the rapid decay of our mixing coefficient profile through a rapid decay across the convective boundary using a small $f_\mathrm{1}$. 
In this way our CBM model represent mixing due to IGWs, and this is the physical process through which in our models the \cdr-pocket forms. 
  The results are shown in \fig{igw:comp}, where $f_\mathrm{2}$=0.24 is the minimum value able to fit the mixing coefficient for IGW.
The curves obtained with $f_\mathrm{2}$=0.25 and $f_\mathrm{2}$=0.26 also well reproduce De03 results for values closer to \cdr-pocket formation regime (10$^{6}$ $\le$ $D_\mathrm{0}$  $\le$ 10$^{8}$).
In this work we used as default $f_\mathrm{1}$=0.014, $D_\mathrm{2}$=10$^{11}$cm$^{2}$s$^{-1}$, $f_\mathrm{2}$=0.25 (see also \tab{tab:model_name}).
On the other hand, the f$_1$ parameter only marginally affects the size of the \cdr-pocket.
In general, by increasing(decreasing)  $f_\mathrm{1}$ the position of the \cdr-pocket is shifted downward(upward) in the He intershell layers. 
This parameter may affect instead the overall TDU efficiency, and thus the amount of C and \spr\ material dredged-up to the surface of the AGB star. 
As previously said, we use  $f_\mathrm{1}$=0.014 as default, consistently with the exponential decay parameter used during the AGB interpulse phase in Pi13.
The robustness of these choices have been tested,
see \S \ref{mesa-mppnp:calc}.
In \tab{tab:model_name} the 
\emph{clipping} column is given, 
where by \emph{clipping} we mean the limitation of the mixing length to the length of the convection zone which is adopted from \MESA\ revision 3713 onwards. 
Therefore, the stellar models M3.z2m2.he07, M2.z2m2.he07, M3.z1m2.he07 and M2.z1m2.he07 in \tab{tab:model_name}, 
are calculated by using \MESA\ rev. 4219 but without \emph{clipping}. 

We recommend as best \MESA\ simulation setup the one used in he07 models, compared to the 
\emph{clipping} models, although the final nucleosynthesis products are similar. 
This point is discussed in detail in the Appendix.


For the first time, we explore the effect of mixing due to molecular diffusion. 
Such mixing may dilute the \cdr-pocket with \nvi\ from above during the long inter-pulse period.
We assume that the molecular diffusivity is equal to the molecular viscosity, because both of them are proportional to a product of
the mean free path and mean velocity of the same particles. On the contrary, we do not consider the radiative viscosity as a component of the microscopic diffusivity, because it describes the exchange of momentum between photons and particles, therefore it is proportional to the photon mean free path and the speed of light. The default MESA revision used for this work allows to include radiative viscosity as microscopic diffusion term, according to \citet{morel:02}. For this work, also according to \cite{alecian:05}, we consider the molecular viscosity term, using the following expression \citep[][]{spitzer:62}:
%
\beq\
\nu_\mathrm{mol} = 2.21\times 10^{-15}\,(1+7X)\,\frac{T^{5/2}\times A^{1/2}}{\rho \times Z^{4} \times Log\Lambda}.
\eeq\
where $\Lambda$ is the Coulomb integral, with value ranging from 15 to 40 depending on the composition of the stellar layers. 
With the present implementation, the impact of molecular diffusion on final surface elemental abundances is $\lesssim$ 5\%. On the contrary, the impact on $s$-process nucleosynthesis is severe if the controversial implementation from \citet{morel:02} is adopted, strongly increasing the \nvi\ diffusion into the \isotope[13]{C} pocket and completely suppressing the s-process production by the \isotope[13]{C}($\alpha$,n)\isotope[16]{O} neutron source.
While we may rule out the implementation by \cite{morel:02} \citep[for more details we refer to the discussion in][]{alecian:05}, the role of molecular diffusion during the AGB phase deserves further investigation.


\subsection{Nucleosynthesis Post-Processing Calculations -- MPPNP}
\label{sec:mppnp}

For the \spr\ nucleosynthesis we used the multi-zone post-processing code \mppnp\ \citep[][hereafter Pi13]{pignatari:13}.
The stellar structure evolution data for all zones at all time steps are saved, and then processed with the \mppnp\ code.
The network can include up to about 5000 isotopes between H and Bi, and more than 50000
nuclear reactions. A dynamical network defines the number of species and reactions
considered in each zone individually, based on the strength of nucleosynthesis flows producing and destroying each isotope.
Nuclear reaction rates are collected from different data sources, including the
European NACRE compilation \citep[][]{angulo:99} and
\cite{iliadis:01}, or more recent if available \citep[e.g.,][]{fynbo:05,kunz:02,imbriani:05}. For the \isotope[13]{C}($\alpha$,n)\isotope[16]{O} and \isotope[22]{Ne}($\alpha$,n)\isotope[25]{Mg} rates we use \cite{heil:08} and \cite{jaeger:01}, respectively. For experimental neutron capture rates of stable isotopes and available rates for unstable isotopes we
use mostly the Kadonis compilation version 0.3 (see \citet{dillmann:14} and \url{http://www.kadonis.org}).  Exceptions relevant for this work are the neutron-capture cross sections of \isotope[90,92,93,94,95,96]{Zr}: we used instead the new rates by \cite{lugaro:14}, calculated based on recent experimental measurements. 
For stellar $\beta$-decay and electron-capture weak rates we use \cite{fuller:85}, \cite{oda:94}, \cite{langanke:00} and \cite{goriely:99}, according to the mass region. Rates are taken from JINA reaclib library \citep[][]{cyburt:10} if not available from one of the resources mentioned above.

\clearpage

\section{Stellar models  - CBM in the He intershell and the \cdr-pocket}
\label{mesa-mppnp:calc}

In this chapter we summarize the relevant CBM features adopted in our simulations for the AGB evolution at the He-intershell boundaries, and we present the main properties of the AGB models, 
which are listed in \tab{tab:model_name}.   In this table, model names contain the following information: The initial mass is given by the number following the initial capital \emph{M}.
Initial metallicity is given by what follows the $z$. Considering M3.z2m2 as an example, \emph{M3} means that this is a 3 \msun\ model, \emph{z2m2} is to be read as Z=2$\times$10$^{-2}$, 
where \emph{m2} means \emph{minus two} referring to the exponent to be applied.


\subsection{CBM at the bottom of the convective TP}
\label{int:nuc}


Based on hydrodynamics simulations of the AGB He flash, He07 suggested the presence of CBM at the bottom of the PDCZ zone. 
Furthermore, He07 obtained that convective motions induce a rich spectrum of IGW in the neighboring stable layers.
For the stellar models M3.z2m2.he07, M2.z2m2.he07, M3.z1m2.he07 and M2.z1m2.he07 we adopt the CBM parameterization by He07. For the analogous models without the He07 setup, we use instead a larger f$_1$ value, obtaining similar He, C and O abundances in the He intershell.
For instance, the M3.z2m2.he07 model shows a final He, C and O of 55\%, 29\% and 16\% respectively, compared to 48\%, 31\% and 13\% of model M3.z2m2. 

We do not present here AGB models exploring the D$_2$ and f$_2$ parameters. The parameter f$_2$ has a negligible impact on the evolution and composition of the He intershell with D$_2$=10$^{5}$.  
The parameters D$_2$ and f$_2$ become relevant only for D$_2$$\gtrsim$10$^{7}$ cm$^2$s$^{-1}$, two orders of magnitude higher than the indications by He07 results.
Therefore, at the bottom of the PDCZ a single exponential-decay parameterization would be enough to include CBM in 1D stellar models.

\subsection{CBM at the bottom of the convective envelope during TDU: the formation of the \cdr-pocket}
\label{c:poc}

The CBM below the convective envelope during each TDU 
all along the AGB phase causes a decreasing profile of protons in the He-intershell material, due to a finite amount of proton diffusion from the convective envelope into the He intershell. 
This profile is the product of the physics mechanisms triggering the CBM, and will directly impact on crucial features of the radiative \cdr-pocket.
The value of the H/Y(\czw) ratio (where Y(\czw) is the molar fraction of
\czw\ in the He intershell) defines the boundary 
between the \cdr-pocket and the \nvi-pocket above. The proton capture rates involved in the production and in the depletion of \cdr\ in these stellar radiative layers and the amount of \czw\ define where the condition X(\cdr)$>$X(\nvi) is satisfied \citep[e.g.,][]{lugaro:03a,goriely:04,cristallo:09}.
The \nvi-pocket is also \cdr\ rich, but the neutrons made by the \cdr($\alpha$,n)\ose\ reaction are mostly captured by the poison reaction \nvi(n,p)\cvi, thus drastically reducing the \spr\ efficiency \citep[e.g.,][]{gallino:98,cristallo:15}. 
With our nuclear-reaction rates choice, 
the upper boundary of the \cdr-pocket is given by H/Y(\czw) $\sim$ 0.4. 
During the TDU, this ratio is obtained for a mixing coefficient D $\sim$ 10$^{7}$cm$^{2}$s$^{-1}$.
See for comparison with other models the discussion in \cite{lugaro:03a}, \cite{goriely:04} and \cite{cristallo:09}.
For H/Y(\czw) $\lesssim$ 0.4 the \cdr-pocket forms, with a decreasing abundance of \cdr\ moving toward the center of the star. The \spr\ production in He-intershell layers with concentration of \cdr\ $\lesssim$10$^{-3}$ becomes negligible. 
The size of the \cdr-pocket (i.e. the \cdr-rich mass region with X(\cdr)$>$X(\nvi) and X(\cdr)$>$10$^{-3}$) is crucial for the \spr\ production. 

\begin{figure}[htbp]
\begin{center}
\includegraphics[scale=0.3]{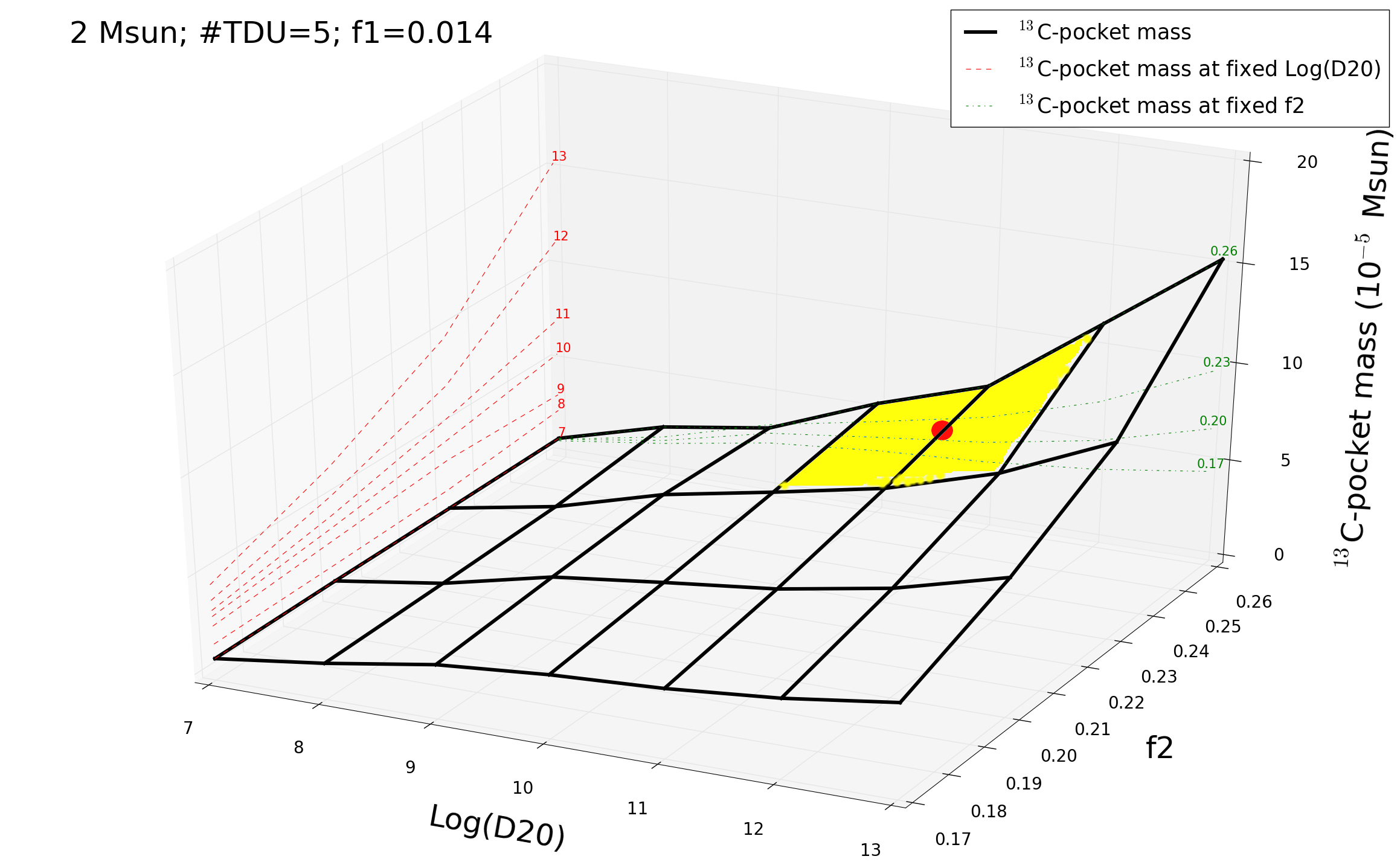}
\end{center}
\caption{\cdr-pocket size as a function of the CBM parameters parameters associated to the 5th TDU event. The red dot represents the \cdr-pocket size obtained by our best fit of De03 results (see \fig{igw:comp}). The yellow area provide an estimation of the uncertainty deriving these parameters (see text for details).}
\label{3d:5}
\end{figure}

We analyzed the impact of the D$_2$ and f$_2$ parameters on the size of the \cdr-pocket.
In \fig{3d:5} the \cdr-pocket size resulting from the model M2.z2m2 is shown as a function of D$_2$ and f$_2$ after the 5th TDU. 
In order to produce the results of this test, we have recalculated the stellar structure from the end of the previous convective TP until the formation of the \cdr-pocket. In these calculations,
we explored the parameter range $10^{7}$$\lesssim$D$_2$$\lesssim$ $10^{13}$, and 0.17$\lesssim$f$_2$$\lesssim$0.29. All the other stellar parameters were not changed.
The typical \cdr-pocket size obtained by using the IGW value from  De03 is $\sim$ 7-8 $\times$10$^{-5}$ M$_{\odot}$.
The size of the \cdr-pocket tends to increase with increasing of D$_2$ and f$_2$, up to a size of 1.5 $\times$10$^{-4}$ M$_{\odot}$ with the largest D$_2$ and f$_2$ values . 
The colored area represents the range of f$_2$ still giving an acceptable fitting of De03 calculations, and of D$_2$ assuming an uncertainty of one order of magnitude.
Within this range, the \cdr-pocket size is varying between 4$\times$10$^{-5}$ and 1.2$\times$10$^{-4}$M$_{\odot}$. We added two AGB models to our set, 
M2.z2m2.hCBM and M3.z1m2.hCBM (\tab{tab:model_name}), with D$_2$=10$^{12}$cm$^{2}$s$^{-1}$ and f$_2$=0.27 where the impact of a larger \cdr-pocket within the mentioned uncertainty range is explored. 
The same investigation has been performed at the 3rd TDU of the same model, giving consistent results. 

\begin{figure}
\centering
\resizebox{10.3cm}{!}{\rotatebox{0}{\includegraphics{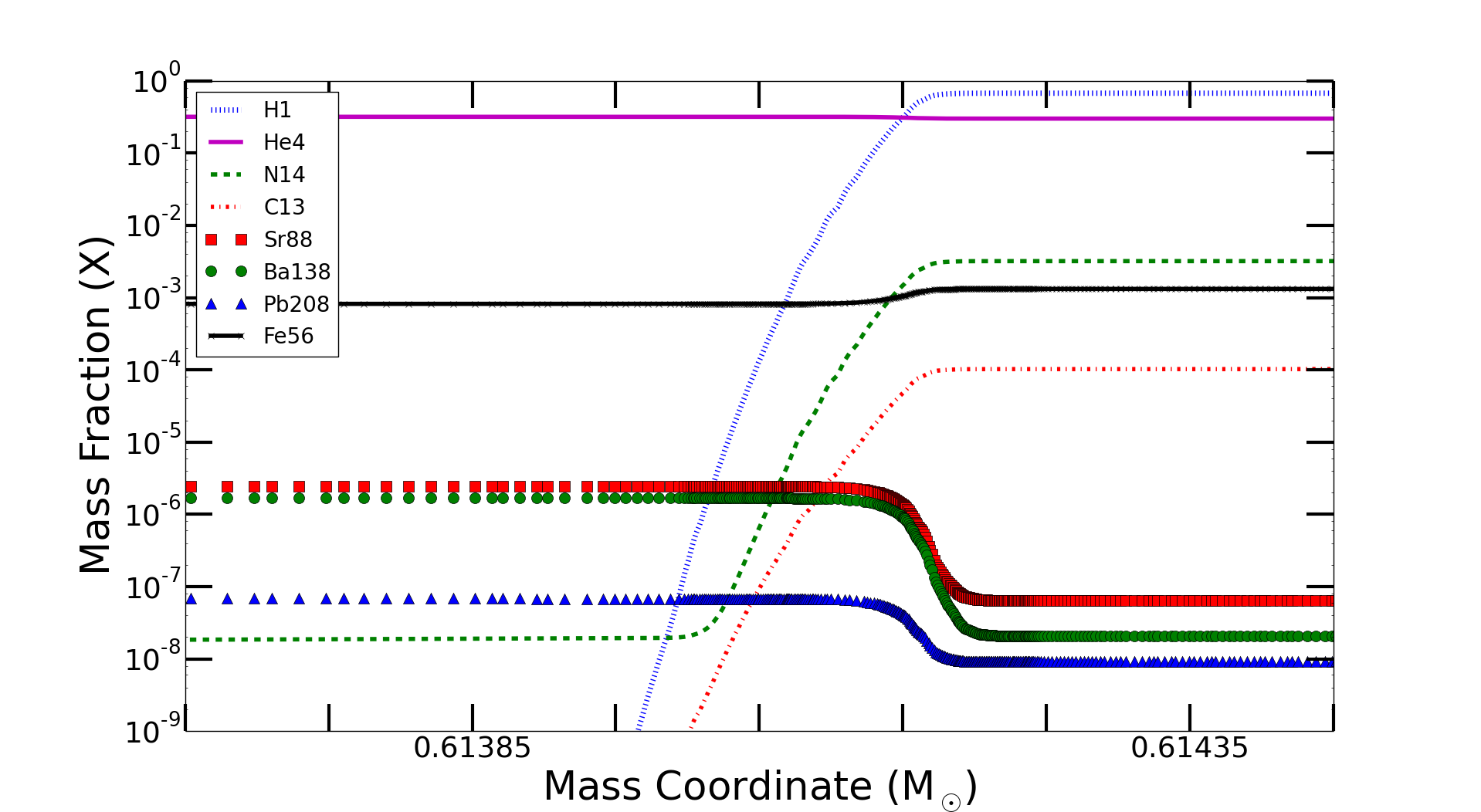}}}
\resizebox{10.3cm}{!}{\rotatebox{0}{\includegraphics{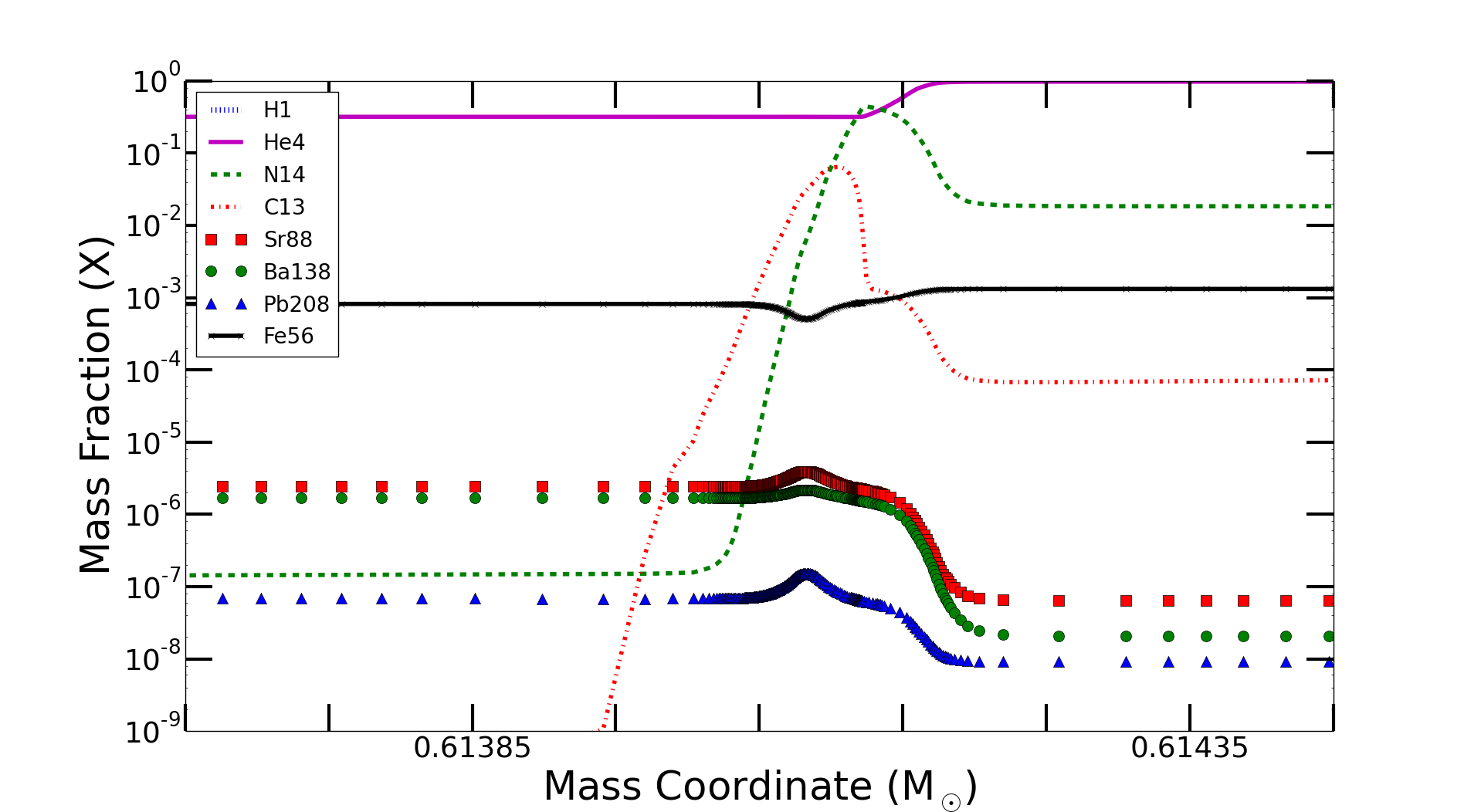}}}
\resizebox{10.3cm}{!}{\rotatebox{0}{\includegraphics{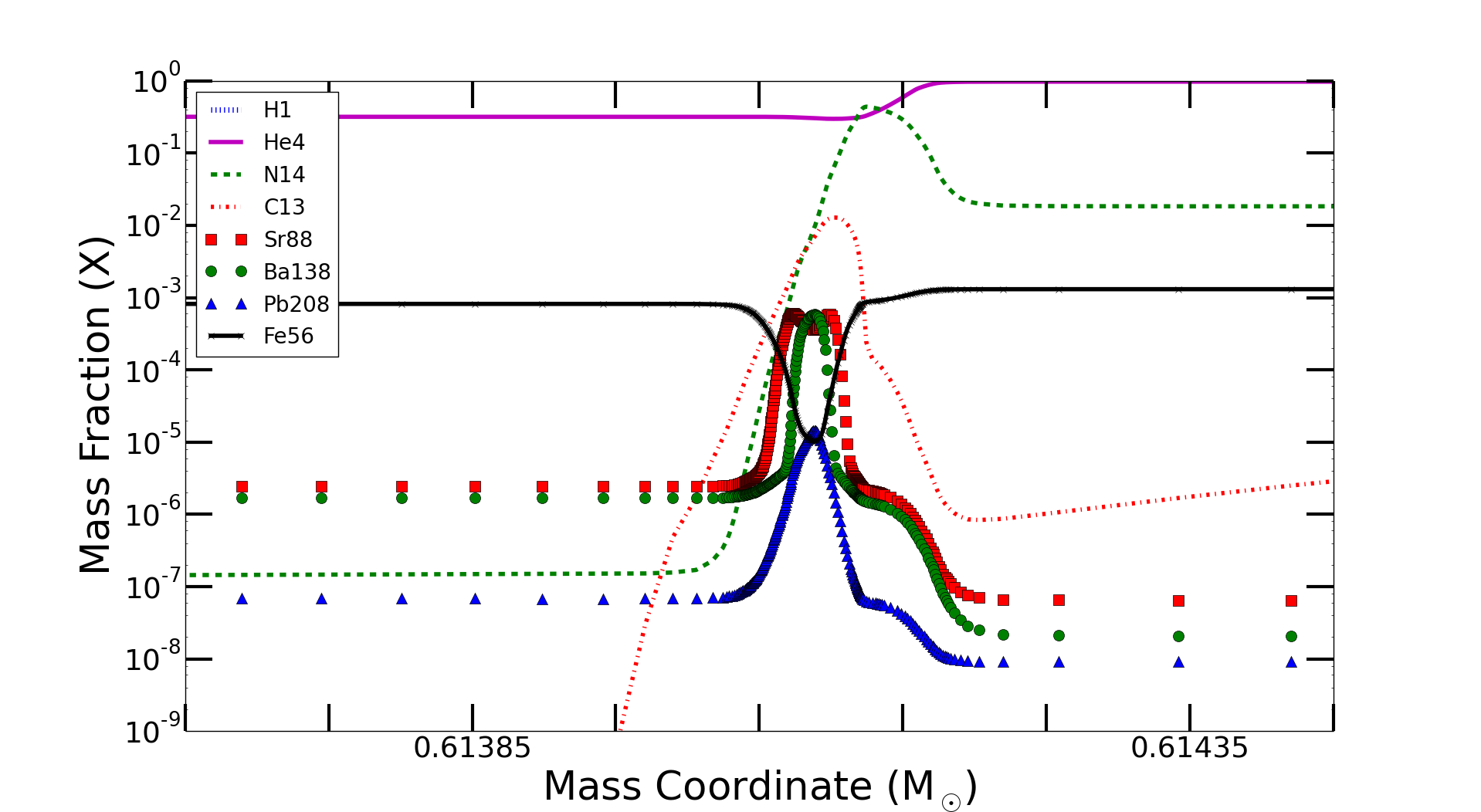}}}
\caption{Three different steps of \cdr-pocket evolution in M3.Z2m2 are shown. We provide the abundances of H, \hevi, \ose, \cdr and \nvi, 
\fese, and \spr\ isotopes at the neutron magic peaks N=50 (\srac), N=82 (\baac) and N=126 (\blac). 
The top panel refers to the moment of maximum penetration of the TDU, 
which is followed by the radiative burning of the \cdr-pocket with the consequent neutron release and \spr\ nucleosynthesis (middle and bottom panels).}
\label{fig:c13poc_snuc_form}
\end{figure}

In \fig{fig:c13poc_snuc_form} we report three snapshots of the abundance profiles of indicative species from model M3.z2m2, showing the maximum penetration of H in the He intershell during the 5th TDU, the following \cdr-pocket when the \cdr($\alpha$,n)\ose\ starts to be activated, depleting \isotope[56]{Fe} and making \spr\ species, and close to the end of the AGB interpulse period, when \cdr\ has been consumed. The following convective TP will mix convectively the \spr\ products in the He intershell and the next TDU will enrich the surface with these newly produced heavy elements.

\subsection{AGB stellar models: summary of their main features}
\label{int:agb_models_all}

\begin{figure}
\centering
\resizebox{13.3cm}{!}{\rotatebox{0}{\includegraphics{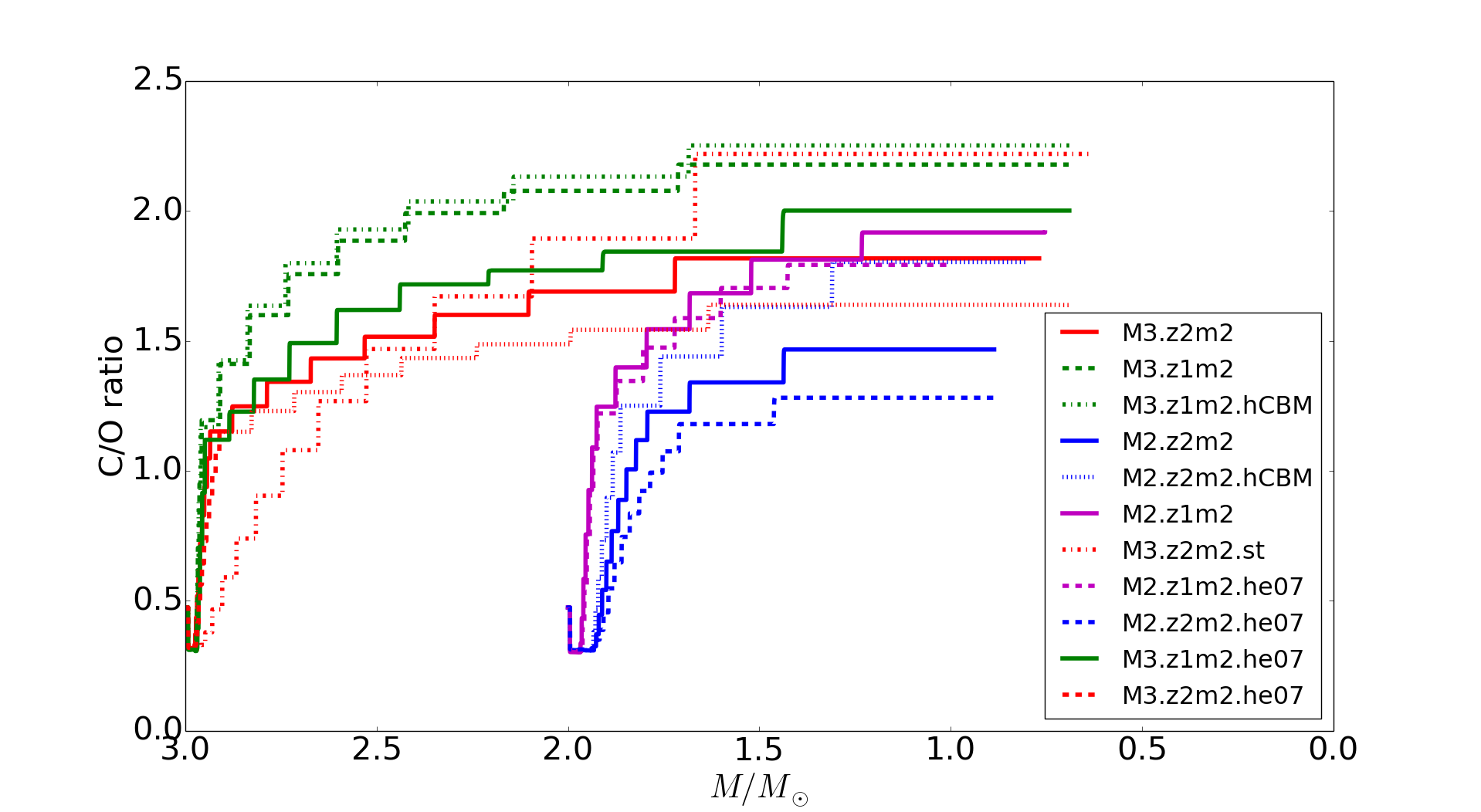}}}
\caption{The evolution of the C/O surface ratio is shown with respect to the total stellar mass for the AGB models indicated. 
}
\label{fig:TAGBprop_up}
\end{figure}

\begin{figure}
\centering
\resizebox{13.3cm}{!}{\rotatebox{0}{\includegraphics{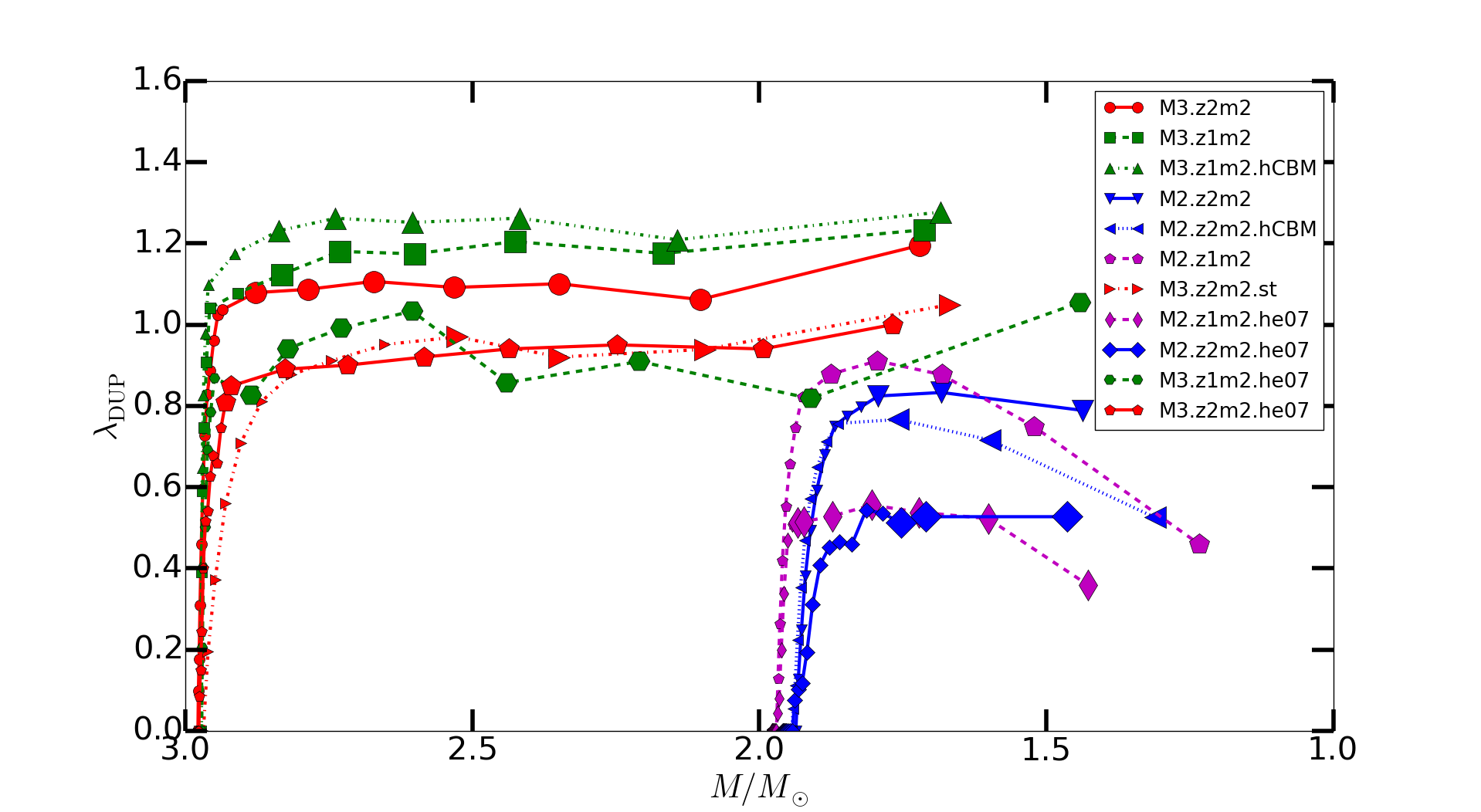}}}
\caption{the $\lambda$$_{DUP}$ parameter is shown with respect to the total stellar mass for the AGB models indicated. Symbols are reported for each convective TP.
}
\label{fig:TAGBprop_middle}
\end{figure}

\begin{figure}
\centering
\resizebox{13.3cm}{!}{\rotatebox{0}{\includegraphics{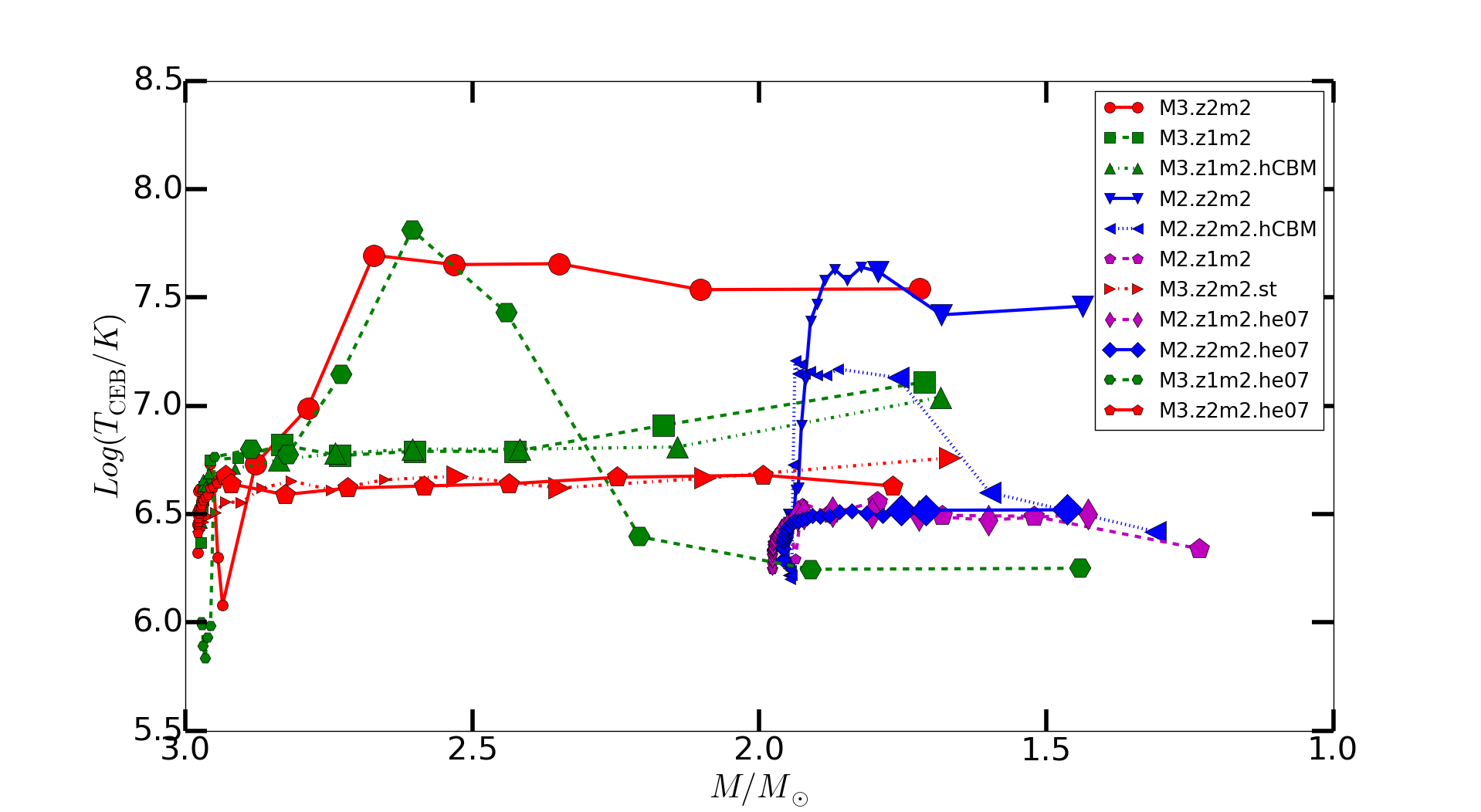}}}
\caption{The temperature at the bottom of the convective envelope during the deepest extend of TDU in logarithmic scale, $T_{CEB}$, is shown with respect to the total stellar mass for the AGB models indicated. Symbols are reported for each convective TP.
}
\label{fig:TAGBprop_down}
\end{figure}

In the previous two sections we have discussed the CBM setup used to calculate the AGB stellar models listed in \tab{tab:model_name}.
The main properties of these AGB models are summarized in \tab{tab:model_prop} and \ref{tab:m3z2m2_TPprop}.
  The number of thermal pulses goes from 13 for model M3.z1m2, to 27 for model M2.z2m2.he07. The model reaching the highest temperature at the bottom of the AGB envelope is M3.z1m2.he07, 
while the coldest model is M2.z1m2.he07. The total mass dredged up goes from 3.243 10$^{-2}$\msun for model M2.z2m2.he07 to 1.298 10$^{-1}$\msun for model M3.z2m2. 
In \fig{fig:TAGBprop_up}, we show the evolution of the C/O ratio at the stellar surface during the AGB evolution.
All these models become C rich at the end of their AGB evolution, and the surface C/O ratio evolves similarly.
The he07 models show a C/O ratio lower by about 0.2, that corresponds to an average departure of 10$\%$ from their corresponding $clipped$ models, 
which is mostly due to a lower $\lambda$ $_{DUP}$ dredge-up parameter during the AGB phase.  
The parameter $\lambda$$_{DUP}$ is shown in \fig{fig:TAGBprop_middle}
and is defined as: 
\begin{equation}
\lambda = \frac{\Delta M_{DUP}}{\Delta M_{H}}
\end{equation}
where $\Delta M_{H}$ is the growth of the H-free core after each TP and $\Delta M_{DUP}$ is the dredged up mass.
As expected we obtain more efficient TDUs (i.e., higher $\lambda$$_{DUP}$) with decreasing of the initial 
metallicity and increasing initial mass \citep[see,][]{lattanzio:89}. The total mass dredged up $M_D$ and the maximum mass dredged up $\Delta M_{Dmax}$ 
increase with initial mass (\tab{tab:model_prop}).
In \fig{fig:TAGBprop_down}, we show the temperature at the bottom of the convective envelope during the deepest extend of TDU ($T_{CEB}$).
  In general, models with Z=0.02 show larger temperatures $T_{CEB}$ compared to models at Z=0.01. 
This is due to the anti-correlation between the largest temperature at the bottom of the He-flash convective zone ($T_{FBOT}$) and $T_{CEB}$: the higher the TP luminosity, 
the more the He intershell will expand causing colder TDUs ($T_{CEB}$ and $T_{FBOT}$ for all the AGB models and all the TPs are provided in \tab{tab:m3z2m2_TPprop}).
We also confirm the strong dependence of the interpulse period with the core-mass as already discussed by \citet{paczynski:74}. 
This is obtained not only along the evolution of single models, but also comparing results between different models. 
The envelope mass is not important for this, 
since our 3 M$_{\odot}$ models have almost the same interpulse period as our 2 M$_{\odot}$ models when core masses the same.
The extension of the different TP episodes reflect the intershell thickness instead, being larger in 2 M$_{\odot}$ models and smaller in 3 M$_{\odot}$ ones
as expected.
Finally, all our models experience a large mass-loss increase as the Bl\"ocker wind coefficient $\eta_{B}$ is artificially increased when the star becomes C-rich, 
 
mimicking in this way the effect of higher opacities in such regime (see discussion in \S \ref{sec:mesa}). 
Another consequence of the higher value of $\eta_{B}$, is the occurrence of a super-wind regime 

after the last TDU event of each model, leading to the loss of a envelope mass ranging from about 0.7 to 1 M$_{\odot}$ and finally leaving the degenerate CO core surrounded by the He-intershell.
In order to simulate the last TPs, we modify the opacity to prevent convergence problems related to the iron opacity peak at the bottom of the envelope.
Indeed, when the star is approaching the end of the TP AGB, close to stripping the envelope from the CO core 
unstable pulsation due to the opacity-mechanism from the Fe-group opacity bump at T around 2$\times$10$^{5}$ K in a zone right-below the surface  set in up.
This can be also seen in large and irregular variations of effective temperature and luminosity in the HR diagram.
This effect was identified by \citet{dziembowski:93} to explain $\beta$ Chepheids pulsations, also determining that a typical solar metal content suffices to account for the pulsation.
Our stellar models calculations manage to advance this stage after several thousand timesteps, eventually with no success.
In order to get through this phase, we confirm that lowering the opacity to prevent the iron bump may help \citep[][]{jeffery:06,lau:12}, but for our purpose this last phase is not important, since the mass loss is so large that none or very little \sprn\ production could still happen before the entire envelope is lost.

\clearpage

\section{Post-processing nucleosynthesis calculations and comparison with observations}
\label{sec:postpro}


In this section we discuss the nucleosynthesis results of our post-processing calculations,
and we compare them with observations and stellar yields from other authors.
The abundances for all the isotopes up to Bi have been calculated using the post-processing tool \mppnp\ (\S \ref{sec:tools}). 
In addition to the stellar models in \tab{tab:model_name}, we performed additional post-processing calculations on the same stellar structures, but using different reaction rate networks. The complete list of these models is given in \tab{tab:model_name_network_test}. 
In particular, we tested the impact of the \isotope[14]{N}(n,p)\isotope[14]{C} reaction rate (models labeled with {\it ntest}, where the default rate is multiplied by a factor of two).
The \isotope[14]{N}(n,p)\isotope[14]{C} is the main neutron poison in the \isotope[13]{C}-pocket. While there are several experimental results beyond 20 keV \citep[][and references therein]{wallner:12}, 
there is only one available so far at energies $\sim$ 8 keV, typical for the \isotope[13]{C}-pocket \citep[][]{koehler:89}. Above 20 keV, independent experiments obtain rates changing within a factor of three.
The Zr neutron capture cross section have been updated by a number of studies in recent years \citep[][and references therein]{tagliente:12,lugaro:14}. In particular, \cite{lugaro:14} provided a new evaluation of the \isotope[95]{Zr}(n,$\gamma$)\isotope[96]{Zr} cross section based on the measurements on neighbor Zr species, which is more than a factor of two lower compared to older rates \citep[e.g.,][]{bao:00}.  This rate is important for the \spr\ branching point at \isotope[95]{Zr}, leading to the production of \isotope[96]{Zr}. Zr isotopic ratios are observed in presolar SiC mainstream grains from AGB stars \citep[][]{barzyk:06}. 
They provide an important diagnostic for the thermodynamics conditions at the bottom of the He-intershell during convective TPs \citep[e.g.,][]{lugaro:03a}. 
Therefore, we have tested the impact of this reaction on the \spr\ Zr products reducing the \isotope[95]{Zr}(n,$\gamma$)\isotope[96]{Zr} rate by a factor of two.

We did not consider in this work the uncertainties of other reaction rates that impact \spr\ nucleosynthesis predictions in AGB stars, such as the \isotope[22]{Ne}($\alpha$,n)\isotope[25]{Mg} \citep[see e.g.,][]{gallino:98,pignatari:05,karakas:06,liu:14b,bisterzo:15}.

\begin{figure}[htbp]
\centering
\resizebox{7.3cm}{!}{\rotatebox{0}{\includegraphics{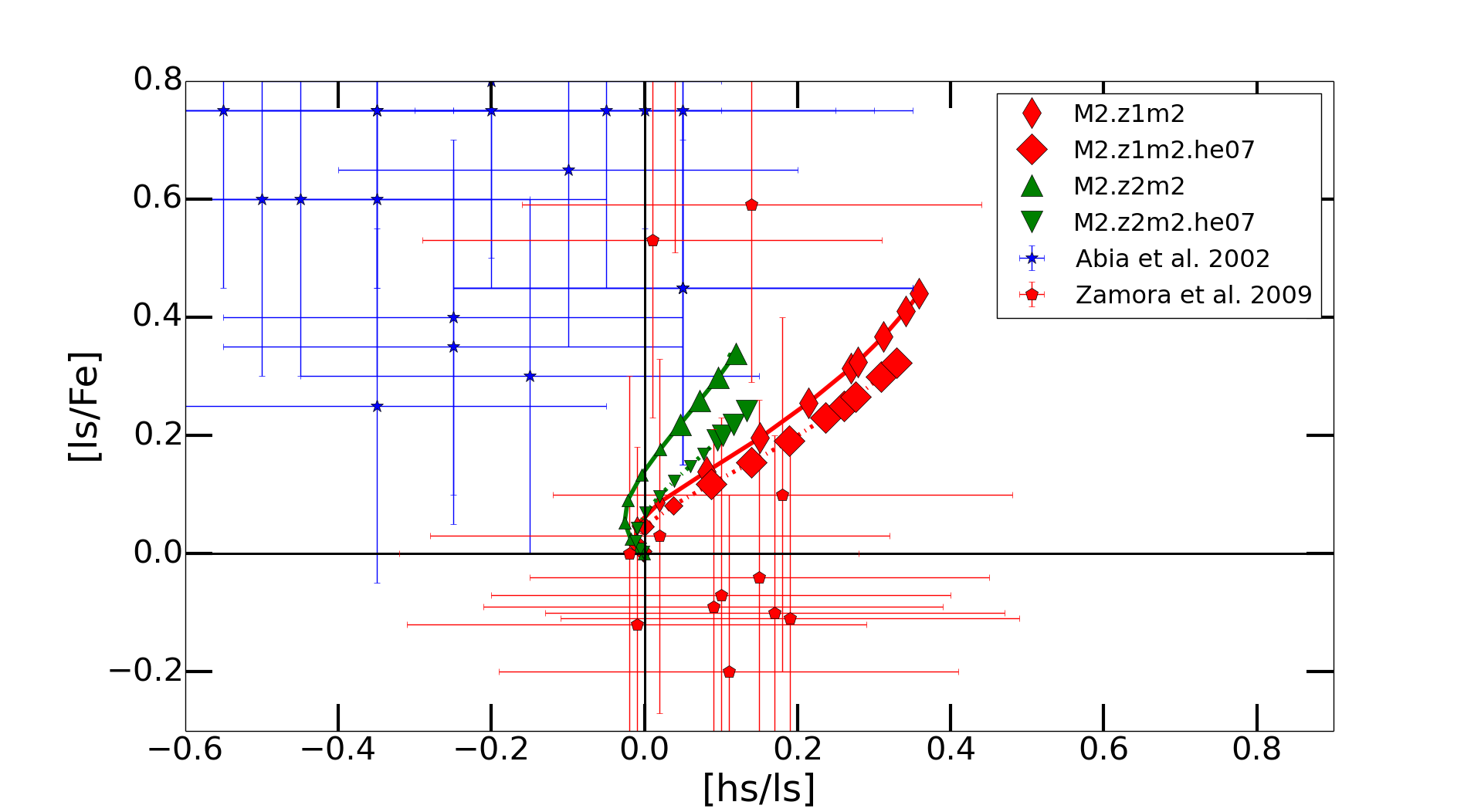}}}
\resizebox{7.3cm}{!}{\rotatebox{0}{\includegraphics{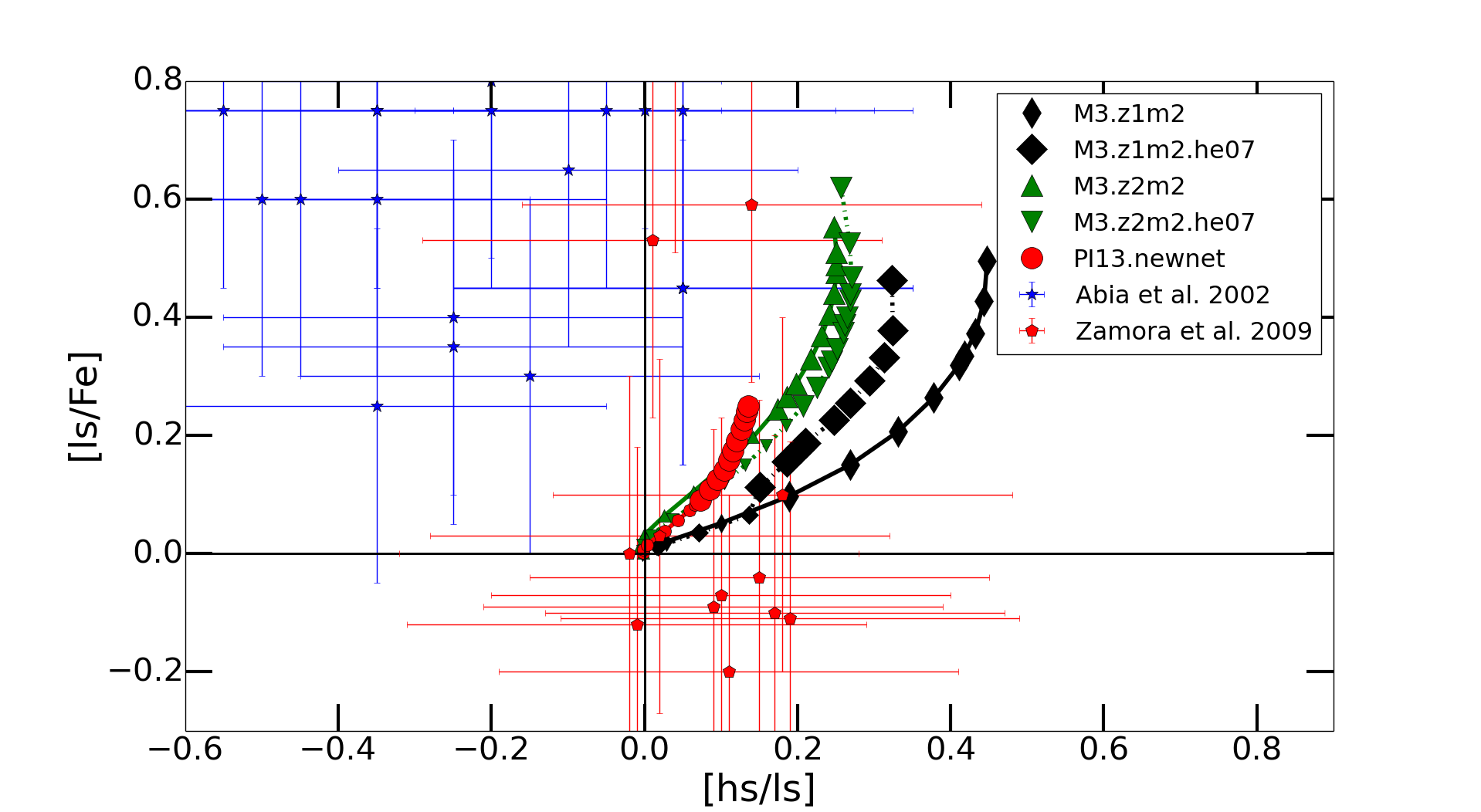}}}
\resizebox{7.3cm}{!}{\rotatebox{0}{\includegraphics{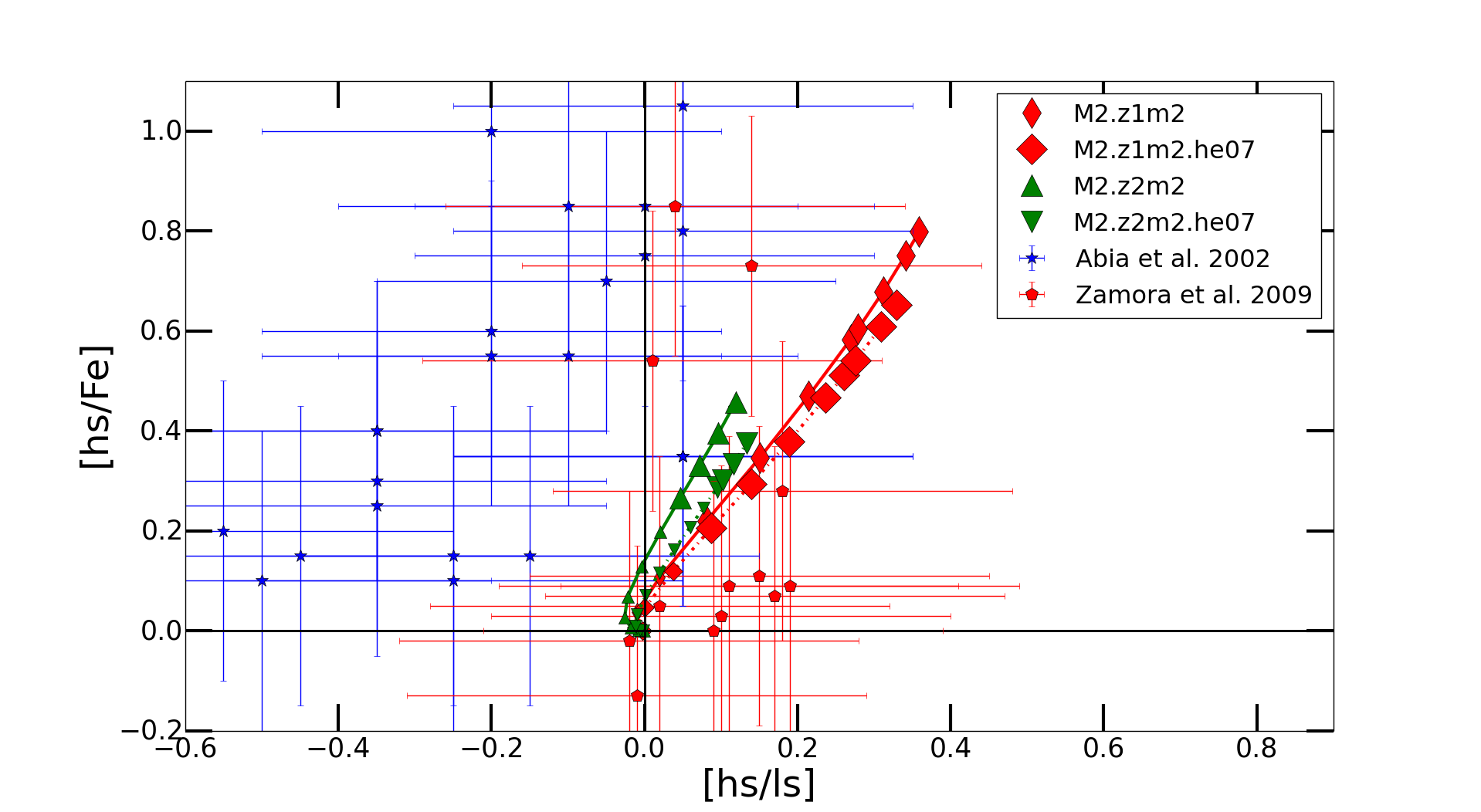}}}
\resizebox{7.3cm}{!}{\rotatebox{0}{\includegraphics{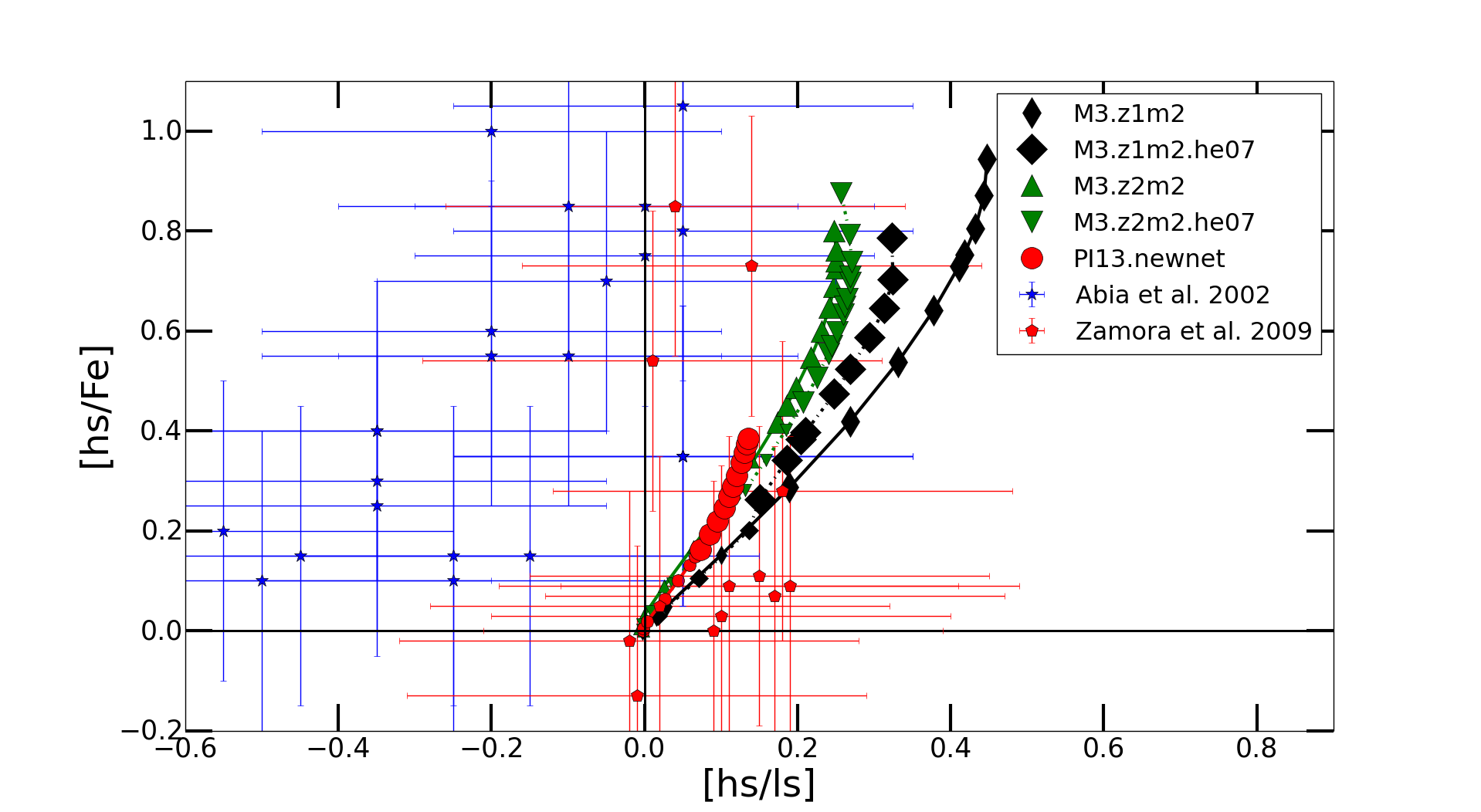}}}
\caption{The evolution of the [ls/Fe], [hs/Fe] and [hs/ls] ratios during the AGB evolution are shown for the models M2.z1m2, M2.z2m2, M2.z1m2.he07 and M2.z2m2.he07 (left panels) and PI13.newnet, M3.z1m2 and M3.z2m2, M3.z1m2.he07 and M3.z2m2.he07 (right panel). Also the comparison with observational data from \citet{abia:02} and \citet{zamora:09} is provided.}
\label{fig:summary_obs_1} 
\end{figure}

\begin{figure}[htbp]
\centering
\resizebox{7.3cm}{!}{\rotatebox{0}{\includegraphics{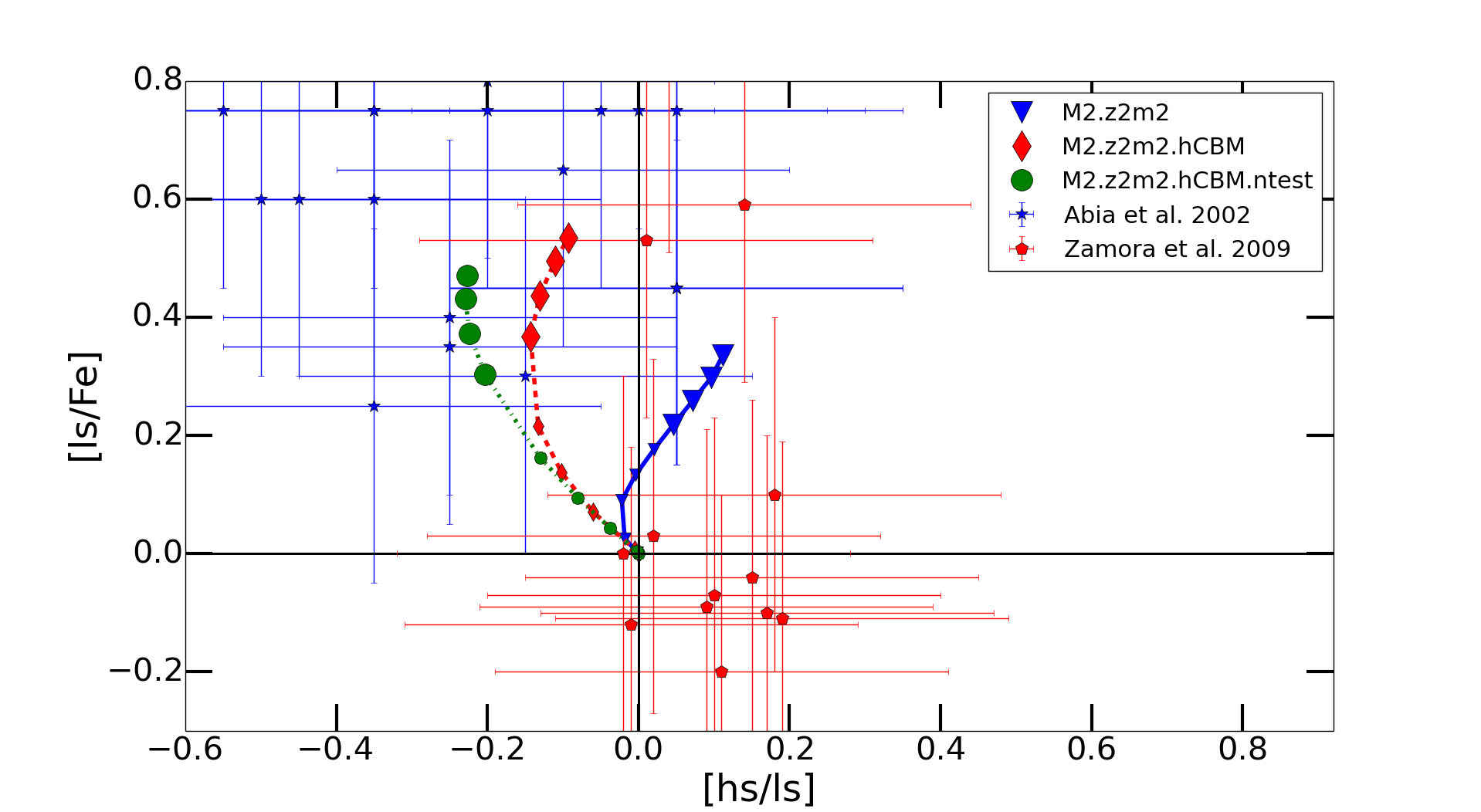}}}
\resizebox{7.3cm}{!}{\rotatebox{0}{\includegraphics{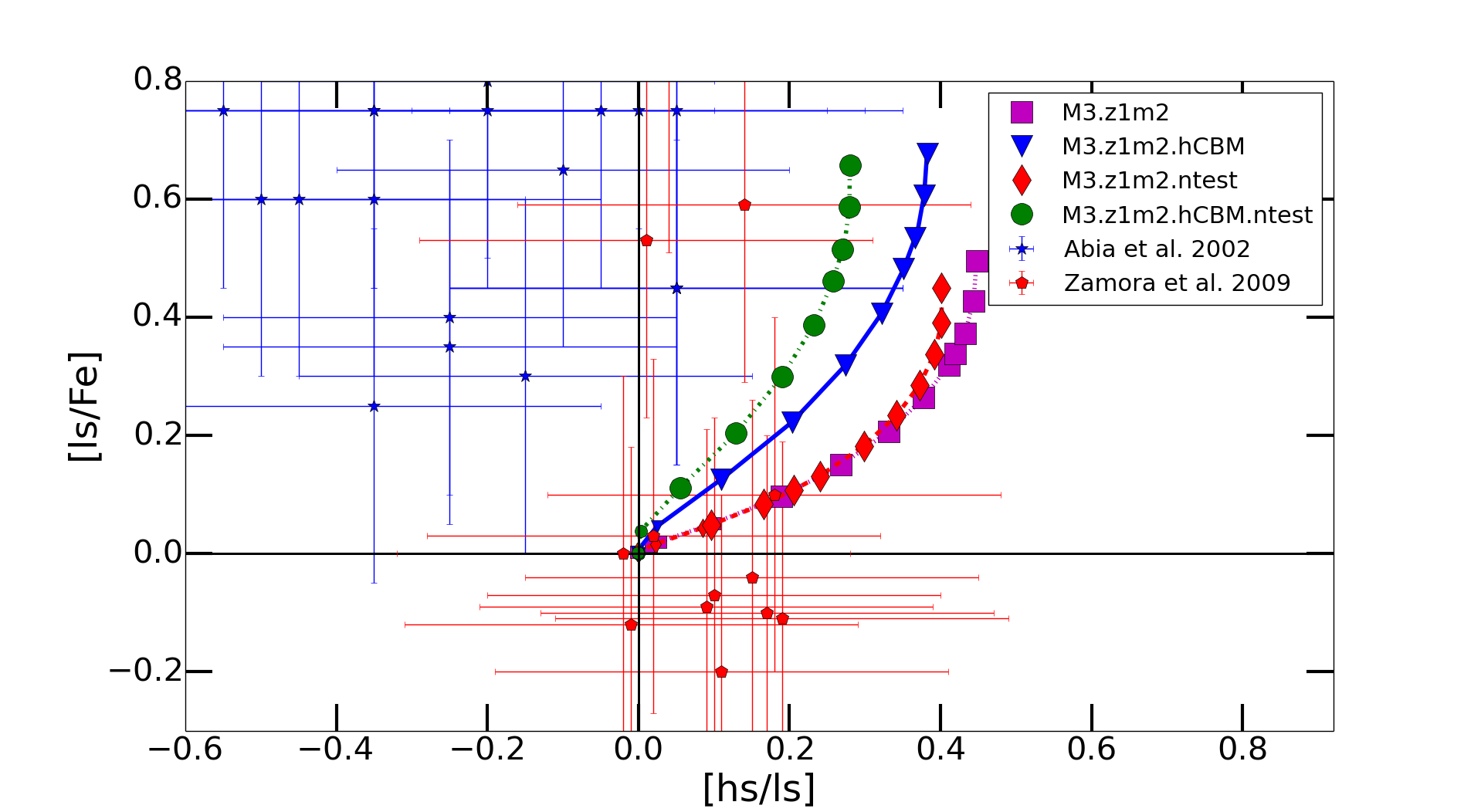}}}
\resizebox{7.3cm}{!}{\rotatebox{0}{\includegraphics{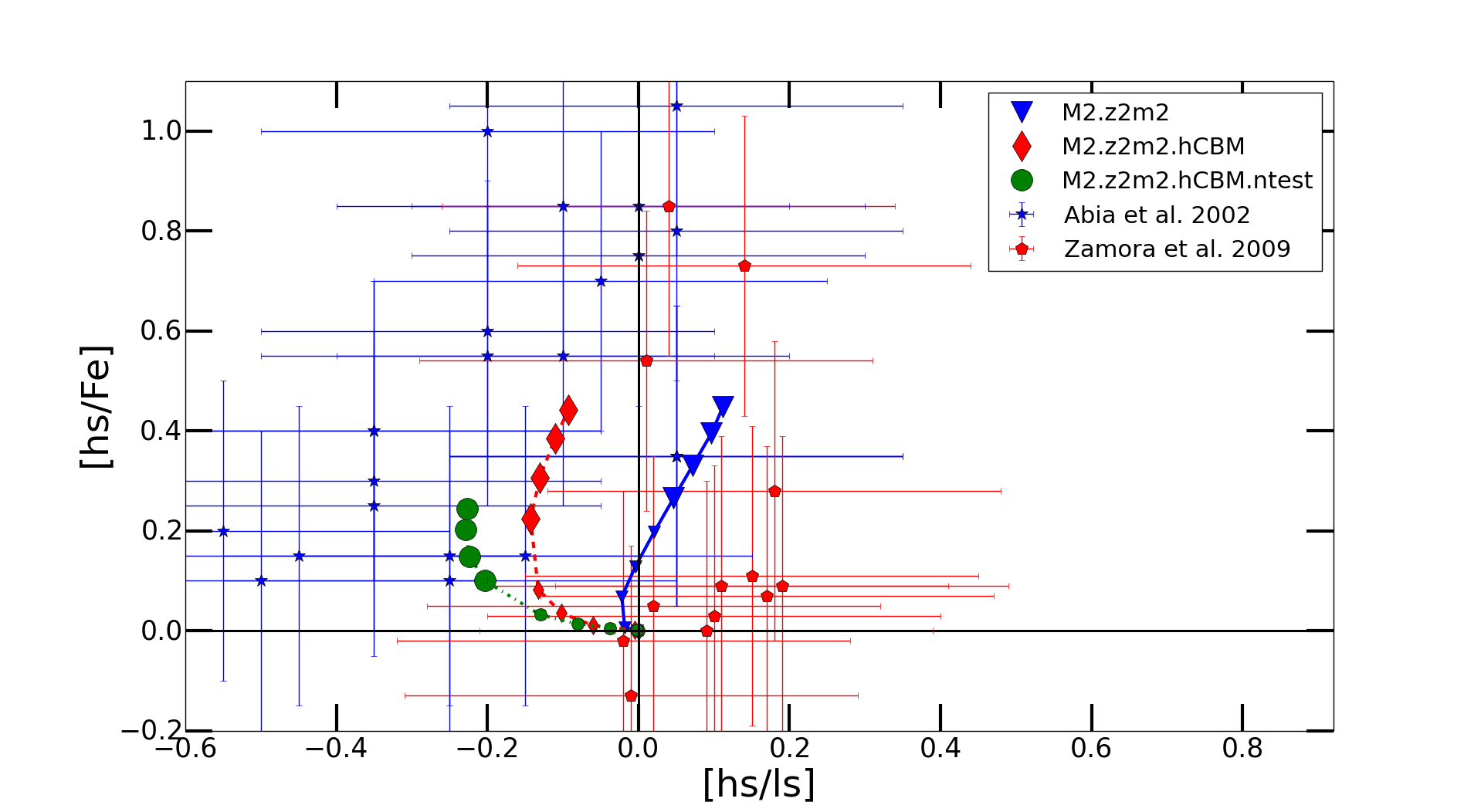}}}
\resizebox{7.3cm}{!}{\rotatebox{0}{\includegraphics{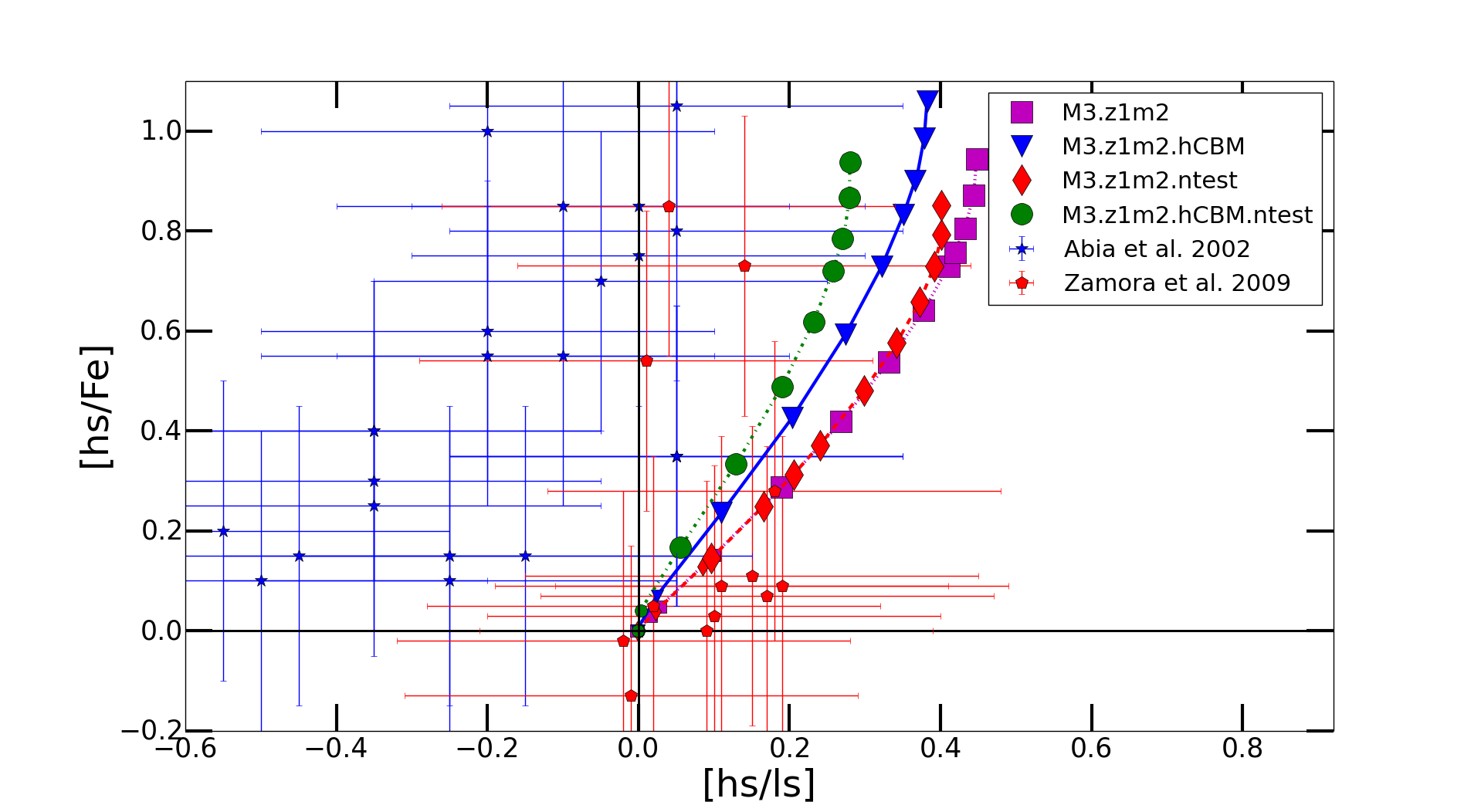}}}
\caption{As in \fig{fig:summary_obs_1}, but the abundances obtained in reference model M2.z2m2 are compared with the models M2.z2m2.hCBM and M2.z2m2.hCBM.ntest; the results of the model M3.z1m2 are compared with the models M3.z1m2.ntest, M3.z1m2.hCBM and M3.z1m2.hCBM.ntest. Also the comparison with observational data from \citet{abia:02} and \citet{zamora:09} is provided.}
\label{fig:summary_obs_2}
\end{figure}

In \S \ref{mesa-mppnp:calc} we described the new CBM parameterization adopted at the boundaries of the He intershell to calculate the AGB stellar models discussed here. 
We have seen from \fig{fig:TAGBprop_up} that all the AGB models become C-rich before the end of the AGB phase, with final 1.4 $\lesssim$ C/O $\lesssim$ 2.4.
In Figs. \ref{fig:summary_obs_1} and \ref{fig:summary_obs_2}, we show the evolution of the \spr\ indices during the AGB evolution \citep[][]{luck:91}   compared to observations of surface abundances of Carbon stars \citep{abia:02,zamora:09}, 
where [ls/Fe] is representative of the surface abundance of \spr\ elements at the neutron shell closure N=50 (ls elements = Sr, Y, Zr), and [hs/Fe] of the elements at N=82 (hs elements = Ba, La, Nd, Sm). 
The ratio [hs/ls] indicates the relative \spr\ production at the two \spr\ neutron-magic peaks, independently from the absolute production of these elements \citep[e.g.][]{busso:01}.
Compared to the model Pi13.newnet, the model M3.z2m2.he07 (and M3.z2m2) 
has a
production more efficient by 0.3-0.4 dex at the two \spr\ peak elements. This is due to the different CBM prescription used at the bottom of the convective envelope during the TDU compared to Pi13.
The IGW model parameterization allows to form \isotope[13]{C} pockets that are a factor of 3-5 larger compared to the overshooting CBM prescription used by Pi13.
On the other hand, the two models have comparable concentrations of \czw\ in the He intershell, allowing to build similar amounts of \isotope[13]{C} in \isotope[13]{C}-pocket layers \citep[][]{lugaro:03a}.
As a consequence, the [hs/ls] ratios are similar within $\sim$ 0.05 dex.
The model M3.z1m2  and the associated test cases show stronger \spr\ enrichment compared to the models with lower mass or higher metallicity. 
In particular, [ls/Fe] $\sim$ 0.7 for model M3.z1m2.hCBM.ntest, and [hs/Fe] $\sim$ 0.95 for M3.z1m2 and M3.z1m2.hCBM.ntest.
The factor driving the difference in the shape of the curves between the 2 and 3 \msun\ models is the larger $\lambda$$_{DUP}$ parameter in the 3 \msun\ models and, 
concerning the Z=0.02 cases, the larger number of TDUs (check \tab{tab:model_prop} and \fig{fig:TAGBprop_middle}).

In \fig{fig:summary_obs_1}, we show the comparison between AGB models with and without clipping, but using the same CBM parameterization at the bottom of TDUs (see Tab. \ref{tab:model_name}). 
The results give similar results within 0.1 dex. Therefore, our $s$-process calculations are not much affected by using these two different setups.
This is because the set of AGB models \textsf{he07} and the analogous models with no clipping but higher $f_\mathrm{1}$ share enhanced C and O abundances in the He intershell (see discussion in \S \ref{ref:models} and \ref{mesa-mppnp:calc}). 
Indeed, as shown by \cite{lugaro:03a}, the amount of \isotope[12]{C} present in the He intershell is a fundamental parameter affecting the neutron exposure in the \isotope[13]{C} pocket.

Most of the models show a final [hs/ls] $>$ 0, with the exception of the models M2.z2m2.hCBM and M2.z2m2.hCBM.ntest, where [hs/ls] = -0.1 and -0.25 respectively.
These models with more efficient IGW CBM than M2.z2m2, host
\isotope[13]{C}-pockets on average 50-70\% larger compared to the
default case.  The resulting \spr\ enrichment in the AGB star envelope
increases by $\lesssim$ 0.2 dex for ls elements and hs elements
(\fig{fig:summary_obs_2}).  In general, a larger
\isotope[13]{C}-pocket allows to have a more gradual decline of \isotope[13]{C}, and to produce lighter elements more efficiently. In general, hCBM models show lower [hs/ls] ratios (i.e., an average lower neutron exposure), compared to their analogous with our default CBM. 

This is interesting, since these variations in the \spr\ abundances are obtained with the same He-intershell conditions. Therefore, while the total amount of \spr\ elements dredged-up in the
AGB envelope is not drastically affected, the uncertainties associated
with the IGW CBM setup in our models affect the relative production at
the Sr peak with respect to the Ba peak.  
According to the
discussion in \S \ref{mesa-mppnp:calc}, the parameters $D_\mathrm{2}$ (i.e. the
point where the IGW mixing efficiency dominates CBM) and $f_\mathrm{2}$ need to
be constrained by future hydrodynamics simulations with an uncertainty
much lower than what we considered here.  

In \fig{fig:summary_obs_2},
we show the cases labeled as {\it reference\_model}.ntest, where the
only difference with respect to their reference models is the
\isotope[14]{N}(n,p)\isotope[14]{C} rate multiplied by a factor of two
(\tab{tab:model_name_network_test}). 
By changing the \isotope[14]{N}(n,p)\isotope[14]{C} rate, the impact is comparable
  to the uncertainty related to the IGW CBM setup. For the default models the rate increase reduces
  the [hs/ls] by about 0.05 dex, while for hCBM models the [hs/ls]
  ratio is reduced by 0.1 dex.  This effect is due to the higher poisoning
  effect of \isotope[14]{N} using the higher \isotope[14]{N}(n,p)\isotope[14]{C} rate, 
reducing the neutron exposure and favoring the production at the Sr peak compared
  to the models using a lower rate.  While the errors given by
  \cite{koehler:89} are much lower than a factor of two, the large
  departure between different experiments at energies larger than 20
  keV requires more experimental analysis.  An accurate determination
  of the \isotope[14]{N}(n,p)\isotope[14]{C} cross section at $\sim$ 8
  keV would allow to better constrain the physics mechanisms driving
  the formation of the \isotope[13]{C} pocket.

\subsection{Comparison with spectroscopic observations of post-AGB H-deficient stars and planetary nebulae}
\label{sub:post_agb}

About 10\% of AGB stars will experience a late pulse or very late thermal pulse event during their post-AGB evolution, becoming H-deficient stars \citep[e.g.,][]{herwig:99c,miller-bertolami:06}. 
Examples are Sakurai's object \citep[e.g.,][and references therein]{herwig:11}, and Fg Sagittae \citep[][]{gonzalez:98}.
The observation of the surface abundances of stars like the PG1159 objects reveal the He-intershell abundances at late AGB stages, 
where the amount of the most abundant elements He, C and O are relics of the AGB stellar evolution and diagnostics for CBM during this earlier phase \citep[e.g.,][]{werner:06,werner:14}.
In particular, the observed range of abundances in mass fractions are 0.3 $<$ He $<$ 0.85, 0.15 $<$ C $<$ 0.6 and 0.02 $<$ O $<$ 0.20. 
The CBM at the bottom of the He-intershell during the convective TPs allows to cover this range of abundances and the largest observed concentrations for C and O, whether the physics mechanism driving the CBM is overshooting \citep[e.g.,][]{herwig:97} or Kelvin-Helmholtz instabilities (this work). \cite{lawlor:06} partially reproduced the observed C and O enrichment in the He intershell, with a maximum O concentration of 5.9\%, by including semi-convection in their calculations. While the observation of C and O in H-deficient stars is affected by uncertainties \citep[e.g.,][]{asplund:99,gallino:11},
there are no published observations questioning the large spread of C and O abundances in post-AGB H-deficient stars, and the largest C and O enrichment that are observed. 
 
\begin{figure}[htbp]
\begin{center}
\includegraphics[scale=0.3]{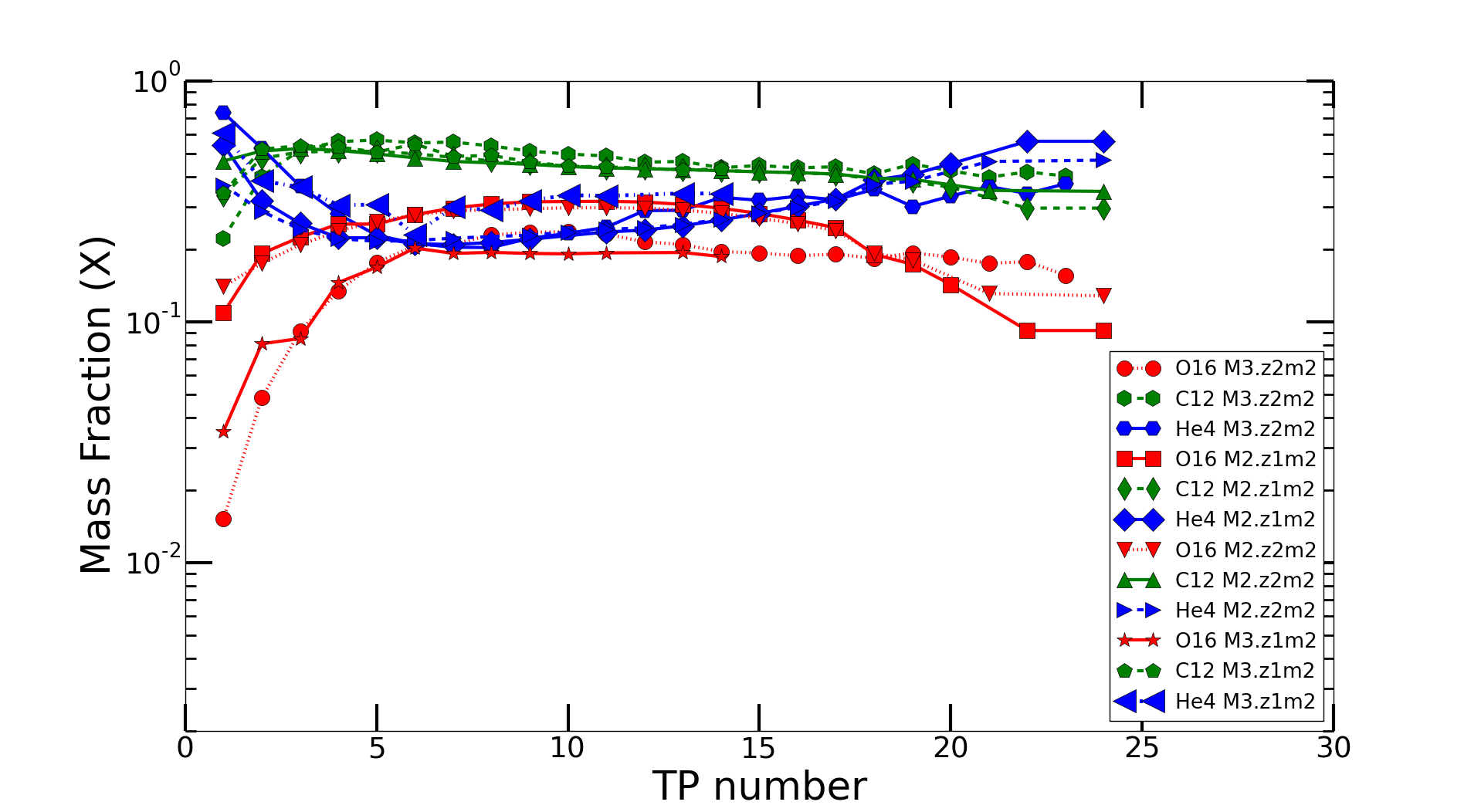}
\includegraphics[scale=0.3]{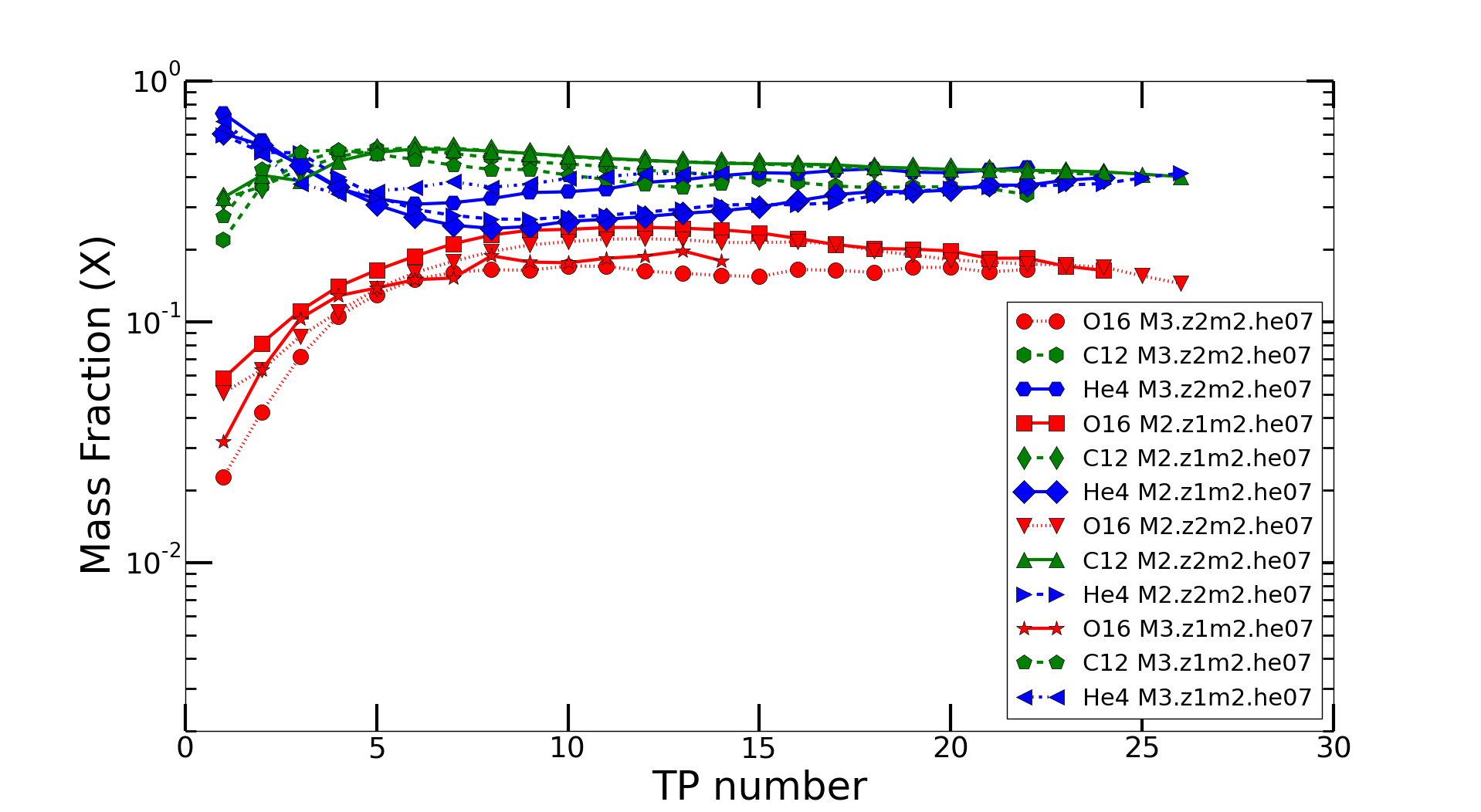}
\end{center}
\caption{He, C and O abundances evolution in the He Intershell as a function of the TP number along the AGB evolution for the AGB models M3.z2m2, M3.z1m2, M2.z2m2 and M2.z1m2 (upper panel), and for M3.z2m2.he07, M3.z1m2.he07, M2.z2m2.he07 and M2.z1m2.he07 (lower panel). 
}
\label{ref_models_nomicro}
\end{figure}

\begin{figure}[htbp]
\begin{center}
\includegraphics[scale=0.3]{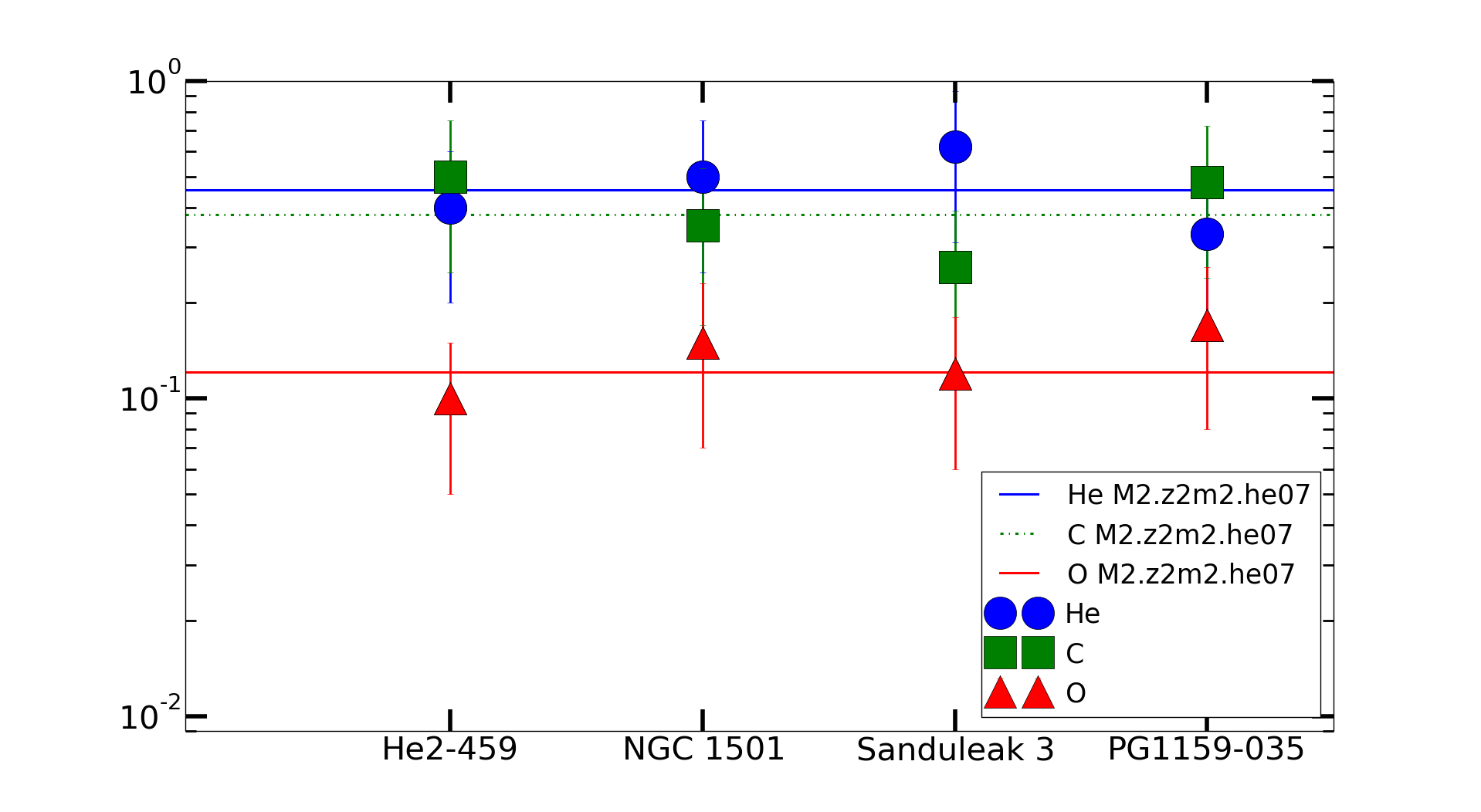}
\end{center}
\caption{He, C and O abundances observed for a sample of H-deficient post-AGB stars classified as PG1159 objects: He2-459, NGC1501, Sanduleak3 and PG1159-035. Observations are given by \cite{werner:06}. Also the final intershell abundances from M2.z2m2.he07 are presented.}
\label{ref_models_nomicro_obs}
\end{figure}

In \fig{ref_models_nomicro}, upper panel, the abundances of He, C and O are shown in the He intershell after each TP for our models M2.z2m2, M3.z2m2, M2.z1m2 and M3.z1m2. In particular, the final C and O abundances are 0.39-0.48 and 0.12-0.18, respectively.
In the lower panel, the same data are given for the models M2.z2m2.he07, M3.z2m2.he07, M2.z1m2.he07 and M3.z1m2.he07. In this case, the final C and O abundances are 0.33-0.41 and 0.13-0.17, respectively. The two sets of AGB models show similar evolution patterns for He-intershell abundances. 
As a comparison, in \fig{ref_models_nomicro_obs} we report the abundances observed for PG1159 stars \citep[][]{werner:06}, 
that are comparable with the final He-intershell abundances shown in \fig{ref_models_nomicro}. 
In particular, in the same plot we show the results from model M2.z2m2.he07 as a representative case of our calculations.

At the end of the post-AGB evolutionary phase, planetary nebulae (PNe) are still carriers of the abundance signatures of the previous AGB phase \citep[][and references therein]{vanwinckel:03}.
The abundances of elements such as O, Cl, Ar have been used in order to identify the initial metallicity of the PN progenitor, assuming that their initial concentrations are not affected by AGB nucleosynthesis. However, evidence for O enrichment have been found first for PNe at low metallicity \citep[e.g.,][]{pequignot:00}, and lately for PNe with metallicities close to solar \citep[][]{rodriguez:11,delgado-inglada:15}.
In particular, \cite{delgado-inglada:15} confirmed that the O enrichment calculated for AGB models including CBM at the bottom of the intershell during the convective TP by Pi13 are compatible with observations of PNe with solar-like metallicity.
Consistently with post-AGB H-deficient stars, another independent confirmation that CBM should be included during the AGB phase comes from observation of O isotopic ratios in C-rich AGB stars \citep[][]{karakas:10}.

\subsection{Comparison to the literature and with spectroscopic data from AGB stars}
\label{spectro:comparison}

%

\begin{figure}[htbp]
\begin{center}
\includegraphics[scale=0.4]{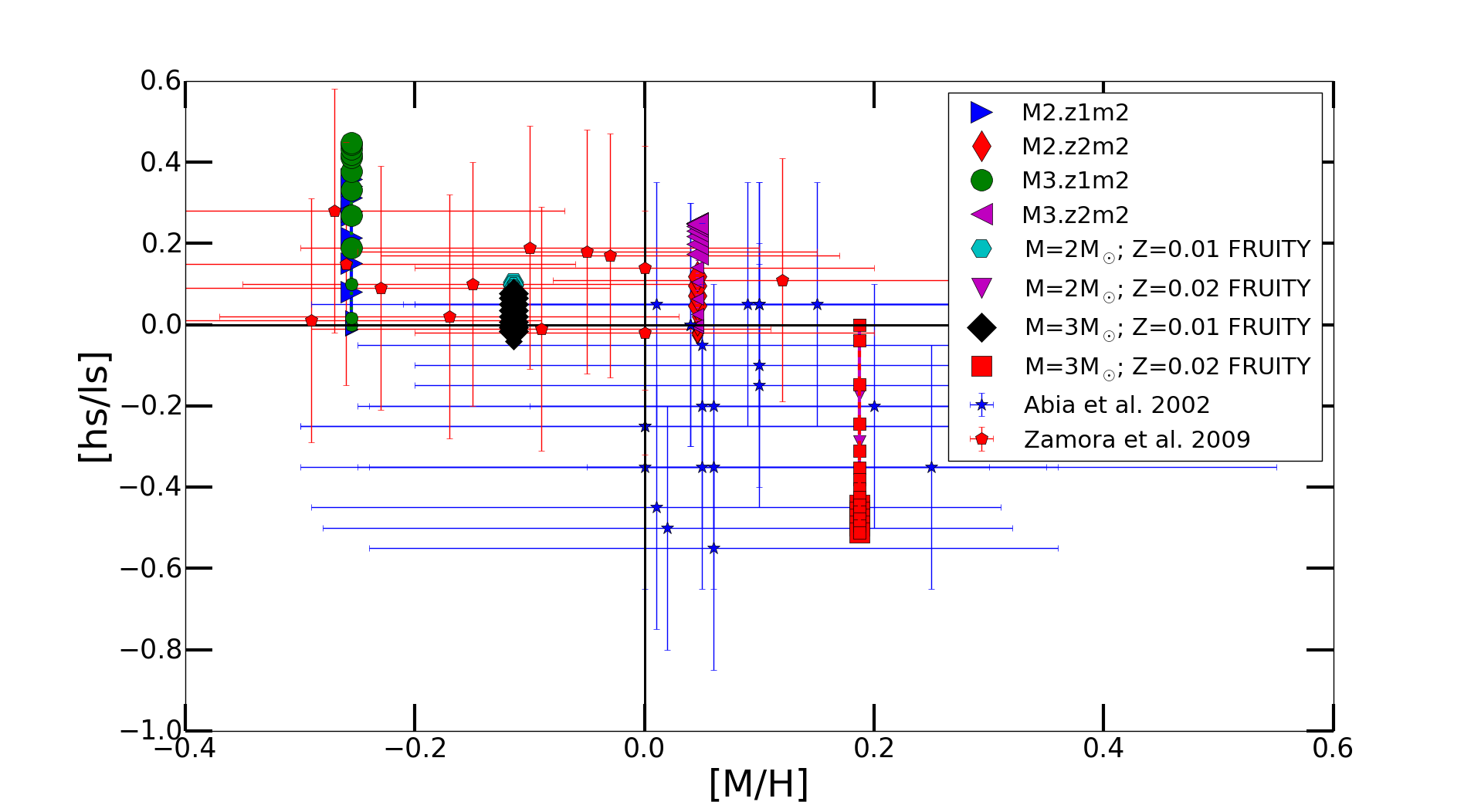}
\end{center}
\caption{Comparison of the [hs/ls] vs [M/H] obtained from our models with observational data from \citet{abia:02} and \citet{zamora:09}.
We also report the AGB calculations from the FRUITY database \citep[][]{cristallo:15b}.}
\label{hsls_feh:spectro}
\end{figure}

In \fig{hsls_feh:spectro}, the [hs/ls] ratio obtained in our models is compared with spectroscopic observations of galactic-disk AGB stars \citep[][]{abia:02,zamora:09}. 
  Both \citet{abia:02} and \citet{zamora:09} derived the $s$-element abundance pattern of Carbon stars.
\citet{abia:02} analyzed N-type stars of nearly solar and super-solar metallicity, while \citet{zamora:09} focused on lower metallicity R-type stars. This is the main reason why data from these two works 
are located in two distinct areas on the [hs/ls] VS [M/H] plane (\fig{hsls_feh:spectro}). They are consistent with each other since the resulting pattern of [hs/ls] decreases with [M/H] as expected as a consequence of the lower number of neutrons captured by each iron seed \citep{busso:01}.
The results for the stellar models with the same initial mass from the FRUITY database are also shown \citep[][]{cristallo:15b}.
The different [M/H] between the two theoretical data sets is due to the different reference solar metals distribution adopted. 

 
In our models, we consider CBM at the bottom of the convective TP, while this is not the case for the models in the FRUITY database shown here for comparison. This implies that we obtain a peak-\cdr\ concentrations the \cdr\ pocket that is about a factor of two larger compared to models without CBM at the bottom of the PDCZ \citep{lugaro:03a}.
This translates into a proportionally larger peak-neutron exposure and in turn yielding a more efficient production of heavier \spr\ elements as seen by a systematically larger [hs/ls] in our models compared to AGB calculations by \cite{cristallo:11}, and in general compared to all models without CBM below the PDCZ \citep[e.g.,][]{bisterzo:11,lugaro:14}. 

Note that it is not only the CBM at the bottom of the intershell during convective TP that defines the evolution of the [hs/ls] ratio at the surface of the AGB star. 
Indeed, the \spr\ nucleosynthesis is also affected by the complex interplay between CBM at both the two He intershell boundaries, and the selection of the nuclear reaction rates. In \fig{fig:summary_obs_2}, we have shown that a different IGW CBM setup at the bottom of the TDU combined with the uncertainty of the \isotope[14]{N}(n,p)\isotope[14]{C} rate might reduce by up to $\sim$0.3 dex the final [hs/ls] ratio. 
The models shown in \fig{hsls_feh:spectro} do not include other relevant physics mechanisms such as rotation and magnetic field. \citet{herwig:03} and \citet{siess:04}, and more recently \cite{piersanti:13}, have shown that by considering rotation in AGB models the final [hs/ls] ratio tends to be reduced, compared to non-rotating models. On the other hand, \citet{herwig:05} discussed the possible interplay between rotation and magnetic field, where the impact of rotation can be partially suppressed by magnetic field. 

Overall, both sets of models in \fig{hsls_feh:spectro} are consistent with observations. This is also due to the large observational uncertainties, reported in the figure.

\begin{figure}[htbp]
\begin{center}
\resizebox{11.4cm}{!}{\rotatebox{0}{\includegraphics{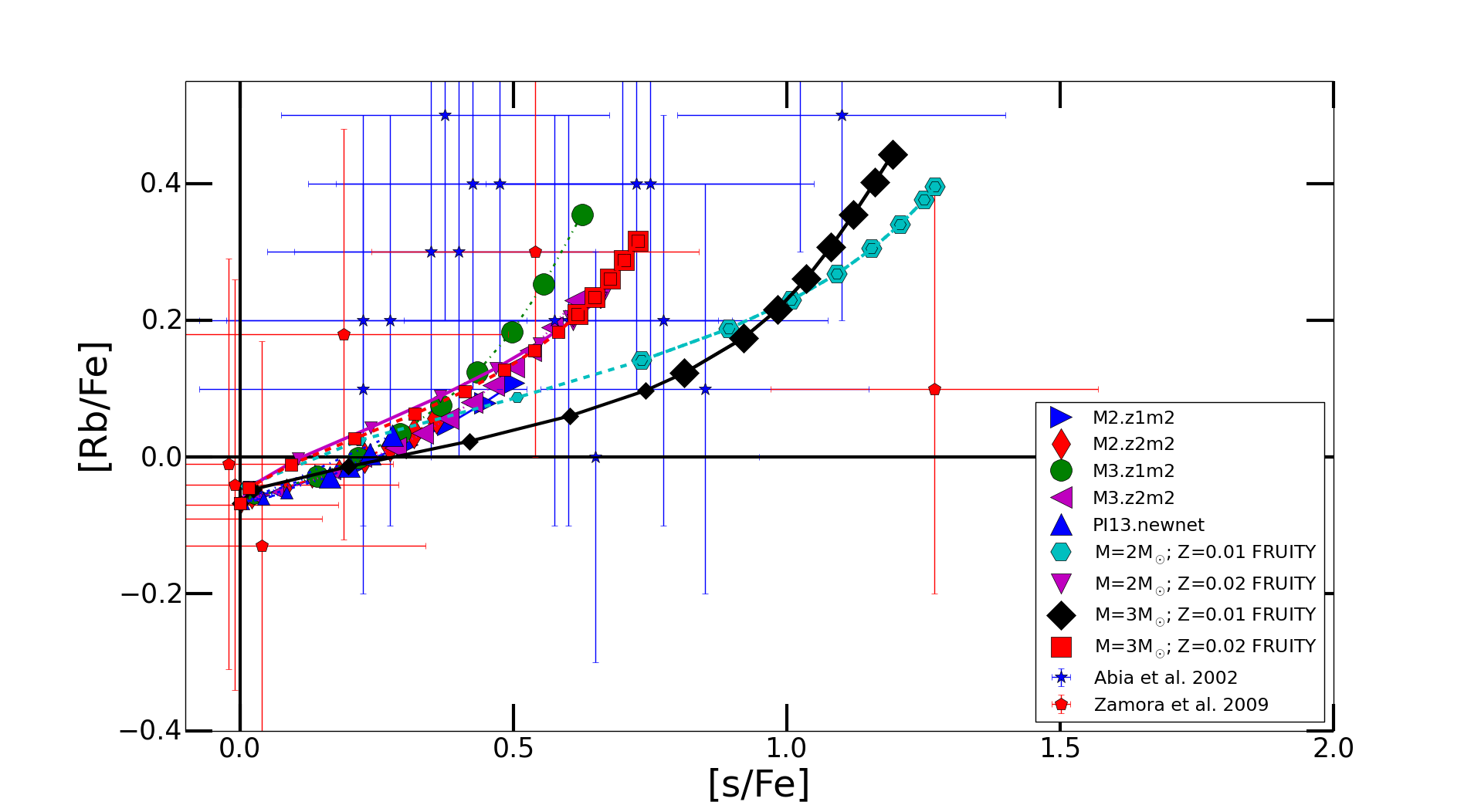}}}
\resizebox{11.4cm}{!}{\rotatebox{0}{\includegraphics{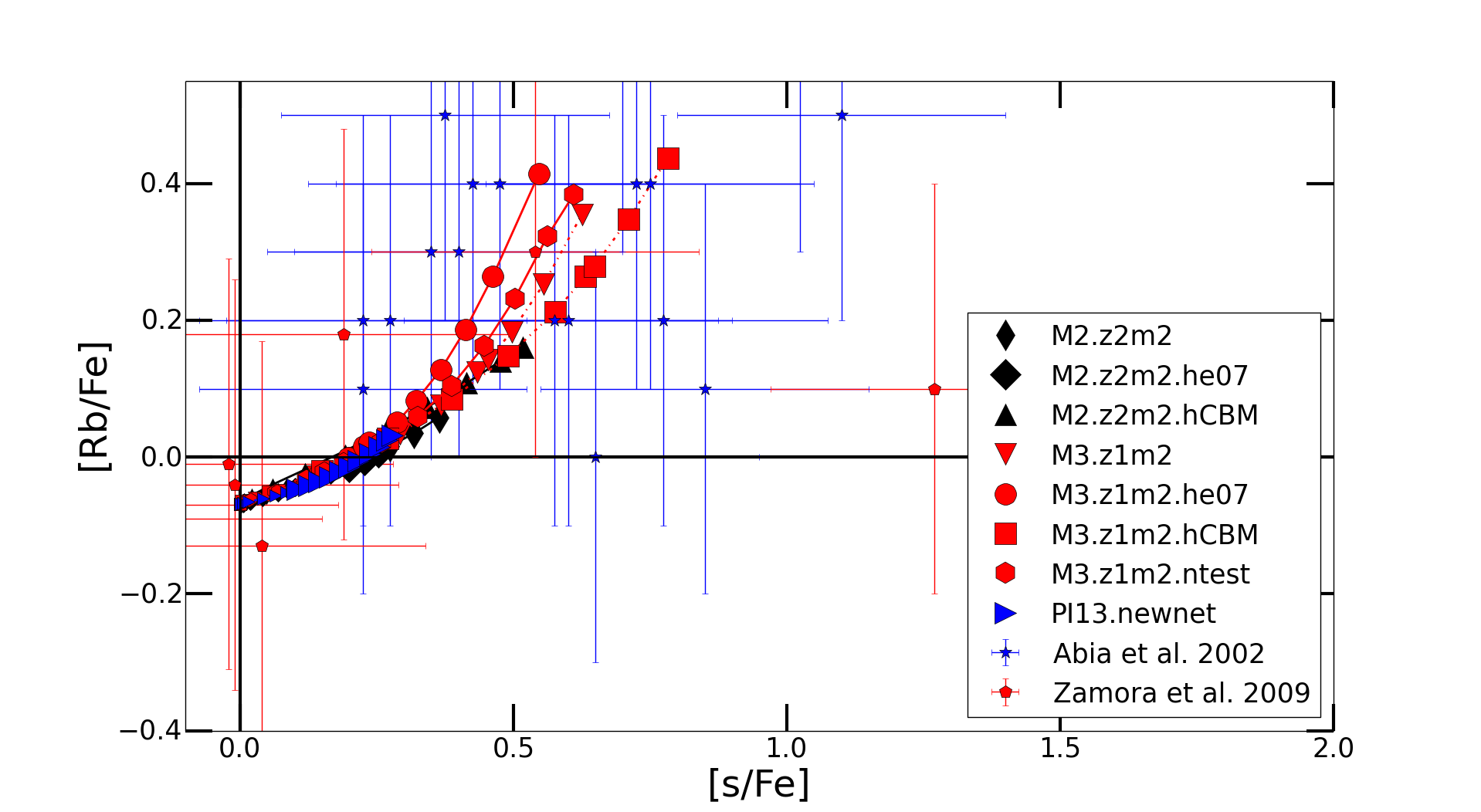}}}
\end{center}
\caption{Upper panel: we report the [Rb/Fe] and [$s$/Fe] ratios obtained from the indicated AGB models, in comparison with a sample of C stars by \citet{abia:02} and \citet{zamora:09}, and with analogous theoretical AGB models by the FRUITY database \citep[][]{cristallo:15b}. 
Only stars with [M/H] $>$-0.3 are considered. 
Lower panel: Additional AGB models from this work are reported in comparison with observations (see the upper panel). }
\label{rb:spectro}
\end{figure}

In \fig{rb:spectro}, we compare our models with spectroscopic observations for [Rb/Fe] and the [s/Fe] ratio, given by the average production at the ls and hs \spr\ neutron-magic peaks. 
The [s/Fe] ratio is a diagnostic for the \spr\ efficiency, and the [Rb/Fe] ratio increases with the increase of the efficiency of the \nezw\ ($\alpha$,n)\mgfu\ reaction during the TP \citep[e.g.,][]{lambert:95}. 
Indeed, Rb is not made efficiently at neutron densities typical of the \isotope[13]{C} pocket, while at the high neutron densities during the TP the nucleosynthesis flows \isotope[84]{Kr}(n,$\gamma$)\isotope[85]{Kr}(n,$\gamma$)\isotope[86]{Kr}(n,$\gamma$)\isotope[87]{Kr}($\beta$$^-$)\isotope[87]{Rb}   and \isotope[84]{Kr}(n,$\gamma$)\isotope[85]{Kr}($\beta$$^-$)\isotope[85]{Rb}(n,$\gamma$)\isotope[86]{Rb}(n,$\gamma$)\isotope[87]{Rb}
accumulate \isotope[87]{Rb}. 
In these conditions, \isotope[87]{Rb} is made more efficiently than \isotope[85]{Rb} and the \spr\ production of Rb is higher, because of the lower neutron capture cross section of \isotope[87]{Rb} compared to \isotope[85]{Rb} \citep[e.g.,][]{abia:01}.  
As for \fig{hsls_feh:spectro}, in \fig{rb:spectro} observational uncertainties pose a serious limitation to the diagnostic power of these observed abundance ratios. 
A large observational scatter is obtained for \spr\ and Rb enrichment. 
On the other hand, it needs to be clarified if such a scatter is just due to observational uncertainties, or if it is instead tracing a real spread of \spr\ nucleosynthesis conditions in the He intershell of AGB stars.

In our models the [s/Fe] ratio ranges between $\sim$ 0.4 dex (M2.z2m2) and 0.8 dex (M3.z1m2.hCBM). 
They all show quite similar theoretical curves in \fig{rb:spectro}, consistent also with results from the FRUITY models at Z=0.02. 
On the other hand, the \spr\ abundance evolution for the models at Z=0.01 by \cite{cristallo:11} shows a larger [s/Fe] up to [s/Fe]$\sim$1.3 dex, with a production of Rb comparable with the models at higher metallicity. 
As already found considering \fig{fig:summary_obs_1}, we obtain similar results for these AGB models and their analogous \textsf{he07} stellar models.
  In the same figure and in \fig{fig:summary_obs_2} our 3 \msun\ models sit right on the highest [hs/ls] region covered by observations, 
as predicted since they are non-rotating model. 
The expected impact of rotation is to  reduce the neutron exposure favoring the production of lighter \spr\ isotopes, potentially allowing to account for all the observed range of the [hs/ls] index.
The model Pi13.newnet has a final [s/Fe] $\sim$ 0.3 and [Rb/Fe] $\sim$ 0.1. The IGW CMB allowed to obtain larger \isotope[13]{C} pockets compared to Pi13, causing a 0.3 dex higher final [s/Fe]. 
Within the observational and stellar uncertainties these models can reproduce the observed range of [s/Fe] (see \fig{rb:spectro}). 
Therefore, IGW provide a suitable mechanism to drive the CBM at the bottom of the TDU and leading to the formation of the radiative \isotope[13]{C} pocket.

\subsection{Comparison with presolar-grains data}
\label{pres:grains}

In this section, we compare the results of our stellar calculations with measurements of isotopic abundances in presolar mainstream SiC grains for Zr and Ba.
Presolar mainstream SiC grains are the most abundant type of presolar SiC grains \citep[e.g.,][]{ott:90,lewis:94,lugaro:03b,zinner:14}. They condensed in the envelope of C-rich AGB stars and were ejected into the surrounding interstellar medium by stellar winds. 
The condition to form in a C-rich environment (i.e., C/O$>$1) is crucial for the formation of C-rich grains.
Thanks to high-precision laboratory measurement of their isotopic composition for heavy elements like Sr, Zr and Ba it is possible to derive fundamental constraints about their parent AGB stars. 
In particular, theoretical stellar simulations can be compared with the conditions in the He intershell inferred by measurement in presolar grains, where the \spr\ is activated in AGB stars \citep[e.g.,][]{lugaro:03a,barzyk:06,avila:12,lugaro:14,liu:14a,liu:14b,liu:15}.  

The measured \isotope[96]{Zr}/\isotope[94]{Zr} ratio in SiC grains is known to be a diagnostic for the activation of the \isotope[22]{Ne}($\alpha$,n)\isotope[25]{Mg} neutron source at the bottom of the convective TPs. This is due to the \spr\ branching point at \isotope[95]{Zr}, which needs 

neutron densities higher than 5$\times$10$^{8}$cm$^{-3}$ to be opened and produce \isotope[96]{Zr} via direct neutron capture on \isotope[95]{Zr} \citep[][]{lugaro:03a}. 
\cite{lugaro:14} identified a positive correlation between the \isotope[92]{Zr}/\isotope[94]{Zr} and \isotope[29]{Si}/\isotope[28]{Si} ratios, suggesting that the observed spread of \isotope[92]{Zr}/\isotope[94]{Zr} is a signature of the initial metallicity of the AGB progenitor.
\cite{liu:14b} suggested that this ratio can also be used to constrain the internal structure of the \isotope[13]{C}-pocket.
The same methodology is adopted by \cite{liu:15} by comparing theoretical predictions with new grain measurements for Sr and Ba. In particular, we compare our AGB calculations with newly measured \isotope[88]{Sr}/\isotope[86]{Sr} and \isotope[138]{Ba}/\isotope[136]{Ba} ratios to derive information about  the \isotope[13]{C}-pocket  shape and size.

\begin{figure}[htbp]
\begin{center}
\resizebox{8.6cm}{!}{\rotatebox{0}{\includegraphics{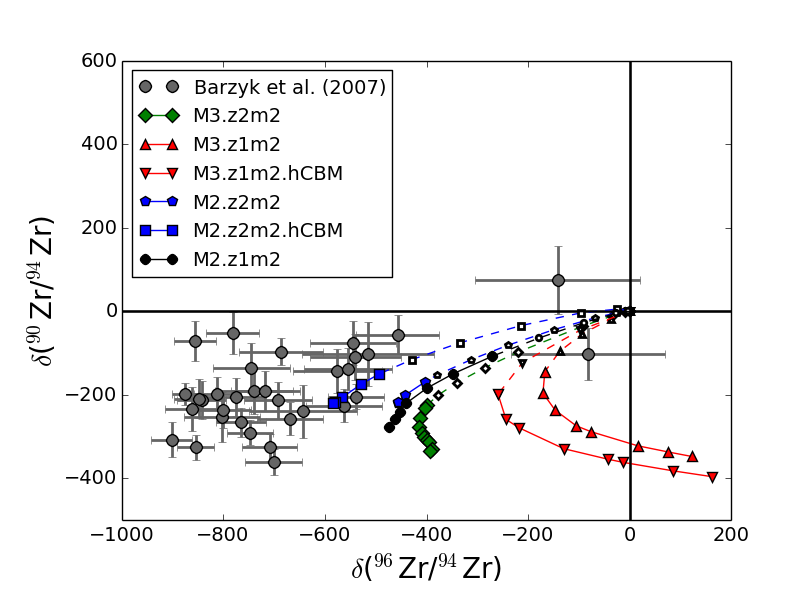}}}
\resizebox{8.6cm}{!}{\rotatebox{0}{\includegraphics{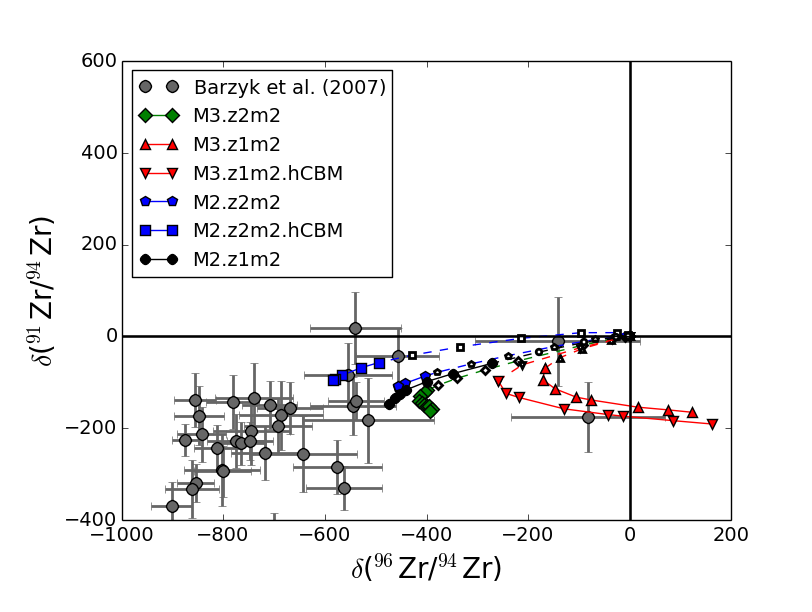}}}
\resizebox{8.6cm}{!}{\rotatebox{0}{\includegraphics{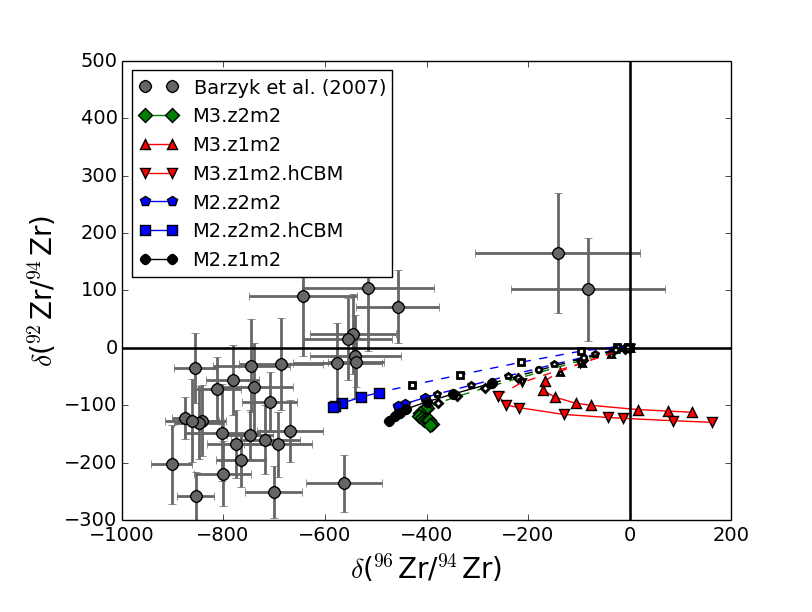}}}
\end{center}
\caption{Upper panel: The evolution of $\delta$(\isotope[90]{Zr}/\isotope[94]{Zr}) and $\delta$(\isotope[96]{Zr}/\isotope[94]{Zr}) ratios in the AGB envelope is shown for our AGB models. Large full markers indentified the abundances at each TP once C$>$O at the surface, while small empty markers identify the occurrence of TPs before the AGB models become C rich. For comparison, the measurements from presolar SiC grain of type mainstream and error bars are reported \citep[][]{barzyk:06}.
Middle panel: $\delta$(\isotope[91]{Zr}/\isotope[94]{Zr}) and $\delta$(\isotope[96]{Zr}/\isotope[94]{Zr}) for the same models in the upper panel. Lower panel: $\delta$(\isotope[92]{Zr}/\isotope[94]{Zr}) and $\delta$(\isotope[96]{Zr}/\isotope[94]{Zr}) again for the same models. 
} 
\label{fig:zr_ratios}
\end{figure}

\begin{figure}[htbp]
\begin{center}
\resizebox{8.6cm}{!}{\rotatebox{0}{\includegraphics{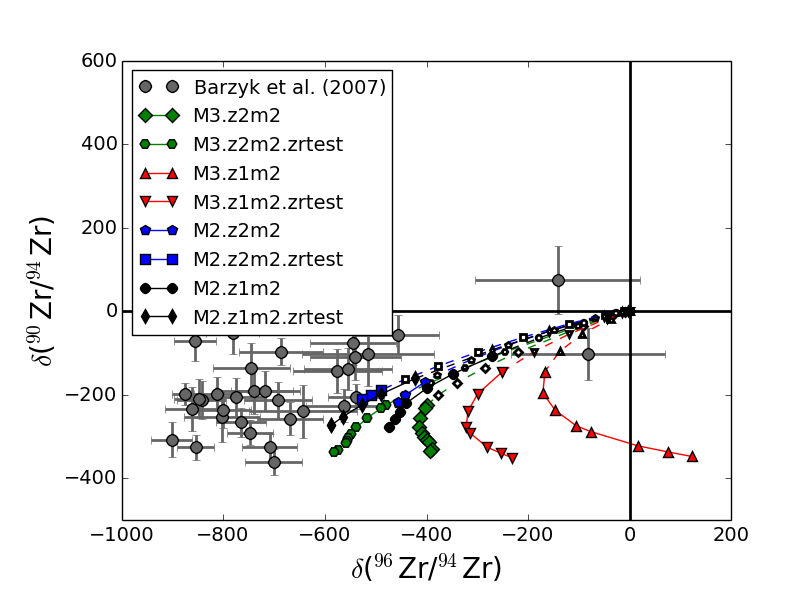}}}
\resizebox{8.6cm}{!}{\rotatebox{0}{\includegraphics{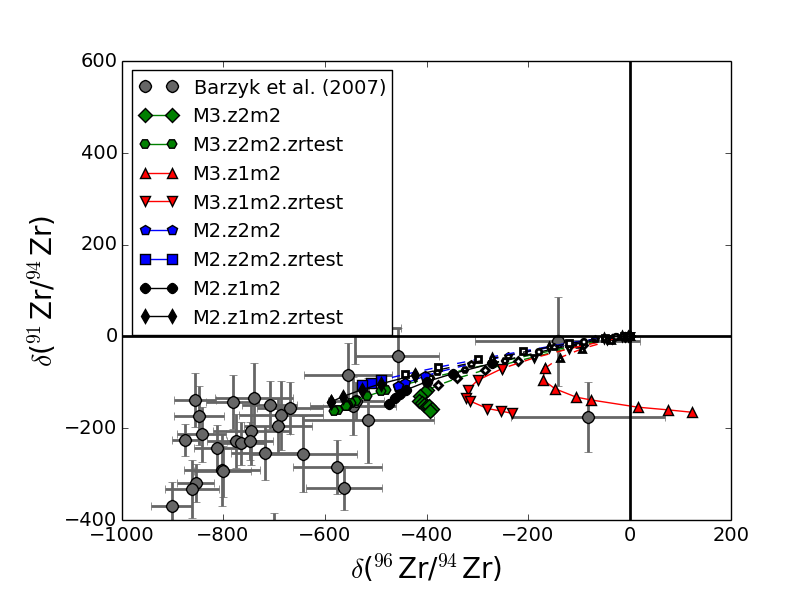}}}
\resizebox{8.6cm}{!}{\rotatebox{0}{\includegraphics{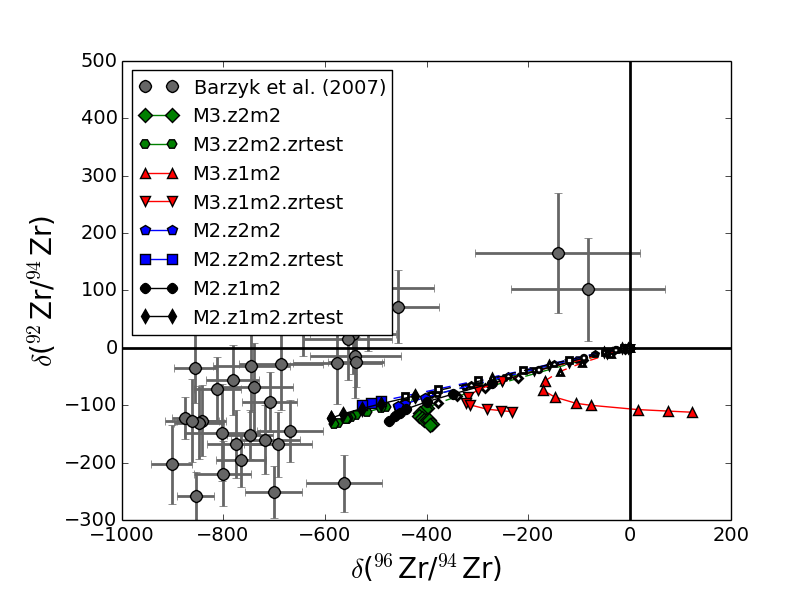}}}
\end{center}
\caption{As in figure \ref{fig:zr_ratios}, but the results are shown for the models calculated by dividing the \isotope[95]{Zr}(n,$\gamma$)\isotope[96]{Zr} reaction rate by a factor of two. 
}
\label{fig:zr_ratios_zr95test}
\end{figure}

In \tab{tab:model_deltaZrBa} the final isotopic ratios obtained in the He intershell and in the AGB envelope are given for our AGB models.
In Figs. \ref{fig:zr_ratios} and \ref{fig:zr_ratios_zr95test} the evolution of the Zr abundances at the stellar surface during the AGB evolution is shown. In Fig. \ref{fig:zr_ratios}, the models cover a large range of \isotope[96]{Zr}/\isotope[94]{Zr} ratios, with 200$\permil$ $\gtrsim$ $\delta$(\isotope[96]{Zr}/\isotope[94]{Zr}) $\gtrsim$ -600$\permil$. The $\delta$ here indicates deviations of the given isotopic ratio from the average solar system value in parts per thousand. The factors with the largest impact on this quantity are the temperature at the bottom of the PDCZ, which is correlated to the CBM description at the bottom of such zone, and the neutron-capture reactions rates on Zr isotopes. Compared to Pi13 and results by \cite{lugaro:03b} obtained for AGB models including CBM during the convective TPs, the negative $\delta$-values are mostly due to the new \isotope[95]{Zr} MACS by \cite{lugaro:14} (see also Fig. \ref{comp:ppd}).
Our models reproduce the observed scatter of $\delta$(\isotope[90]{Zr}/\isotope[94]{Zr}), while a relevant fraction of grains with low $\delta$(\isotope[91]{Zr}/\isotope[94]{Zr}) and $\delta$(\isotope[92]{Zr}/\isotope[94]{Zr}) ratios are not reproduced. As discussed by \cite{liu:14b}, Zr isotopic ratios can be used to test size and properties of the \isotope[13]{C} pocket. 
In our models the \isotope[13]{C} pocket is made after each TDU consistently with the IGW CBM adopted to calculate the stellar structure. 
On the other hand, the IGW CBM implementation was made by a simple fitting of the De03 simulations. 
This allows us to provide a good indication of the size of the \isotope[13]{C} pocket due to IGW CBM, but the detailed shape needs to be better constrained by multi-dimensional hydrodynamics simulations. 
Furthermore, rotation and magnetic field are two fundamental physics ingredients still missing in our models, that will affect the \isotope[13]{C} pocket properties $after$ its formation \citep[for rotation, e.g.,][]{herwig:03,piersanti:13} and eventually the \spr\ Zr isotopic ratios \citep[][]{liu:15}. 
Therefore, a crucial step forward to challenge the scenario in which IGW CBM is the physics mechanism responsible for the formation of the \isotope[13]{C} pocket, will be to calculate how the pocket is modified by rotation and magnetic field before and during the \spr\ production.

Grains with $\delta$(\isotope[96]{Zr}/\isotope[94]{Zr})$<$-900 $\permil$ are not reproduced by baseline AGB models \citep[][]{liu:14b,lugaro:14}. With our models in \fig{fig:zr_ratios}, we confirm the increasing trend of the \isotope[96]{Zr}/\isotope[94]{Zr} ratio with increasing initial mass and with decreasing initial metallicity \citep[][]{lugaro:03a,liu:14b,lugaro:14}. 
However, our AGB models cannot reproduce grains with $\delta$(\isotope[96]{Zr}/\isotope[94]{Zr})$<$-600 $\permil$. In Fig. \ref{fig:zr_ratios_zr95test}, we show the impact on our results of the \isotope[95]{Zr}(n,$\gamma$)\isotope[96]{Zr} neutron capture cross section. The cross section provided by \cite{lugaro:14} was reduced by a factor of two. In general, the use of the reduced rate allows to decrease the final \isotope[96]{Zr}/\isotope[94]{Zr} ratio by $\delta$ $\sim$200$\permil$. Therefore, while the new \isotope[95]{Zr}(n,$\gamma$)\isotope[96]{Zr} cross section helped to alleviate the overproduction of \isotope[96]{Zr} compared to \isotope[94]{Zr}, the entire observed range is not yet reproduced. From the nuclear physics point of view, the other reaction rate relevant for the \isotope[95]{Zr} branching is the rate of the neutron source \isotope[22]{Ne}($\alpha$,n)\isotope[25]{Mg}.  
Once the combined uncertainties of the \isotope[95]{Zr}(n,$\gamma$)\isotope[96]{Zr} and \isotope[22]{Ne}($\alpha$,n)\isotope[25]{Mg} rates will be fully constrained by experiments, the \isotope[96]{Zr}/\isotope[94]{Zr} will be a crucial diagnostic to constrain our simulations.
By comparing Fig.~\ref{fig:zr_ratios} with Fig.~\ref{fig:zr_ratios_zr95test}, the impact of the \isotope[95]{Zr}(n,$\gamma$)\isotope[96]{Zr} are comparable with the variations between models M2.z2m2 and M2.z2m2.hCBM. 
The difference between these two models shows the impact of the uncertainty associated with the IGW CMB implementation in our models. This is due to the fact that the model M2.z2m2.hCBM tends to have \isotope[13]{C} pockets larger 
than the model M2.z2m2. 
This means that the \isotope[13]{C}($\alpha$,n)\isotope[16]{O} (producing \isotope[94]{Zr} but not \isotope[96]{Zr}) has a relatively much larger contribution than the \isotope[22]{Ne}($\alpha$,n)\isotope[25]{Mg} (eventually producing also \isotope[96]{Zr}) in hCBM models. 
Therefore, the \isotope[13]{C}-pocket properties may also affect the \isotope[96]{Zr}/\isotope[94]{Zr} ratio.

\begin{figure}[htbp]
\begin{center}
\resizebox{8.6cm}{!}{\rotatebox{0}{\includegraphics{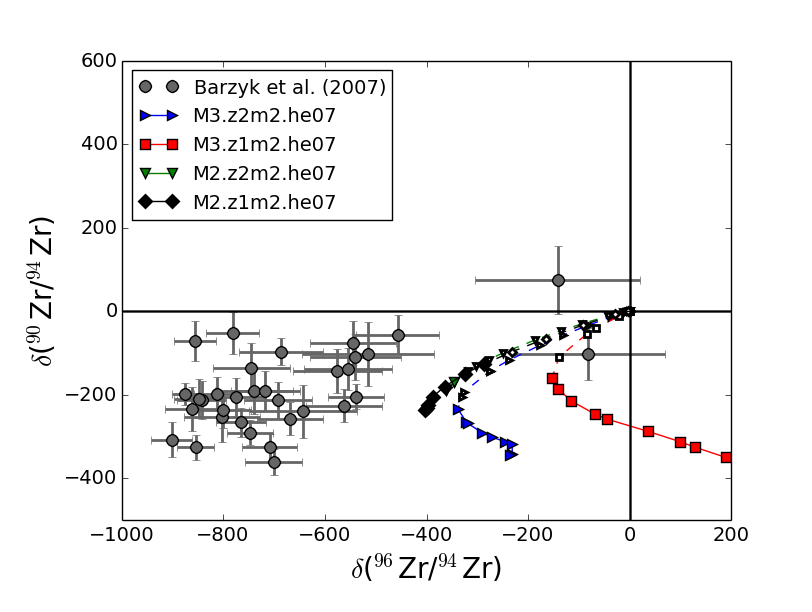}}}
\resizebox{8.6cm}{!}{\rotatebox{0}{\includegraphics{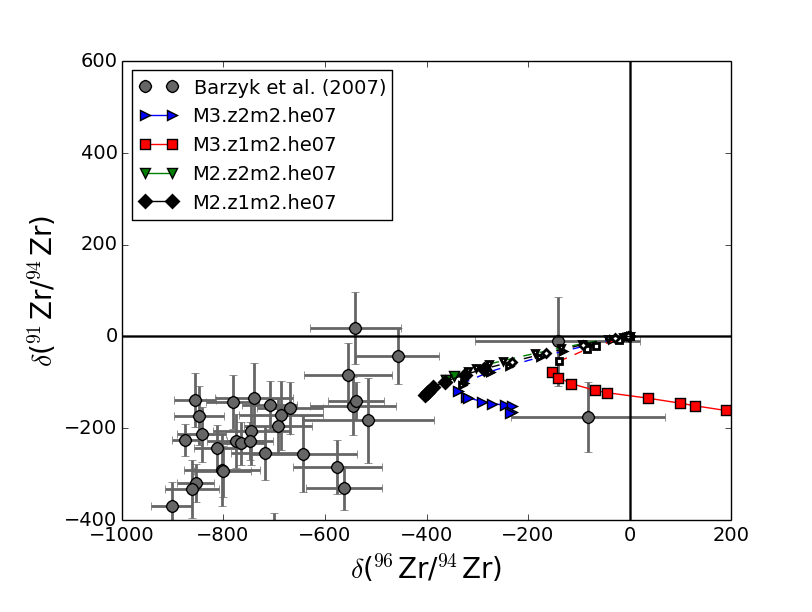}}}
\resizebox{8.6cm}{!}{\rotatebox{0}{\includegraphics{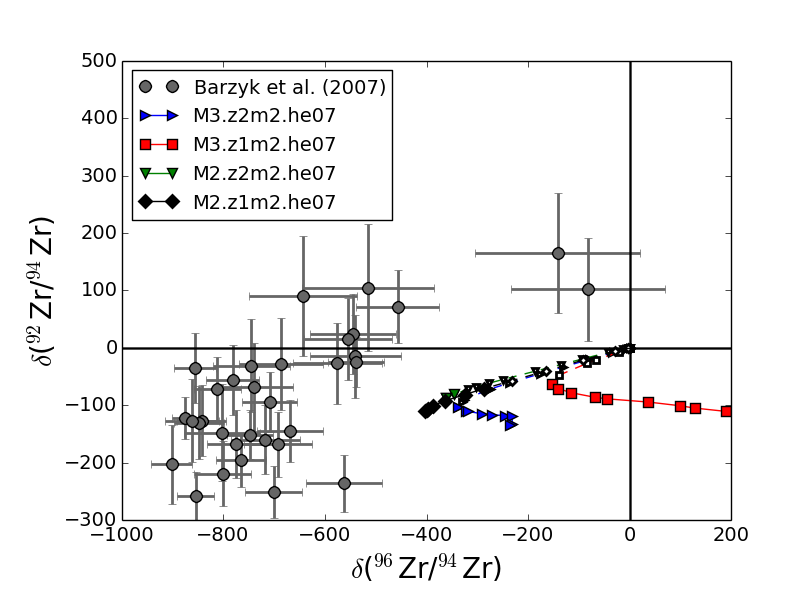}}}
\end{center}
\caption{As in figure \ref{fig:zr_ratios}, but the results are shown for the models calculated with the He07 CBM prescriptions. 
}
\label{fig:zr_ratios_he07}
\end{figure}

If we compare our $clipped$ and \textsf{he07} sets of AGB models, in general the values evolve in a similar way. 
The only exception is between M3.z2m2 and M3.z2m2.he07 models, since the final $\delta$(\isotope[96]{Zr}/\isotope[94]{Zr}) values in M3.z2m2.he07 is higher by $\delta$ $\sim$ 100$\permil$. 
In Fig.~\ref{fig:zr_ratios_he07} we do the same comparison for the \textsf{he07} models. In this case, our models do not reproduce $\delta$(\isotope[96]{Zr}/\isotope[94]{Zr}) values lower than $\sim$-400$\permil$.

From \tab{tab:model_deltaZrBa}, the final surface abundance for most of the models is representative of the He-intershell abundances, with the tendency to show a milder departure from the solar composition in the AGB envelope compared to the He intershell, due to the dilution with the pristine stellar composition. Concerning the \isotope[96]{Zr}/\isotope[94]{Zr} ratio, this trend is maintained for both positive and negative $\delta$-values. 
For instance, the model M3.z1m2.hCBM has final $\delta$(\isotope[96]{Zr}/\isotope[94]{Zr}) equal to +631$\permil$ and +162$\permil$. 
On the other hand, the model M2.z2m2.hCBM shows $\delta$ = -741$\permil$ and -584$\permil$ in the He intershell and in the AGB envelope, respectively. 
The model with the lowest $\delta$-values is M2.z1m2.zrtest, with -831 $\permil$ and -613 $\permil$. 
More efficient TDUs, or a larger number of them would have eventually allowed to reach lower final $\delta$(\isotope[96]{Zr}/\isotope[94]{Zr}) values. 

If we look carefully at theoretical evolution curves in Figs.~\ref{fig:zr_ratios} and \ref{fig:zr_ratios_zr95test}, all the models with initial mass M=3\msun\ show a signature of efficient \isotope[96]{Zr} production due to the \isotope[22]{Ne}($\alpha$,n)\isotope[25]{Mg} activation at the bottom of the convective TP, eventually leading to positive $\delta$-values. 
This picture is consistent with \cite{lugaro:03a} and Pi13, where CBM at the bottom of the convective TPs leads to a stronger \isotope[22]{Ne}($\alpha$,n)\isotope[25]{Mg} activation due to the larger temperatures compared to models without CBM. On the other hand, the new \isotope[95]{Zr}(n,$\gamma$)\isotope[96]{Zr} cross section strongly reduces the production of \isotope[96]{Zr}. 
Therefore, according to our simulations, AGB models with initial mass M$\lesssim$2\msun\ can have at the same time negative $\delta$(\isotope[96]{Zr}/\isotope[94]{Zr}), and C and O concentrations in the He intershell consistent with post-AGB stars and planetary nebula observations.
However, for the $2\msun$ stellar models the degree of pollution of the AGB envelope with He-intershell material seems not to be high enough to explain the abundances for all the presolar grains.

\begin{figure}[htbp]
\begin{center}
\resizebox{8.6cm}{!}{\rotatebox{0}{\includegraphics{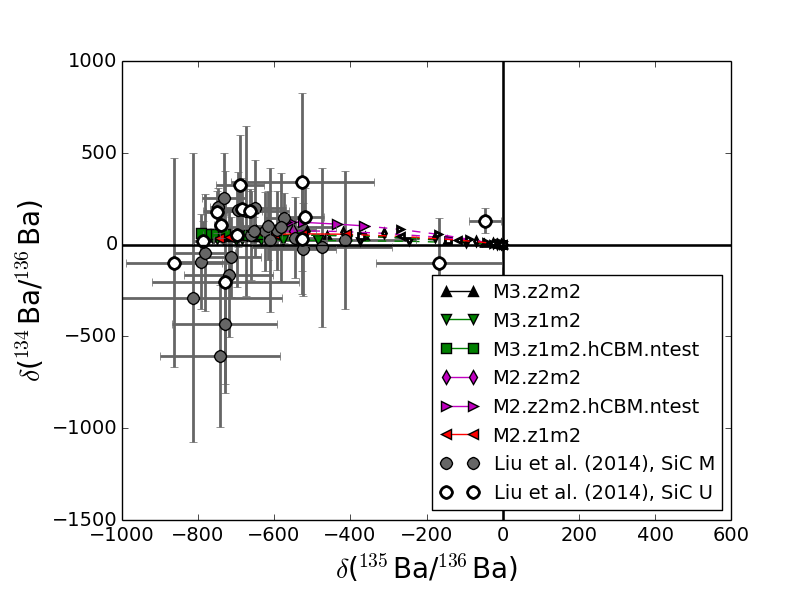}}}
\resizebox{8.6cm}{!}{\rotatebox{0}{\includegraphics{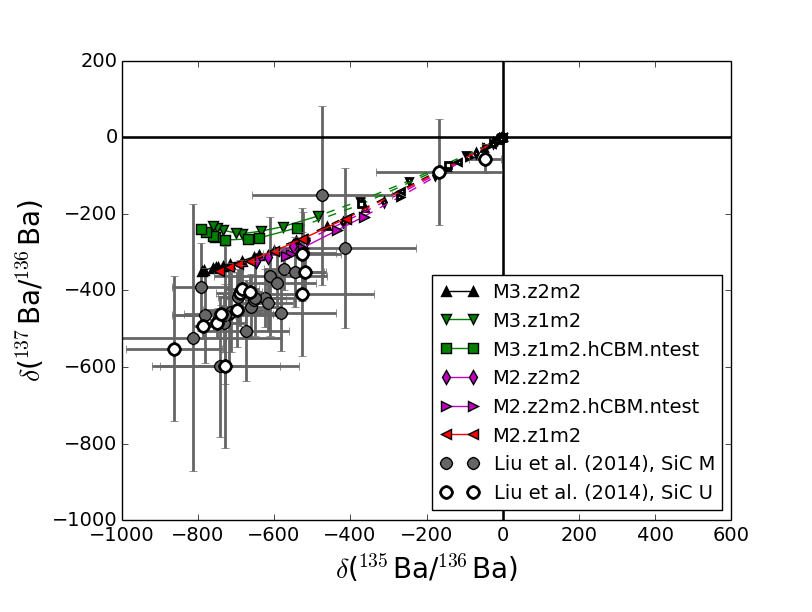}}}
\resizebox{8.6cm}{!}{\rotatebox{0}{\includegraphics{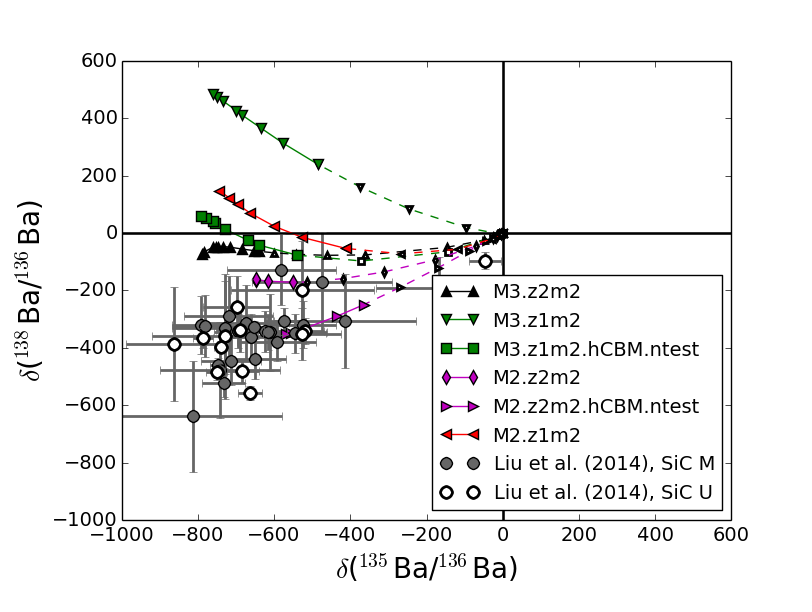}}}
\end{center}
\caption{Upper panel: the evolution of $\delta$(\isotope[134]{Ba}/\isotope[136]{Ba}) and $\delta$(\isotope[135]{Ba}/\isotope[136]{Ba}) ratios in the AGB envelopes is shown for our models. For comparison, the measurements from presolar SiC grain of type mainstream and error bars are reported \citep[][]{liu:14a}. 
Middle panel: As for the upper panel, for $\delta$(\isotope[137]{Ba}/\isotope[136]{Ba}) and $\delta$(\isotope[135]{Ba}/\isotope[136]{Ba}) ratios.
Lower panel: As for the upper panel, for $\delta$(\isotope[138]{Ba}/\isotope[136]{Ba}) and $\delta$(\isotope[135]{Ba}/\isotope[136]{Ba}) ratios.
}
\label{fig:ba_ratios}
\end{figure}

\begin{figure}[htbp]
\begin{center}
\resizebox{8.6cm}{!}{\rotatebox{0}{\includegraphics{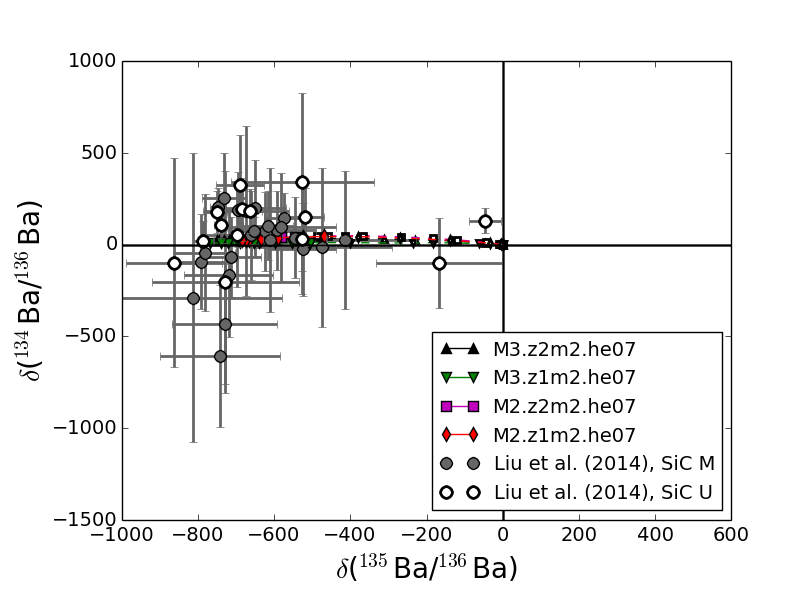}}}
\resizebox{8.6cm}{!}{\rotatebox{0}{\includegraphics{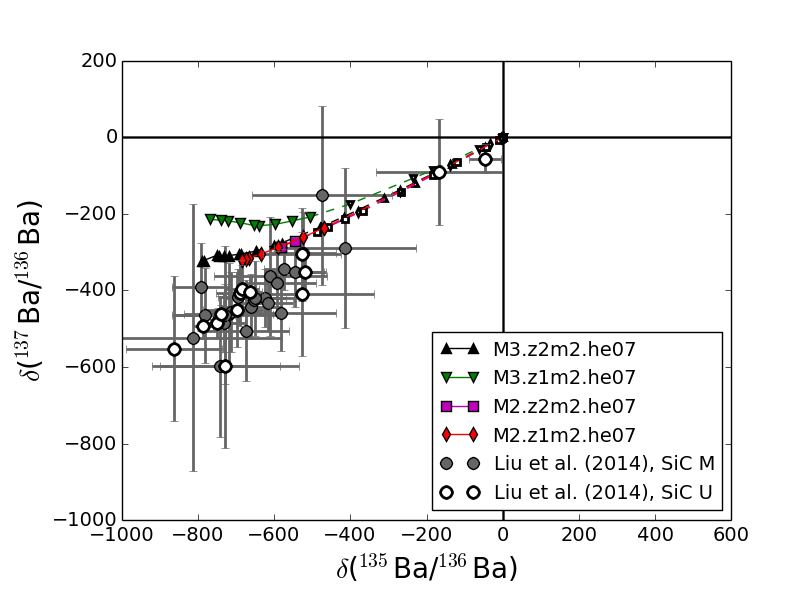}}}
\resizebox{8.6cm}{!}{\rotatebox{0}{\includegraphics{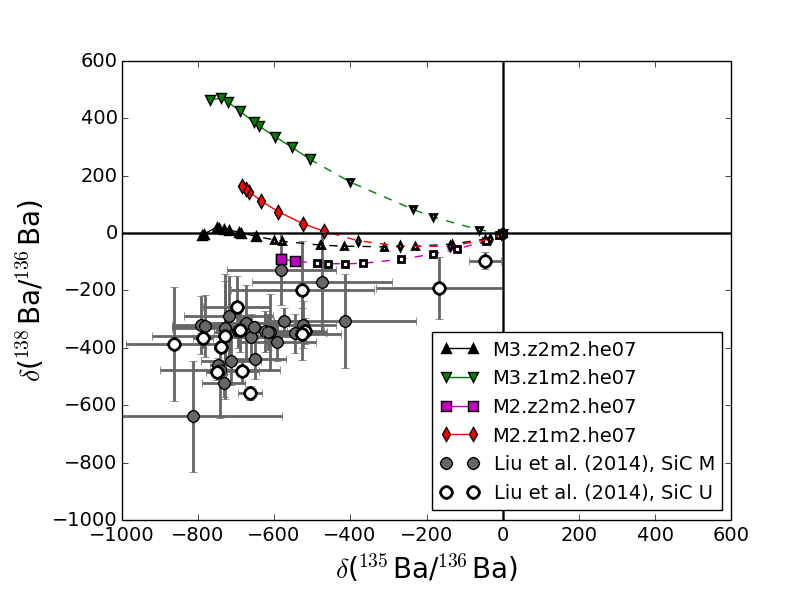}}}
\end{center}
\caption{Same as in Fig. \ref{fig:ba_ratios}, but the results are shown for the models calculated with the He07 CBM prescriptions.
}
\label{fig:ba_ratios_he07}
\end{figure}

In Fig.~\ref{fig:ba_ratios}, the Ba isotopic ratios in our calculations are compared with observations. 
The \isotope[138]{Ba}/\isotope[136]{Ba} ratio decreases with increasing metallicity and with decreasing stellar mass as a consequence of the lower neutron exposure (because of mass conservation and higher \isotope[12]{C} content in the intershell respectively \citep{lugaro:03a}). 
Furthermore, as also indicated by \cite{liu:14a} and \cite{liu:15}, the shape of the \isotope[13]{C} pocket is affecting the results. 
The uncertainty of the \isotope[14]{N}(n,p)\isotope[14]{C} rate is also relevant for the Ba isotopic ratios, since it is the main neutron poison in the \isotope[13]{C} pocket. 
In Fig.~\ref{fig:ba_ratios}, we compare the results for the models M3.z1m2, M3z1m2.hCBM and M3.z2m2.hCBM.ntest 
(\tab{tab:model_name} and \ref{tab:model_name_network_test}). 
With the exception of the grains with the lowest $\delta$(\isotope[138]{Ba}/\isotope[136]{Ba}) and $\delta$(\isotope[135]{Ba}/\isotope[136]{Ba}), the observed range is reproduced by our models within the uncertainties, and the same conclusion can be reached considering $\delta$(\isotope[137]{Ba}/\isotope[136]{Ba}).
In Fig.~\ref{fig:ba_ratios_he07} we show the same kind of comparison as in Fig.~\ref{fig:ba_ratios}, but this time for our \textsf{he07} models, showing comparable results.

To conclude, our models still present some possible limitations in the comparison with presolar grains data, although they do a much better work compared to previous models adopting CBM at the bottom of the PDCZ. 
In order to perform a more detailed comparison with presolar mainstream SiC grains, we need to calculate also AGB models with initial mass lower than M = 2\msun.
Finally, we believe that AGB models including rotation (and  magnetic field) may also have an important impact in this discussion. 
At least rotation affects the \isotope[13]{C} pocket history once the \isotope[13]{C} pocket has formed \citep[][]{piersanti:13} reducing the neutron exposure and favoring the production of light $s$-isotopes like \isotope[94]{Zr}, eventually reducing the $\delta$(\isotope[96]{Zr}/\isotope[94]{Zr}). 
Presolar grains are likely carrying the signature of these effects \citep[e.g.,][]{liu:15}.

\clearpage

\section{Conclusions}
\label{sec:conclusions}

In this work we have presented eleven new AGB stellar models with initial mass M = 2\msun\ and 3\msun, and initial metallicity Z = 0.01 and 0.02.
Additionally, we calculated seven other complete stellar runs using the same stellar structures, but using different rates for the reactions 
\isotope[14]{N}(n,p)\isotope[14]{C} and \isotope[95]{Zr}(n,$\gamma$)\isotope[96]{Zr}.
For the first time, these models study the impact of the following physics ingredients on AGB stellar evolution and nucleosynthesis: the Convective-Boundary-Mixing (CBM) at the bottom of the convective TPs according to \cite{herwig:07} simulations, 
the CBM below the TDU driven by Internal-Gravity-Waves (IGW) according to \cite{denissenkov:03}, 
and the molecular diffusion in the stellar layers where the radiative \isotope[13]{C} pocket is forming and evolves.

The main results are the following. 
Our AGB models show final \isotope[12]{C} and \isotope[16]{O} abundances in the He intershell in the order of 30-50\% and 10-20\%, respectively. These results are consistent with previous AGB simulations where overshooting was assumed to be the dominant CBM mechanism at the bottom of the PDCZ
\citep[e.g.,][]{herwig:00,lugaro:03a,pignatari:13}. 
The main reason is that the second shallower CBM term due to IGW found by \cite{herwig:07} has only a marginal impact on the He intershell during the AGB evolution. Therefore, we confirm that the CBM at the bottom of the PDCZ  can be well represented in 1D models with a single exponential-decay of the mixing efficiency, as done in previous works.  

We assume that CBM at the bottom of the convective envelope during TDU is driven by IGW instabilities, by fitting our CBM parameterization with simulations by \cite{denissenkov:03}.
We obtain radiative \isotope[13]{C} pockets with size of about 10$^{-4}$\msun \citep[consistently with][calculations]{denissenkov:03}. In particular, the default CBM setup below the TDU used in our calculations are the following: $f_\mathrm{1}$=0.014, $f_\mathrm{2}$=0.25 and $D_\mathrm{2}$=10$^{11}$ cm$^2$s$^{-1}$. 
We show that the parameter $f_\mathrm{1}$ does not affect the size of the \isotope[13]{C} pocket, which is instead dominated by the $f_\mathrm{2}$ and $D_\mathrm{2}$ parameters, i.e. by IGW.
We also provide an uncertainty study of the CBM setup on the \isotope[13]{C}-pocket size and on the \spr\ production. Since IGW appears like a suitable physics mechanism to explain the formation of the 
\isotope[13]{C} pocket, the original study by \cite{denissenkov:03} used as a guidance in our work needs be confirmed and improved by future 3D hydrodynamics simulations.

At the end of the AGB evolution we obtain an \spr\ production 0.36 \textless [$s$/Fe] \textless 0.78 and -0.23 \textless [hs/ls] \textless 0.45, which is consistent with spectroscopic observations of C-rich AGB stars. 
We explored the impact on our results of the uncertainty of the \isotope[14]{N}(n,p)\isotope[14]{C} rate. 
We showed that according to our models the increase by a factor of two of the mentioned rate at a relevant energy of $\sim$8 keV reduces the final [hs/ls] by 0.05-0.1 dex.
Similar variations are obtained by using different IGW CBM parameters. 
Therefore, the \isotope[14]{N}(n,p)\isotope[14]{C} rate needs to be constrained with an uncertainty much lower than a factor of two in order to better study the physics mechanisms responsible for the formation of the \isotope[13]{C} pocket.

We have compared our models with different types of observations, including isotopic measurements in presolar mainstream SiC grains. For this specific comparison we choose to focus our analysis on the heavy elements Zr and Ba. 
We highlight few potential limitations of our present AGB models, that needs to be explored in more details in the future. 
In particular, within the mass range considered we do not produce low enough \isotope[96]{Zr}/\isotope[94]{Zr} and \isotope[135]{Ba}/\isotope[136]{Ba} ratios as observed in all grains. 
On the other hand, present AGB models are getting much closer to fit the grain data than previous works where the CBM at the bottom of the PDCZ was used, in particular for the \isotope[96]{Zr}/\isotope[94]{Zr} ratio. The main reason of this improvement is due to the new nuclear reaction rates in the Zr region, with a much lower \isotope[95]{Zr} neutron capture cross section reducing the production of \isotope[96]{Zr} in the convective TPs.
The AGB models with initial mass M = 2\msun\ do not show any relevant signature of \isotope[96]{Zr} production, while in the models with M = 3\msun\ carry the signature of the \spr\ branching at  \isotope[95]{Zr}. Stellar models with M $<$ 2\msun should be produced in order to perform a detailed comparison with mainstream SiC presolar grains.
 
%
%
Furthermore, \cite{piersanti:13} and \cite{liu:15} showed that a physics mechanism like rotation might affect the main properties and the nucleosynthesis in the \isotope[13]{C} pocket $after$ its formation. 
This is due to the slow mixing of material (including \isotope[14]{N} and the \spr\ seed \isotope[56]{Fe}) from stellar layers located above the pocket into the thin regions where the \spr\ takes place. 
Therefore, the measurements in presolar grains may give an insight about the physics mechanisms crucial for the formation \isotope[13]{C} pocket, but also the physics affecting the pocket along its evolution before the \isotope[13]{C}($\alpha$,n)\isotope[16]{O} neutron source runs out of fuel.

In order to use presolar grain data to answer the question of what the physical mechanisms for the formation of the \cdr-pocket are, AGB stellar models need to take into account processes with a delayed impact like rotation, magnetic field and molecular diffusion.
This might be challenging, but thanks to future guidance from multi-dimensional hydrodynamics simulation it will be possible in the next few years, making AGB stars a unique laboratory to study different physical mechanisms in stellar environments and disentangle their relative effects.

Qualitatively, the same effect of rotation is triggered by molecular diffusion. 
We have shown that with the implementation adopted in this work the impact on the final \spr\ abundances is marginal. 
However, by using the default \MESA\ which adopts the controversial implementation from \citet{morel:02}, the \spr\ nucleosynthesis in the \isotope[13]{C} pocket would have been suppressed. 

\acknowledgments 
NuGrid acknowledges significant support from NSF grants PHY 02-16783
and PHY 09-22648 (Joint Institute for Nuclear Astrophysics, JINA), NSF grant PHY-1430152 (JINA Center for the Evolution of the Elements) and
EU MIRG-CT-2006-046520. The continued work on codes and in disseminating
data is made possible through funding from STFC and EU-FP7-ERC-2012-St
Grant 306901 (RH, UK), and NSERC Discovery grant (FH, Canada), and from the "Lendulet-2014" Programme of the Hungarian Academy of Sciences (MP, Hungary). UB and MP acknowledges support from SNF (Switzerland). NuGrid data is served by Canfar/CADC. RT is supported by NASA Headquarters under the NASA Earth and Planetary Science Fellowship Program through grant NNX12AL85H and was partially supported by the NASA Cosmochemistry Program through grant NNX09AG39G (to A. M. Davis).

\clearpage

\section{APPENDIX: Impact of the new \MESA\ revision and of the new nuclear reaction network}
\label{ref:models}

\begin{figure}[htbp]
\centering
\includegraphics[scale=0.3]{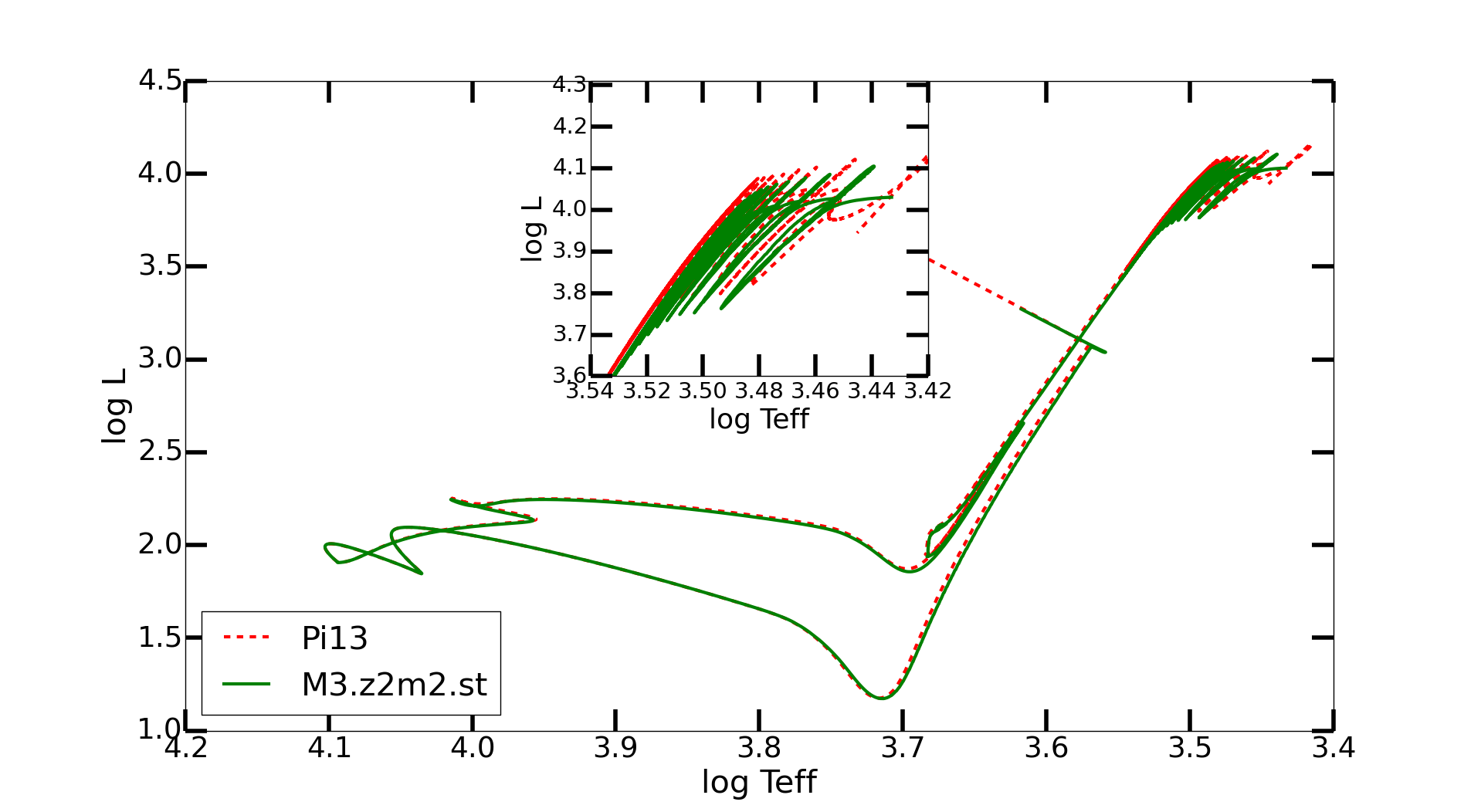}
\caption{HR diagrams for M3.z2m2.st and the analogous model calculated with \MESA\ rev. 3372 (as in \citet{pignatari:13})}
\label{comp:hrd}
\end{figure}

\begin{figure}[htbp]
\centering
\resizebox{11.4cm}{!}{\rotatebox{0}{\includegraphics{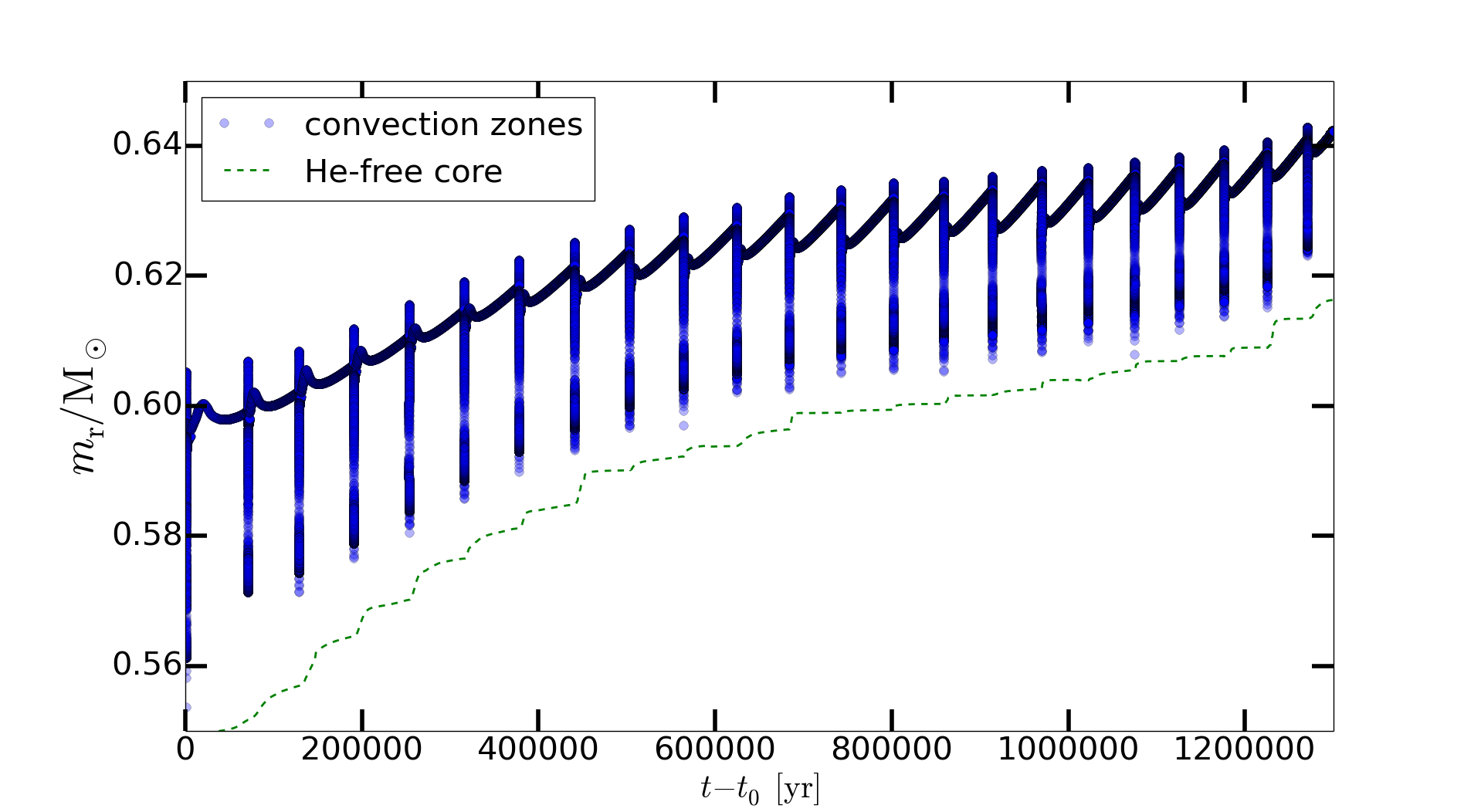}}}
\resizebox{11.4cm}{!}{\rotatebox{0}{\includegraphics{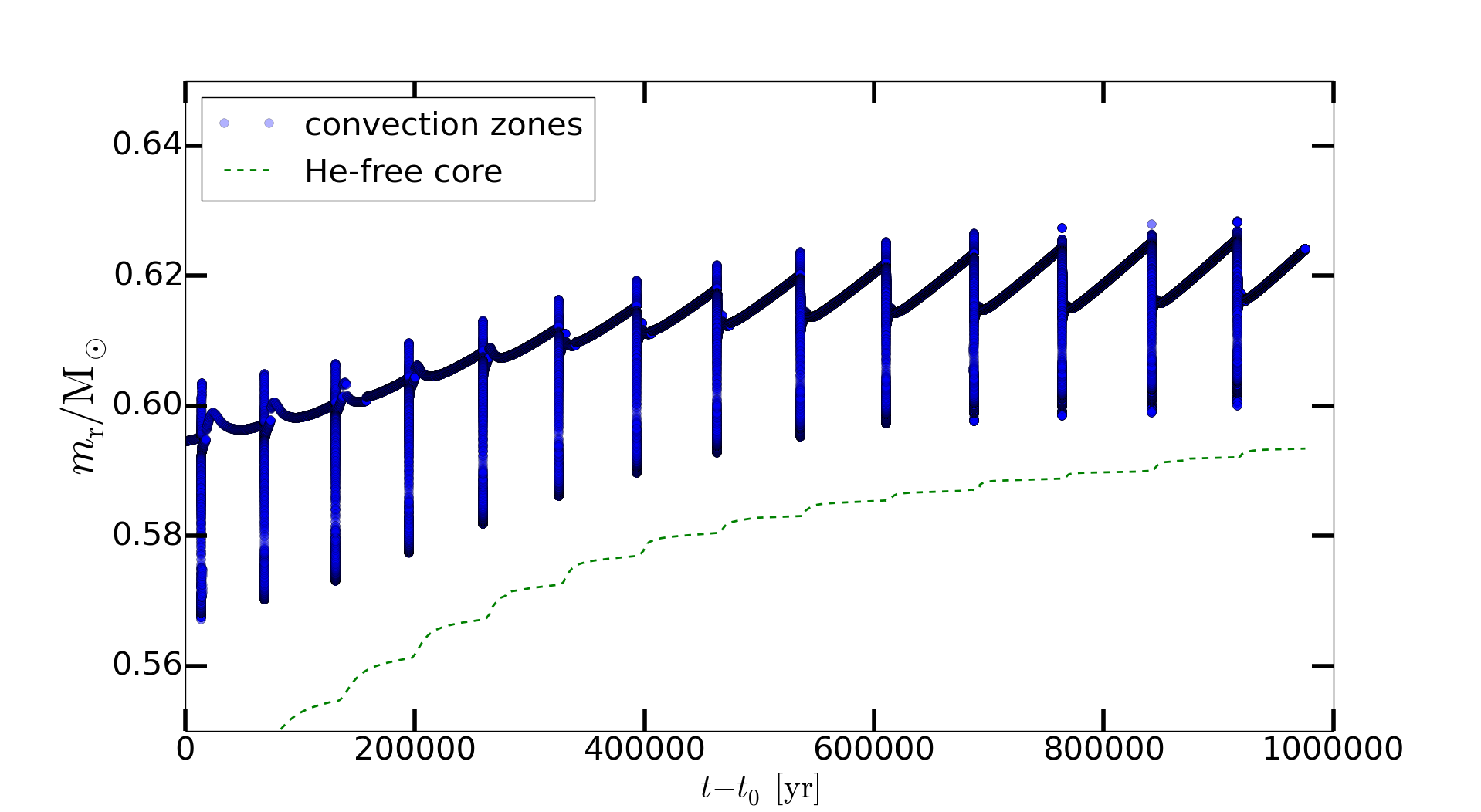}}}
\caption{
  Upper panel: Kippenhahn diagrams of the Pi13 $3\msun$ case at solar
  metallicity calculated with rev. 3372. The whole AGB phase is
  presented zoomed in the He-intershell. Lower panel: As in the upper
  panels, but for model M3.z2m2.st.}
\label{comp:kippe}
\end{figure}

For our previous stellar AGB models (Pi13) we have adopted the \MESA\ rev.\,3372, in this work we use rev.\,4219.  
We compared the results between these two different revisions. 
In \fig{comp:hrd} and \ref{comp:kippe} the HR and Kippenhahn diagrams, respectively, of two models with initial mass $M = 3\msun$ and initial metallicity $Z=0.02$, from pre-\ms\ to the tip of the AGB phase, are shown: the model M3.z2m2.st (see \tab{tab:model_name} for more details), calculated with the \MESA\ 
rev. 4219, with its analogous stellar model from Pi13.
All other model assumptions, such as mass loss, nuclear reaction rates, CBM parameters, time and spatial resolution, opacities and outer boundary choice are the same. 

\begin{figure}[htbp]
\centering
\includegraphics[scale=0.3]{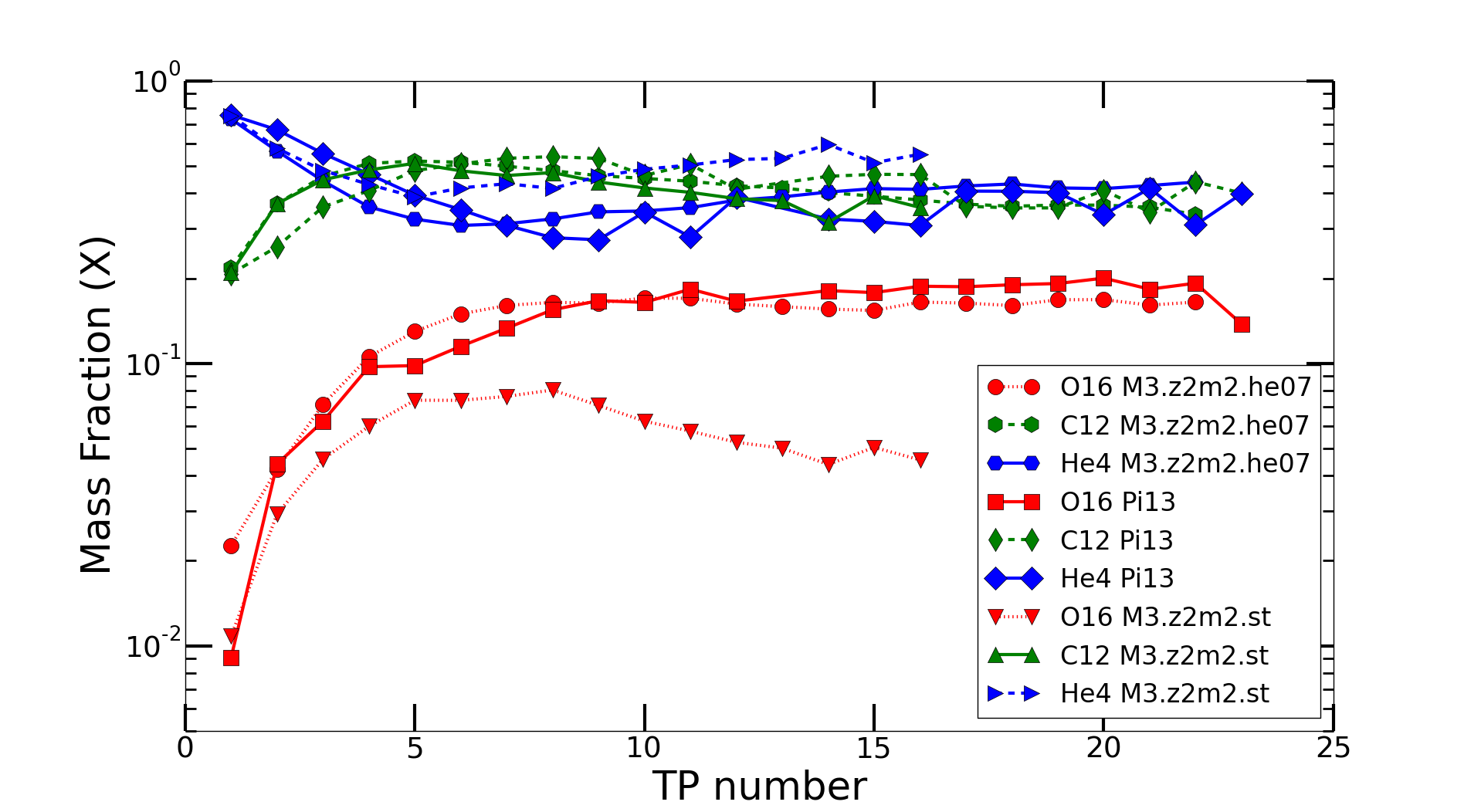}
\caption{He, C and O abundances evolution in the He Intershell as a function of the TP number along the AGB evolution of M3.z2m2.st and the analogous model calculated with \MESA\ rev. 3372 (Pi13). We also included M3.z2m2.he07 model to get the impact of mixing-length clipping during the TP comparing it with M3.z2m2.st (see text for more details).}
\label{comp:int}
\end{figure}

In \fig{comp:hrd}, the evolution in the HR diagram are extremely similar until the start of the AGB phase. Then the two models give different results. The different behavior is observed also in the C/O ratio at the surface during the AGB phase (see \fig{comp:kippe}). 
This is due to specific modifications adopted since \MESA\ rev.\, 3713, which are related to the handling of convection zones of the order of 10$^{-3}$ \msun\ or less, where the radial extent of the zone is so small that the  mixing length is larger than the size of the zone. We refer to these code modifications 
as \emph{clipping}. Therefore, from rev.\, 3713 on (including rev.\, 4219) the mixing length is limited to be smaller than the height of the zone.  
The main impact for our analysis is that small convection zones, which form under and separately from the big PDCZ during TP event using \MESA\ rev.\,3372, will be more weakly mixed because of mixing scale length being limited to the size of the zone, and the He-intershell tends to be less enriched in O than with older \MESA\ revisions. 
This is shown in \fig{comp:int}, 
where the evolution of He, C and O abundances in the He-intershell are shown for model M3.z2m2.st, M3.z2m2.he07 and the corresponding model in Pi13.  M3.z2m2.st is the only model including clipping (see \tab{tab:model_name} for more detail about model parameters).
The \hevi\ abundance in the He-intershell of model M3.z2m2.st is 30\% higher compared to Pi13 and M3.z2m2.he07, while \isotope[12]{C} and \isotope[16]{O} are smaller. 
On the other hand, M3.z2m2.he07 is similar to the results of Pi13, showing a good agreement all along the AGB evolution.
For M3.z2m2.st the final mass fractions of \isotope[4]{He}, \isotope[12]{C} and \isotope[16]{O} in the He intershell are 0.55, 0.35 and 0.045, respectively, while for the 3 \msun\ star model adopting the older \MESA\ revision the mass fractions are 0.40, 0.40 and 0.15 respectively. Finally, we obtain 0.44, 0.34, and 0.16 for the model M3.z2m2.he07. Therefore, in models with clipping like M3.z2m2.st, an f parameter larger by a factor of 2.4 at the PDCZ is needed to arrive at the same intershell abundance enhancement of O and C compared to model M3.z2m2.he07, without clipping.

The clipping is the main source of the differences seen in \fig{comp:int}. This detail of how small convection zones are treated has significant implications for the evolution of the inter-shell abundances of TP-AGB stars. 
This is affecting the parameterization of physics mixing mechanisms in 1D models, and it is not clear a priori what is the best solution. 
However, hydrodynamics simulations presented in He07 give an indication that the clipping implementation used in \MESA\ revisions 4219 should not be used, 
as we did in the set of AGB models labeled 
\textsf{he07} to simulate the CBM physics at the He intershell convective boundary.
In He07, the mixing parameters extrapolated for the parameterization in 1D models f$_1$,f$_2$ and D$_2$ should be considered more as upper limits, since for instance buoyancy due to stabilizing
chemical gradients, that might work against the mixing and reduce the size of the diffusion coefficients, was ignored. 
For this reason, the possibility that the f parameter at the bottom of the PDCZ is in fact smaller cannot be excluded.
Instead, in order to obtain similar C and O concentrations in the He intershell, the models with the clipping require an f$_1$ larger than the upper limit given by hydrodynamics simulations. 
Based on these considerations, we recommend the set of AGB models labeled \textsf{he07} as the most representative. In the paper we still consider models with clipping and enhanced f1.
Although these models do not have an ideal CBM setup at the bottom of convective TPs, their results are still valuable to study \spr\ predictions and their dependence on mixing assumptions.
Indeed, we will see in the next sections that with similar He, C and O abundances in the He intershell, similar nucleosynthesis results are obtained during the AGB phase.


There are further differences between the models calculated with the different \MESA\ revisions (see \fig{comp:kippe}). With the 
revision 4219 less TPs take place compared to the 
revision 3372.
The models Pi13 has 23 TPs, with 19 TDU events, while M3.z2m2.st 17 and 14 respectively 
TDUs are more efficient in M3.z2m2.st compared to the older revision. 
Point (1) and (2) are connected, since more efficient TDUs allow the AGB envelope to become C-rich earlier, 
and therefore to be consumed by stellar winds at earlier times.
The final surface C/O numeric ratio reached in the 3\msun\ star model by Pi13, M3.z2m2.he07 and M3.z2m2.st is 1.7, 1.6 and 2.2 respectively.

Compared to Pi13, for the present work we have adopted an updated nuclear reaction network, including a few different neutron capture reaction rates.
In particular, for this work we used the new cross sections for neutron captures on \isotope[20,21,22]{Ne} by \cite{heil:14}, \isotope[62,63]{Ni} by \cite{lederer:14}, 
and \isotope[90,91,92,93,94,95,96]{Zr} \citep[][and references therein]{tagliente:12,lugaro:14}. The only exception is model M3.z2m2.st, 
that was calculated using the same nuclear reaction network of Pi13.
While none of the rates mentioned above have a relevant impact on stellar evolution or on the total $s$-process production, 
the new Zr cross sections affect the $s$-process branching at \isotope[95]{Zr} during convective TPs. For this reason, 
we also provide here below the results for the 3\msun\ star model by Pi13, 
but using the same nuclear reaction network adopted for this work (model Pi13.newnet). 

\begin{figure}[htbp]
\centering
\resizebox{10.6cm}{!}{\rotatebox{0}{\includegraphics{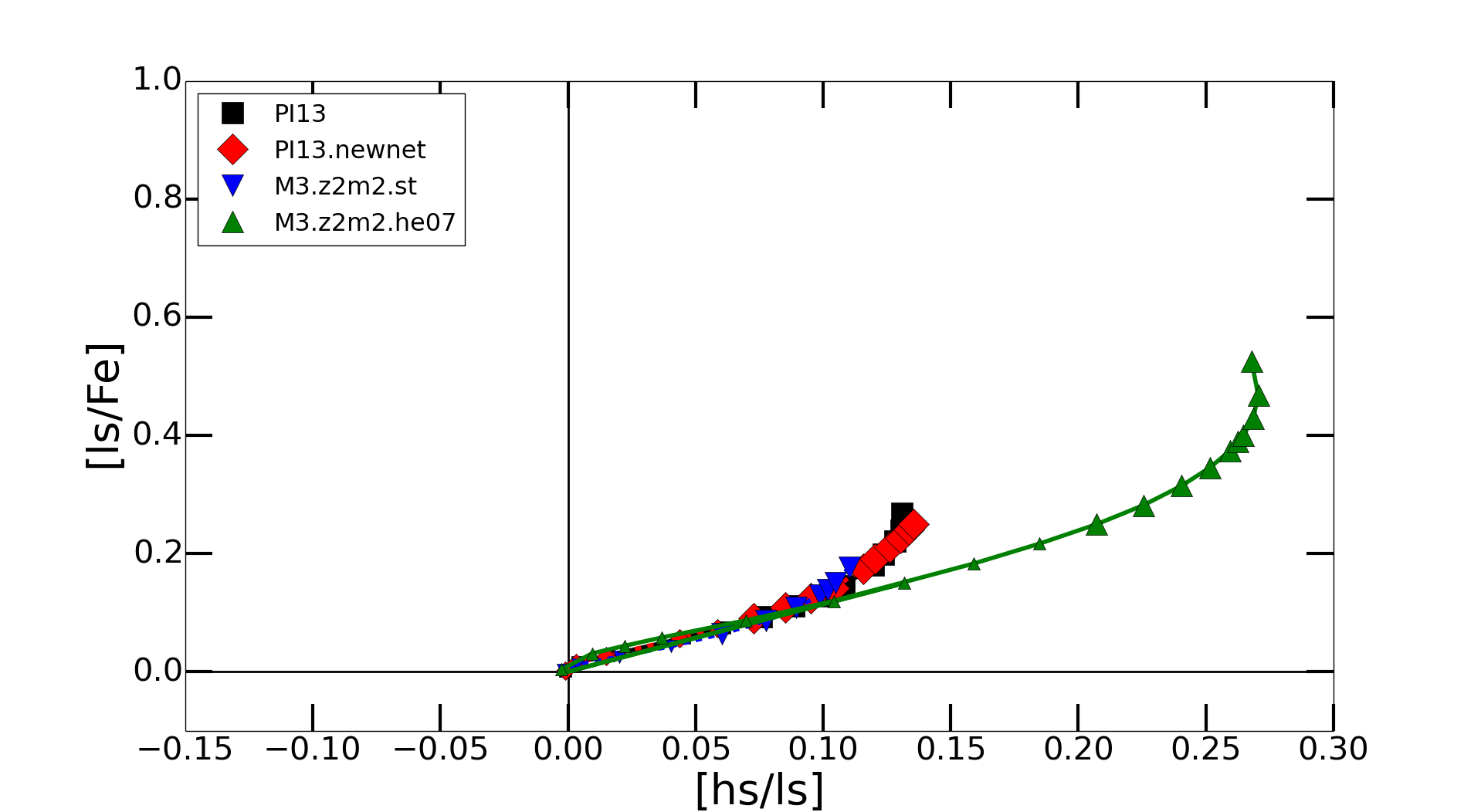}}}
\resizebox{10.6cm}{!}{\rotatebox{0}{\includegraphics{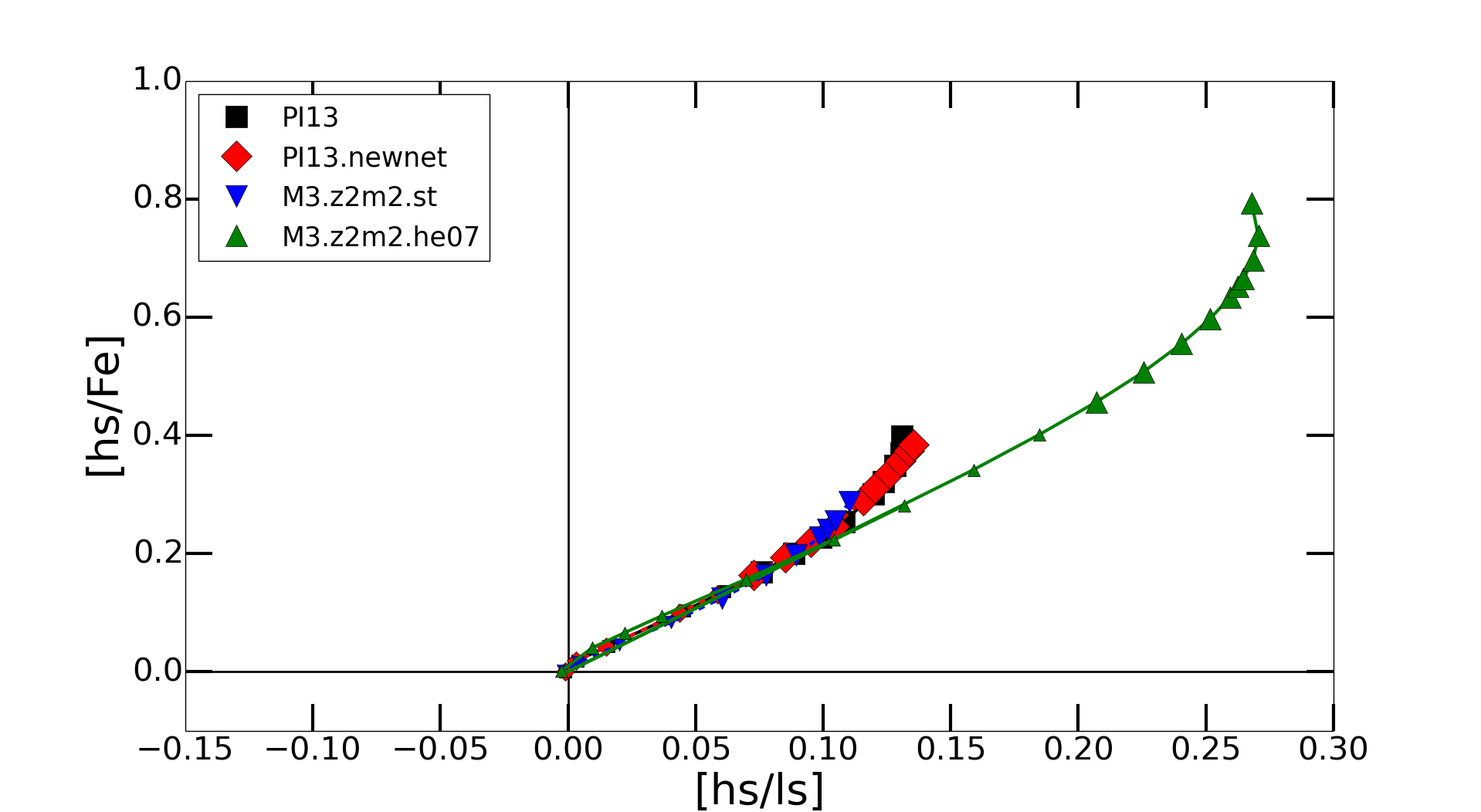}}}
\resizebox{10.6cm}{!}{\rotatebox{0}{\includegraphics{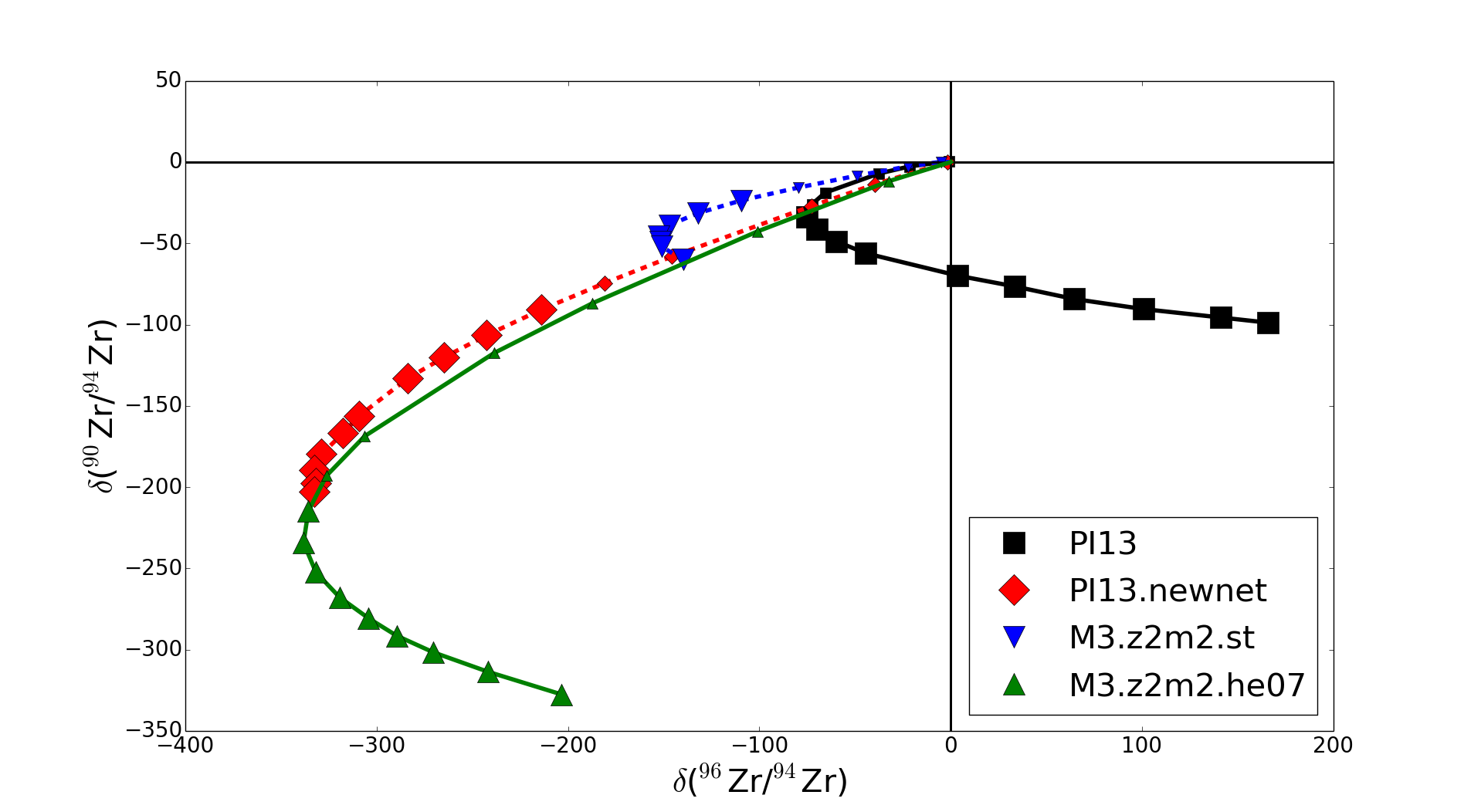}}}
\caption{Comparison between the nucleosynthesis products of the same models in figure \ref{comp:int} and model PI13.newnet. 
The evolution of the [ls/Fe] ratio (upper panel) and of the [hs/Fe] ratio (middle panel) are shown in comparison with the [hs/ls] ratio. 
In particular, each marker represents a TP during the AGB phase. Larger markers are used when the surface C/O ratio exceeds 1. 
In the lower panel the evolution of $\delta$(\isotope[90]{Zr}/\isotope[94]{Zr}) and $\delta$(\isotope[96]{Zr}/\isotope[94]{Zr}) ratios
are shown for the same models in the previous panels. The isotopic ratios are shown in $\delta$=((ratio/solar)-1)$\times$1000.
}
\label{comp:ppd}
\end{figure}

\fig{comp:ppd} shows the differences arising from the nucleosynthesis calculations of these four models. 
Due to the lower number of TDUs, the M3.z2m2.st shows a smaller \spr\ enrichment at both the Sr peak and the Ba peaks,
only partially compensated by the larger TDU efficiency.
On the other hand we obtain similar [hs/ls] ratios, defining it as the average logarithmic ratio normalized to solar ([hs/ls]=log(hs/ls)$-$log(hs/ls)$_{\odot}$, 
a similar definition is given to the [ls/Fe] and [hs/Fe] indices). 
The model M3.z2m2.he07 shows a much larger $s$-process enrichment compared to the other two models. This is due to the different CBM implementation adopted at the bottom of the convective envelope during TDU.


The evolution of the Zr isotopic ratios shows strong differences. 
The use of new Zr neutron capture cross sections (and in particular of the \isotope[95]{Zr} cross section, 
that is more than a factor of two lower than the rate used by Pi13) allows to obtain much lower \isotope[96]{Zr}/\isotope[94]{Zr} ratios, 
compared to the results of Pi13 model and Pi13.newnet. On the other hand, M3.z2m2.st (adopting the same nuclear reaction network of Pi13) 
shows milder $s$-process signatures compared to Pi13 and Pi13.newnet models, 
due to the lower amount of TPs and to the lower temperatures obtained at the bottom of convective TPs. 
This is an effect of the larger \isotope[4]{He} abundance in the He intershell of M3.z2m2.st, allowing the He-burning activation at lower temperatures (see Fig.~\ref{comp:int} and previous discussion). 
The new Zr cross sections have an impact on the final Zr isotopic rations comparable to the differences related to stellar model uncertainties.
The \isotope[96]{Zr}/\isotope[94]{Zr} ratio is considered an indicator of the \isotope[22]{Ne}($\alpha$,n)\isotope[25]{Mg} efficiency at the bottom of convective TPs \citep[e.g.,][]{lugaro:03a,bisterzo:15}. 


Concluding, we 
showed that the main source of these differences is coming from the different handling of small convective zones in the default setup of the two revisions. A priori it is not clear what the best implementation for 1D models is. However, hydrodynamics simulations clearly indicate that the no clipping setup (\tab{tab:model_name}) should be favored.
Thanks to the example of the Zr isotopes, we have seen that nuclear uncertainties are also crucial: their relevance can be comparable to stellar uncertainties.

\clearpage

\bibliography{astro}

\clearpage

\begin{table}
\begin{center}
\caption{List of AGB stellar models and their relevant parameters:
  initial mass, initial metallicity and CBM parameterization.  The CBM
  parameterization can be given by a single exponential decreasing
  profile (sf), as in Pi13, or by a double exponential
  decreasing profile (df) adopted in this work, with or without
  limiting the mixing length to the size of the convections zones
  (clipping).  The CBM parameters are given below the PDCZ (f1, D2 and
  f2) and below the envelope convection during the TDU (f1*, D2* and f2*).  }
\begin{tabular}{lccccccccccc}
\hline
name & mass [M$_{\odot}$] &  metallicity &  CBM & f1 & D2 & f2 & f1* & D2* & f2* & clipping \\
\hline
\hline
M3.z2m2.st & 3.0  & 0.02 	& sf & 0.008 & -  & - & 0.126 & -  & - & yes\\ 		
M3.z2m2	& 3.0 & 0.02	& df & 0.024 & 10$^{5}$ & 0.14 & 0.014 & 10$^{11}$  & 0.25 & yes\\ 	
M3.z1m2	& 3.0 & 0.01	& df & 0.024 & 10$^{5}$ & 0.14 & 0.014 & 10$^{11}$  & 0.25 & yes \\ 
M2.z2m2	& 2.0 & 0.02	& df & 0.024 & 10$^{5}$ & 0.14 & 0.014 & 10$^{11}$ & 0.25 & yes\\ 
M2.z1m2	& 2.0 & 0.01	& df & 0.024 & 10$^{5}$ & 0.14 & 0.014 & 10$^{11}$  & 0.25 & yes\\ 
M3.z1m2.hCBM & 3.0 & 0.01 & df & 0.024 & 10$^{5}$ & 0.14 & 0.014 & 10$^{12}$  & 0.27 & yes\\ 
M2.z2m2.hCBM & 2.0 & 0.02 & df & 0.024 & 10$^{5}$ & 0.14 & 0.014 & 10$^{12}$  & 0.27 & yes\\ 	
M3.z2m2.he07 & 3.0 & 0.02 & df & 0.010 & 10$^{5}$ & 0.14 & 0.014 & 10$^{11}$  & 0.25 & no\\ 	
M3.z1m2.he07 & 3.0 & 0.01 & df & 0.010 & 10$^{5}$ & 0.14 & 0.014 & 10$^{11}$  & 0.25 & no\\ 
M2.z2m2.he07 & 2.0 & 0.02 & df & 0.010 & 10$^{5}$ & 0.14 & 0.014 & 10$^{11}$ & 0.25 & no\\ 
M2.z1m2.he07 & 2.0 & 0.01 & df & 0.010 & 10$^{5}$ & 0.14 & 0.014 & 10$^{11}$  & 0.25 & no\\ 
\noalign{\smallskip}
\hline
\end{tabular}
\label{tab:model_name}
\end{center}
\end{table}

\clearpage

\begin{center}
\tiny\begin{table}
\caption{AGB stars properties.} 
\scalebox{0.7}{
\begin{tabular}{cccccccccccccc}
\hline
Name & $M_{ini}$ & $Z_{ini}$ &  $m_c$       & $logL_{\ast}$ & $R_{\ast}$  & $N_{TP}$ & $N_{TDUP}$ & $t_{TPI}$  & $\Delta M_{Dmax}$ & $M_D$      & $t_{ip}$ & $M_{lost}$ & $logT_{PDCZ,max}$ \\
& {[$M_{\odot}$]} &  &  {[$M_{\odot}$]}  &  {[$L_{\odot}$]}   & {[$R_{\odot}$]} &         &                &  {[$10^6 yr$]} & {[$10^{-2}M_{\odot}$]} & {[$10^{-2}M_{\odot}$]} & {[$10^{3}yr$]}     & {[$M_{\odot}$]} &  {[$K$]} \\
\hline 
M2.z1m2      & 2.00   & 0.01 & 0.495  & 3.47 & 169 &  24    &  12 &  1.265E+03 &  0.8       &  6.348         &  164.5  & 1.38      &  8.476  \\
M2.z2m2      & 2.00   & 0.02 & 0.515  & 3.59 & 229 &  24    &  12 &  1.357E+03 &  0.7       &  5.563         &  112.6  & 1.34      &  8.394  \\
M3.z1m2      & 3.00   & 0.01 & 0.640  & 3.97 & 308 &  13    &  12 &  4.092E+02 &  1.2       &  9.324         &  57.7   & 2.33      &  8.480  \\
M3.z2m2      & 3.00   & 0.02 & 0.588  & 3.89 & 302 &  21    &  18 &  4.798E+02 &  1.3       &  12.983        &  67.6   & 2.36      &  8.487  \\
M2.z2m2.hCBM & 2.00   & 0.02 & 0.514  & 3.58 & 223 &  21    &  12 &  1.357E+03 &  0.7       &  4.897         &  122.5  & 1.35      &  8.487  \\
M3.z1m2.hCBM & 3.00   & 0.01 & 0.645  & 3.98 & 310 &  12    &  11 &  4.125E+02 &  1.4       &  9.874         &  58.8   & 2.33      &  8.488  \\
M3.z2m2.st   & 3.00   & 0.02 & 0.593  & 3.87 & 300 &  14    &  11 &  4.835E+02 &  1.0       &  7.188         &  69.4   & 2.35      &  8.400  \\
M2.z1m2.he07 & 2.00   & 0.01 & 0.497  & 3.48 & 170 &  25    &  13 &  1.279E+03 &  0.4       &  3.748         &  146.3  & 1.36      &  8.460  \\
M2.z2m2.he07 & 2.00   & 0.02 & 0.510  & 3.58 & 223 &  27    &  14 &  1.406E+03 &  0.4       &  3.243         &  108.0  & 1.32      &  8.463  \\
M3.z1m2.he07 & 3.00   & 0.01 & 0.647  & 3.99 & 312 &  15    &  14 &  4.127E+02 &  0.7       &  6.426         &  46.3   & 2.30      &  8.247  \\
M3.z2m2.he07 & 3.00   & 0.02 & 0.592  & 3.85 & 281 &  23    &  19 &  4.818E+02 &  0.8       &  7.129         &  58.4   & 2.34      &  8.471  \\

\hline
\multicolumn{13}{l}{$M_{ini}$: Initial stellar mass.} \\
\multicolumn{13}{l}{$Z_{ini}$: Initial metallicity.} \\
\multicolumn{13}{l}{  $m_c$: H-free core mass at the first TP.}\\
\multicolumn{13}{l}{ $L_{\ast}$: Approximated mean Luminosity.}\\
\multicolumn{13}{l}{ $R_{\ast}$ : Approximated mean radius.}\\
\multicolumn{13}{l}{ $N_{TP}$: Number of TP's.}\\
\multicolumn{13}{l}{ $N_{TDUP}$ : Number of TP's with TDUP.}\\
\multicolumn{13}{l}{ $t_{TPI}$: Time at first TP.} \\
\multicolumn{13}{l}{ $\Delta M_{Dmax}$: Maximum dredged-up mass after a single TP.} \\
\multicolumn{13}{l}{ $M_D$: Total dredged-up mass of all TPs.} \\
\multicolumn{13}{l}{ $t_{ip}$ : Average interpulse duration of TPs.} \\
\multicolumn{13}{l}{$M_{lost}$: Total mass lost during the evolution.} \\
\multicolumn{13}{l}{$T_{PDCZ,max}$: Maximum temperature during the TPAGB phase.}\\
\end{tabular}
}
\label{tab:model_prop}
\end{table}
\end{center}

\clearpage

\begin{center}
\tiny\begin{longtable}{ccccccccccc}
\caption{TP-AGB evolution properties of stellar models presented in this work.}\\
\hline
 TP & $DUP_\lambda$ & $t_{TP}$ & $T_{FBOT}$ & $T_{HES}$ & $T_{HS}$ & $T_{CEB}$ & $m_{FBOT}$ & $m_{HTP}$ & $m_{D,max}$ & $M_{\ast}$\\
 &  & [$yrs$] & [$K$] & [$K$] & [$K$] & [$K$] & [$M_{\odot}$] & [$M_{\odot}$] & [$M_{\odot}$] & [$M_{\odot}$] \\
\hline
\multicolumn{11}{c}{M2.z1m2} \\
\hline
1 &0.00  & 0.00E+00  &  8.31&  8.15&  7.09    &  6.25 &  0.4452 &  0.4948       &  0.4961         &  1.978 \\
2 &0.00  & 7.43E+05  &  8.36&  8.15&  7.14    &  6.31 &  0.4574 &  0.5056       &  0.5063         &  1.978 \\
3 &0.00  & 1.15E+06  &  8.38&  8.16&  7.16    &  6.27 &  0.4677 &  0.5131       &  0.5138         &  1.978 \\
4 &0.00  & 1.33E+06  &  8.37&  8.15&  7.17    &  6.33 &  0.4721 &  0.5165       &  0.5174         &  1.978 \\
5 &0.00  & 1.50E+06  &  8.38&  7.75&  7.15    &  6.33 &  0.4758 &  0.5208       &  0.5215         &  1.977 \\
6 &0.00  & 1.68E+06  &  8.41&  7.76&  7.16    &  6.33 &  0.4808 &  0.5261       &  0.5267         &  1.977 \\
7 &0.00  & 1.86E+06  &  8.41&  7.78&  7.18    &  6.36 &  0.4873 &  0.5319       &  0.5324         &  1.977 \\
8 &0.00  & 2.02E+06  &  8.43&  7.79&  7.25    &  6.37 &  0.4948 &  0.5381       &  0.5385         &  1.976 \\
9 &0.00  & 2.18E+06  &  8.42&  7.79&  7.27    &  6.37 &  0.5029 &  0.5444       &  0.5447         &  1.975 \\
10&0.00  & 2.33E+06  &  8.44&  7.79&  7.31    &  6.40 &  0.5114 &  0.5508       &  0.5511         &  1.974 \\
11&0.00  & 2.47E+06  &  8.43&  7.80&  7.32    &  6.39 &  0.5198 &  0.5572       &  0.5574         &  1.972 \\
12&0.00  & 2.60E+06  &  8.45&  7.80&  7.57    &  6.41 &  0.5280 &  0.5636       &  0.5636         &  1.970 \\
13&0.13  & 2.72E+06  &  8.44&  7.79&  7.64    &  6.41 &  0.5362 &  0.5699       &  0.5693         &  1.967 \\
14&0.26  & 2.83E+06  &  8.46&  7.81&  7.66    &  6.43 &  0.5437 &  0.5758       &  0.5742         &  1.964 \\
15&0.42  & 2.94E+06  &  8.45&  8.11&  7.66    &  6.43 &  0.5504 &  0.5810       &  0.5783         &  1.960 \\
16&0.55  & 3.04E+06  &  8.46&  8.13&  7.67    &  6.45 &  0.5561 &  0.5854       &  0.5815         &  1.954 \\
17&0.66  & 3.14E+06  &  8.47&  8.13&  7.70    &  6.47 &  0.5608 &  0.5890       &  0.5841         &  1.947 \\
18&0.75  & 3.24E+06  &  8.47&  7.93&  7.46    &  6.29 &  0.5647 &  0.5920       &  0.5862         &  1.937 \\
19&0.82  & 3.34E+06  &  8.47&  8.13&  7.71    &  6.55 &  0.5679 &  0.5945       &  0.5877         &  1.925 \\
20&0.88  & 3.43E+06  &  8.47&  8.13&  7.70    &  6.50 &  0.5704 &  0.5964       &  0.5887         &  1.876 \\
21&0.91  & 3.52E+06  &  8.48&  8.12&  7.68    &  6.56 &  0.5723 &  0.5979       &  0.5896         &  1.795 \\
22&0.88  & 3.61E+06  &  8.48&  8.12&  7.68    &  6.49 &  0.5739 &  0.5990       &  0.5907         &  1.682 \\
23&0.75  & 3.70E+06  &  8.46&  8.12&  7.68    &  6.49 &  0.5759 &  0.6001       &  0.5931         &  1.522 \\
24&0.46  & 3.78E+06  &  8.44&  8.21&  7.68    &  6.34 &  0.5800 &  0.6023       &  0.5941         &  1.233 \\
 \\
\hline
\multicolumn{11}{c}{M2.z2m2} \\
\hline
1 &0.00  & 0.00E+00  &  8.23&  8.15&  7.16    &  6.28 &  0.4737 &  0.5145       &  0.5151         &  1.951 \\
2 &0.00  & 3.19E+05  &  8.26&  8.17&  7.17    &  6.37 &  0.4796 &  0.5203       &  0.5209         &  1.951 \\
3 &0.00  & 4.77E+05  &  8.27&  8.16&  7.19    &  6.27 &  0.4824 &  0.5233       &  0.5240         &  1.950 \\
4 &0.00  & 6.11E+05  &  8.30&  8.07&  7.18    &  6.41 &  0.4854 &  0.5269       &  0.5275         &  1.950 \\
5 &0.00  & 7.51E+05  &  8.30&  7.83&  7.22    &  6.39 &  0.4895 &  0.5315       &  0.5320         &  1.950 \\
6 &0.00  & 8.94E+05  &  8.30&  7.84&  7.22    &  6.50 &  0.4949 &  0.5368       &  0.5372         &  1.949 \\
7 &0.00  & 1.03E+06  &  8.30&  7.85&  7.29    &  6.50 &  0.5016 &  0.5425       &  0.5429         &  1.948 \\
8 &0.00  & 1.17E+06  &  8.29&  7.86&  7.30    &  6.46 &  0.5091 &  0.5484       &  0.5487         &  1.947 \\
9 &0.00  & 1.29E+06  &  8.29&  7.86&  7.36    &  6.45 &  0.5168 &  0.5544       &  0.5547         &  1.945 \\
10&0.00  & 1.41E+06  &  8.28&  7.87&  7.34    &  6.49 &  0.5246 &  0.5604       &  0.5606         &  1.943 \\
11&0.00  & 1.52E+06  &  8.28&  7.87&  7.44    &  6.49 &  0.5323 &  0.5664       &  0.5665         &  1.940 \\
12&0.00  & 1.62E+06  &  8.27&  7.85&  7.56    &  6.61 &  0.5399 &  0.5724       &  0.5722         &  1.936 \\
13&0.13  & 1.72E+06  &  8.27&  7.86&  7.73    &  6.62 &  0.5471 &  0.5782       &  0.5776         &  1.932 \\
14&0.25  & 1.81E+06  &  8.27&  7.91&  7.72    &  6.91 &  0.5541 &  0.5837       &  0.5823         &  1.927 \\
15&0.38  & 1.90E+06  &  8.28&  8.05&  7.74    &  7.12 &  0.5602 &  0.5886       &  0.5863         &  1.920 \\
16&0.49  & 1.98E+06  &  8.27&  8.11&  7.72    &  7.39 &  0.5657 &  0.5929       &  0.5897         &  1.911 \\
17&0.59  & 2.06E+06  &  8.27&  8.11&  7.74    &  7.47 &  0.5704 &  0.5965       &  0.5925         &  1.900 \\
18&0.68  & 2.14E+06  &  8.27&  8.11&  7.69    &  7.58 &  0.5743 &  0.5997       &  0.5948         &  1.886 \\
19&0.75  & 2.22E+06  &  8.27&  8.11&  7.69    &  7.63 &  0.5778 &  0.6023       &  0.5967         &  1.869 \\
20&0.77  & 2.30E+06  &  8.29&  8.11&  7.71    &  7.58 &  0.5806 &  0.6045       &  0.5985         &  1.848 \\
21&0.80  & 2.37E+06  &  8.26&  8.11&  7.72    &  7.64 &  0.5831 &  0.6064       &  0.6001         &  1.822 \\
22&0.82  & 2.45E+06  &  8.26&  8.11&  7.70    &  7.62 &  0.5854 &  0.6082       &  0.6015         &  1.793 \\
23&0.83  & 2.52E+06  &  8.25&  8.11&  7.72    &  7.42 &  0.5875 &  0.6097       &  0.6029         &  1.683 \\
24&0.79  & 2.59E+06  &  8.29&  8.21&  7.41    &  7.46 &  0.5894 &  0.6113       &  0.6035         &  1.437 \\
 \\
\hline
\multicolumn{11}{c}{M3.z1m2} \\
\hline
1 &0.00  & 0.00E+00  &  8.39&  8.20&  7.59    &  6.37 &  0.6192 &  0.6397       &  0.6393         &  2.973 \\
2 &0.39  & 4.85E+04  &  8.41&  8.17&  7.68    &  6.54 &  0.6218 &  0.6421       &  0.6411         &  2.972 \\
3 &0.59  & 9.69E+04  &  8.31&  8.16&  7.68    &  6.53 &  0.6243 &  0.6450       &  0.6428         &  2.970 \\
4 &0.75  & 1.48E+05  &  8.42&  8.15&  7.69    &  6.63 &  0.6268 &  0.6477       &  0.6440         &  2.967 \\
5 &0.91  & 2.03E+05  &  8.46&  8.14&  7.70    &  6.64 &  0.6293 &  0.6498       &  0.6445         &  2.963 \\
6 &1.04  & 2.59E+05  &  8.43&  8.13&  7.67    &  6.75 &  0.6310 &  0.6512       &  0.6443         &  2.957 \\
7 &1.08  & 3.18E+05  &  8.46&  8.12&  7.65    &  6.76 &  0.6319 &  0.6517       &  0.6437         &  2.909 \\
8 &1.12  & 3.78E+05  &  8.46&  8.12&  7.63    &  6.82 &  0.6321 &  0.6517       &  0.6427         &  2.832 \\
9 &1.18  & 4.40E+05  &  8.47&  8.12&  7.60    &  6.77 &  0.6318 &  0.6513       &  0.6411         &  2.731 \\
10&1.17  & 5.02E+05  &  8.48&  8.11&  7.60    &  6.79 &  0.6308 &  0.6501       &  0.6395         &  2.601 \\
11&1.20  & 5.65E+05  &  8.45&  8.11&  7.60    &  6.79 &  0.6297 &  0.6489       &  0.6376         &  2.426 \\
12&1.18  & 6.29E+05  &  8.46&  8.11&  7.65    &  6.91 &  0.6280 &  0.6474       &  0.6358         &  2.168 \\
13&1.23  & 6.93E+05  &  8.46&  8.10&  7.65    &  7.11 &  0.6265 &  0.6458       &  0.6336         &  1.713 \\
 \\
\hline
\multicolumn{11}{c}{M3.z2m2} \\
\hline
1 &0.00  & 0.00E+00  &  8.31&  8.19&  7.21    &  6.32 &  0.5644 &  0.5879       &  0.5888         &  2.978 \\
2 &0.00  & 5.72E+04  &  8.37&  8.18&  7.26    &  6.51 &  0.5645 &  0.5903       &  0.5906         &  2.978 \\
3 &0.00  & 1.22E+05  &  8.39&  8.18&  7.30    &  6.45 &  0.5670 &  0.5936       &  0.5937         &  2.978 \\
4 &0.10  & 1.89E+05  &  8.42&  8.17&  7.57    &  6.61 &  0.5706 &  0.5977       &  0.5974         &  2.977 \\
5 &0.18  & 2.57E+05  &  8.43&  8.15&  7.59    &  6.53 &  0.5751 &  0.6021       &  0.6015         &  2.976 \\
6 &0.31  & 3.25E+05  &  8.43&  8.11&  7.67    &  6.62 &  0.5803 &  0.6067       &  0.6052         &  2.974 \\
7 &0.46  & 3.93E+05  &  8.42&  8.14&  7.68    &  6.56 &  0.5853 &  0.6109       &  0.6084         &  2.972 \\
8 &0.60  & 4.59E+05  &  8.44&  8.14&  7.72    &  6.55 &  0.5899 &  0.6145       &  0.6109         &  2.969 \\
9 &0.73  & 5.25E+05  &  8.46&  8.14&  7.71    &  6.64 &  0.5937 &  0.6174       &  0.6127         &  2.966 \\
10&0.83  & 5.91E+05  &  8.44&  8.14&  7.69    &  6.64 &  0.5967 &  0.6196       &  0.6139         &  2.962 \\
11&0.89  & 6.58E+05  &  8.46&  8.13&  7.70    &  6.73 &  0.5990 &  0.6213       &  0.6147         &  2.956 \\
12&0.96  & 7.24E+05  &  8.46&  8.13&  7.69    &  6.66 &  0.6006 &  0.6225       &  0.6150         &  2.950 \\
13&1.02  & 7.92E+05  &  8.45&  7.76&  7.64    &  6.30 &  0.6017 &  0.6233       &  0.6148         &  2.943 \\
14&1.04  & 8.60E+05  &  8.45&  7.51&  7.66    &  6.08 &  0.6021 &  0.6236       &  0.6145         &  2.935 \\
15&1.08  & 9.29E+05  &  8.39&  8.12&  7.66    &  6.73 &  0.6022 &  0.6235       &  0.6138         &  2.877 \\
16&1.09  & 9.98E+05  &  8.49&  8.11&  7.63    &  6.99 &  0.6019 &  0.6233       &  0.6129         &  2.787 \\
17&1.11  & 1.07E+06  &  8.42&  8.09&  7.65    &  7.71 &  0.6014 &  0.6229       &  0.6119         &  2.672 \\
18&1.09  & 1.14E+06  &  8.34&  8.10&  7.67    &  7.66 &  0.6006 &  0.6219       &  0.6109         &  2.531 \\
19&1.10  & 1.21E+06  &  8.33&  8.10&  7.67    &  7.68 &  0.5999 &  0.6212       &  0.6099         &  2.349 \\
20&1.06  & 1.28E+06  &  8.21&  8.10&  7.73    &  7.52 &  0.5991 &  0.6203       &  0.6092         &  2.103 \\
21&1.19  & 1.35E+06  &  8.23&  8.04&  7.79    &  7.54 &  0.5985 &  0.6197       &  0.6079         &  1.721 \\
 \\
\hline
\multicolumn{11}{c}{M2.z2m2.hCBM} \\
\hline
1 &0.00  & 0.00E+00  &  8.31&  8.21&  7.39    &  6.22 &  0.4743 &  0.5141       &  0.5153         &  1.950 \\
2 &0.00  & 2.36E+05  &  8.16&  8.24&  7.42    &  6.27 &  0.4783 &  0.5177       &  0.5187         &  1.950 \\
3 &0.00  & 4.71E+05  &  8.19&  8.18&  7.23    &  6.25 &  0.4840 &  0.5233       &  0.5238         &  1.950 \\
4 &0.00  & 7.16E+05  &  8.28&  8.13&  7.25    &  6.35 &  0.4922 &  0.5308       &  0.5311         &  1.949 \\
5 &0.00  & 8.40E+05  &  8.25&  8.22&  7.36    &  6.55 &  0.4980 &  0.5341       &  0.5347         &  1.949 \\
6 &0.00  & 9.63E+05  &  8.26&  8.10&  7.52    &  6.85 &  0.5021 &  0.5388       &  0.5392         &  1.948 \\
7 &0.00  & 1.09E+06  &  8.27&  8.20&  7.61    &  7.21 &  0.5078 &  0.5440       &  0.5443         &  1.947 \\
8 &0.00  & 1.21E+06  &  8.25&  8.09&  7.49    &  7.41 &  0.5142 &  0.5495       &  0.5497         &  1.945 \\
9 &0.00  & 1.32E+06  &  8.25&  8.11&  7.39    &  7.55 &  0.5212 &  0.5552       &  0.5553         &  1.944 \\
10&0.06  & 1.43E+06  &  8.28&  7.81&  7.62    &  7.61 &  0.5285 &  0.5610       &  0.5608         &  1.941 \\
11&0.11  & 1.54E+06  &  8.28&  7.75&  7.65    &  7.21 &  0.5357 &  0.5667       &  0.5662         &  1.938 \\
12&0.22  & 1.64E+06  &  8.23&  7.76&  7.66    &  7.15 &  0.5426 &  0.5724       &  0.5711         &  1.934 \\
13&0.35  & 1.74E+06  &  8.29&  8.02&  7.62    &  7.19 &  0.5491 &  0.5776       &  0.5753         &  1.928 \\
14&0.47  & 1.83E+06  &  8.28&  8.13&  7.63    &  7.37 &  0.5549 &  0.5823       &  0.5791         &  1.921 \\
15&0.57  & 1.92E+06  &  8.28&  8.14&  7.63    &  7.21 &  0.5599 &  0.5865       &  0.5823         &  1.911 \\
16&0.65  & 2.01E+06  &  8.27&  8.14&  7.63    &  7.07 &  0.5643 &  0.5902       &  0.5851         &  1.899 \\
17&0.71  & 2.11E+06  &  8.27&  8.14&  7.64    &  6.83 &  0.5682 &  0.5934       &  0.5875         &  1.883 \\
18&0.76  & 2.19E+06  &  8.26&  8.13&  7.63    &  6.67 &  0.5717 &  0.5963       &  0.5897         &  1.863 \\
19&0.77  & 2.28E+06  &  8.26&  8.16&  7.68    &  6.32 &  0.5748 &  0.5988       &  0.5918         &  1.759 \\
20&0.72  & 2.37E+06  &  8.47&  8.13&  7.61    &  6.60 &  0.5777 &  0.6011       &  0.5944         &  1.598 \\
21&0.53  & 2.45E+06  &  8.47&  8.08&  7.47    &  6.42 &  0.5815 &  0.6038       &  0.5958         &  1.311 \\
 \\
\hline
\multicolumn{11}{c}{M3.z1m2.hCBM} \\
\hline
1 &0.00  & 0.00E+00  &  8.39&  8.20&  7.56    &  6.46 &  0.6251 &  0.6445       &  0.6437         &  2.972 \\
2 &0.65  & 4.59E+04  &  8.40&  8.17&  7.59    &  6.63 &  0.6271 &  0.6466       &  0.6448         &  2.970 \\
3 &0.83  & 9.31E+04  &  8.41&  8.16&  7.64    &  6.66 &  0.6287 &  0.6489       &  0.6455         &  2.968 \\
4 &0.98  & 1.44E+05  &  8.43&  8.14&  7.65    &  6.63 &  0.6304 &  0.6508       &  0.6456         &  2.965 \\
5 &1.10  & 1.99E+05  &  8.43&  8.13&  7.66    &  6.69 &  0.6318 &  0.6519       &  0.6450         &  2.960 \\
6 &1.17  & 2.57E+05  &  8.46&  8.12&  7.65    &  6.71 &  0.6323 &  0.6523       &  0.6437         &  2.913 \\
7 &1.23  & 3.18E+05  &  8.45&  8.11&  7.62    &  6.75 &  0.6320 &  0.6518       &  0.6418         &  2.837 \\
8 &1.26  & 3.81E+05  &  8.46&  8.11&  7.59    &  6.78 &  0.6308 &  0.6505       &  0.6395         &  2.738 \\
9 &1.25  & 4.46E+05  &  8.47&  8.10&  7.54    &  6.80 &  0.6291 &  0.6490       &  0.6371         &  2.604 \\
10&1.26  & 5.13E+05  &  8.46&  8.10&  7.55    &  6.80 &  0.6272 &  0.6472       &  0.6344         &  2.417 \\
11&1.21  & 5.80E+05  &  8.49&  8.10&  7.55    &  6.81 &  0.6248 &  0.6448       &  0.6322         &  2.143 \\
12&1.28  & 6.47E+05  &  8.46&  8.14&  7.55    &  7.04 &  0.6229 &  0.6427       &  0.6304         &  1.685 \\
 \\
\hline
\multicolumn{11}{c}{M3.z2m2.st} \\
\hline
1 &0.00  & 0.00E+00  &  8.35&  8.19&  7.17    &  6.43 &  0.5689 &  0.5928       &  0.5933         &  2.975 \\
2 &0.00  & 5.48E+04  &  8.37&  8.18&  7.20    &  6.44 &  0.5710 &  0.5951       &  0.5955         &  2.973 \\
3 &0.00  & 1.17E+05  &  8.37&  8.18&  7.25    &  6.46 &  0.5741 &  0.5988       &  0.5988         &  2.968 \\
4 &0.20  & 1.81E+05  &  8.38&  8.17&  7.45    &  6.49 &  0.5787 &  0.6029       &  0.6022         &  2.960 \\
5 &0.37  & 2.45E+05  &  8.41&  8.16&  7.46    &  6.51 &  0.5833 &  0.6071       &  0.6054         &  2.948 \\
6 &0.56  & 3.11E+05  &  8.44&  8.14&  7.46    &  6.56 &  0.5877 &  0.6111       &  0.6079         &  2.930 \\
7 &0.71  & 3.79E+05  &  8.41&  8.13&  7.47    &  6.55 &  0.5915 &  0.6144       &  0.6098         &  2.903 \\
8 &0.81  & 4.49E+05  &  8.42&  8.12&  7.42    &  6.62 &  0.5948 &  0.6172       &  0.6112         &  2.866 \\
9 &0.88  & 5.22E+05  &  8.39&  8.11&  7.44    &  6.65 &  0.5973 &  0.6195       &  0.6123         &  2.815 \\
10&0.91  & 5.96E+05  &  8.47&  8.11&  7.42    &  6.61 &  0.5995 &  0.6214       &  0.6131         &  2.746 \\
11&0.95  & 6.73E+05  &  8.48&  8.10&  7.42    &  6.66 &  0.6013 &  0.6229       &  0.6136         &  2.652 \\
12&0.97  & 7.50E+05  &  8.48&  8.09&  7.42    &  6.68 &  0.6024 &  0.6239       &  0.6139         &  2.527 \\
13&0.92  & 8.28E+05  &  8.39&  8.10&  7.42    &  6.62 &  0.6033 &  0.6247       &  0.6147         &  2.348 \\
14&0.94  & 9.02E+05  &  8.49&  8.10&  7.47    &  6.67 &  0.6045 &  0.6253       &  0.6154         &  2.094 \\
15&1.05  & 9.76E+05  &  8.49&  8.11&  7.48    &  6.76 &  0.6056 &  0.6261       &  0.6262         &  1.668 \\
\hline
\multicolumn{11}{c}{M2.z1m2.he07} \\
\hline
1 &0.00  & 0.00E+00  &  8.30&  8.15&  7.12    &  6.25 &  0.4490 &  0.4975       &  0.4993         &  1.978 \\
2 &0.00  & 5.50E+05  &  8.34&  8.15&  7.12    &  6.29 &  0.4590 &  0.5051       &  0.5061         &  1.978 \\
3 &0.00  & 9.09E+05  &  8.37&  8.16&  7.15    &  6.29 &  0.4674 &  0.5117       &  0.5124         &  1.978 \\
4 &0.00  & 1.10E+06  &  8.35&  8.15&  7.18    &  6.31 &  0.4722 &  0.5149       &  0.5159         &  1.978 \\
5 &0.00  & 1.26E+06  &  8.37&  8.15&  7.18    &  6.33 &  0.4752 &  0.5190       &  0.5198         &  1.977 \\
6 &0.00  & 1.43E+06  &  8.37&  8.13&  7.17    &  6.34 &  0.4805 &  0.5240       &  0.5247         &  1.977 \\
7 &0.00  & 1.60E+06  &  8.39&  7.76&  7.17    &  6.34 &  0.4866 &  0.5296       &  0.5301         &  1.977 \\
8 &0.00  & 1.76E+06  &  8.41&  7.78&  7.20    &  6.36 &  0.4936 &  0.5354       &  0.5359         &  1.976 \\
9 &0.00  & 1.92E+06  &  8.40&  7.79&  7.23    &  6.37 &  0.5013 &  0.5415       &  0.5419         &  1.975 \\
10&0.00  & 2.06E+06  &  8.42&  7.84&  7.23    &  6.38 &  0.5092 &  0.5477       &  0.5480         &  1.974 \\
11&0.00  & 2.20E+06  &  8.42&  8.11&  7.27    &  6.40 &  0.5175 &  0.5541       &  0.5543         &  1.973 \\
12&0.00  & 2.33E+06  &  8.41&  8.13&  7.31    &  6.40 &  0.5257 &  0.5603       &  0.5604         &  1.971 \\
13&0.04  & 2.45E+06  &  8.43&  8.14&  7.53    &  6.41 &  0.5334 &  0.5666       &  0.5665         &  1.968 \\
14&0.08  & 2.56E+06  &  8.41&  8.14&  7.63    &  6.42 &  0.5412 &  0.5727       &  0.5724         &  1.966 \\
15&0.20  & 2.67E+06  &  8.42&  8.14&  7.64    &  6.44 &  0.5487 &  0.5788       &  0.5777         &  1.962 \\
16&0.34  & 2.77E+06  &  8.43&  8.14&  7.68    &  6.45 &  0.5556 &  0.5843       &  0.5822         &  1.957 \\
17&0.47  & 2.87E+06  &  8.45&  8.14&  7.68    &  6.47 &  0.5615 &  0.5893       &  0.5860         &  1.951 \\
18&0.51  & 2.97E+06  &  8.42&  8.14&  7.68    &  6.48 &  0.5668 &  0.5934       &  0.5897         &  1.943 \\
19&0.51  & 3.06E+06  &  8.42&  8.14&  7.68    &  6.49 &  0.5715 &  0.5969       &  0.5932         &  1.934 \\
20&0.51  & 3.14E+06  &  8.45&  8.15&  7.66    &  6.50 &  0.5758 &  0.6005       &  0.5969         &  1.923 \\
21&0.53  & 3.22E+06  &  8.44&  8.14&  7.69    &  6.51 &  0.5802 &  0.6039       &  0.6002         &  1.873 \\
22&0.56  & 3.30E+06  &  8.45&  8.15&  7.65    &  6.51 &  0.5844 &  0.6077       &  0.6036         &  1.804 \\
23&0.54  & 3.37E+06  &  8.44&  8.14&  7.65    &  6.49 &  0.5884 &  0.6106       &  0.6069         &  1.722 \\
24&0.52  & 3.44E+06  &  8.46&  8.15&  7.70    &  6.47 &  0.5925 &  0.6142       &  0.6104         &  1.602 \\
25&0.36  & 3.51E+06  &  8.46&  8.15&  7.70    &  6.50 &  0.5970 &  0.6174       &  0.6222         &  1.428 \\
\hline
\multicolumn{11}{c}{M2.z2m2.he07} \\
\hline
1 &0.00  & 0.00E+00  &  8.23&  8.18&  7.15    &  6.22 &  0.4686 &  0.5097       &  0.5112         &  1.959 \\ 
2 &0.00  & 2.72E+05  &  8.30&  8.14&  7.19    &  6.35 &  0.4722 &  0.5139       &  0.5149         &  1.959 \\
3 &0.00  & 6.42E+05  &  8.33&  8.18&  7.17    &  6.26 &  0.4836 &  0.5230       &  0.5236         &  1.958 \\
4 &0.00  & 7.79E+05  &  8.32&  8.18&  7.20    &  6.42 &  0.4884 &  0.5257       &  0.5265         &  1.958 \\
5 &0.00  & 9.04E+05  &  8.39&  8.20&  7.26    &  6.19 &  0.4906 &  0.5298       &  0.5303         &  1.958 \\
6 &0.00  & 1.04E+06  &  8.26&  8.13&  7.21    &  6.33 &  0.4954 &  0.5343       &  0.5348         &  1.957 \\
7 &0.00  & 1.17E+06  &  8.34&  8.09&  7.25    &  6.48 &  0.5008 &  0.5396       &  0.5399         &  1.956 \\
8 &0.00  & 1.29E+06  &  8.30&  8.10&  7.25    &  6.42 &  0.5076 &  0.5452       &  0.5455         &  1.955 \\
9 &0.00  & 1.41E+06  &  8.22&  8.09&  7.27    &  6.39 &  0.5146 &  0.5508       &  0.5511         &  1.954 \\
10&0.00  & 1.53E+06  &  8.40&  8.16&  7.47    &  6.25 &  0.5217 &  0.5567       &  0.5569         &  1.952 \\
11&0.00  & 1.64E+06  &  8.39&  8.15&  7.35    &  6.66 &  0.5292 &  0.5625       &  0.5627         &  1.949 \\
12&0.00  & 1.74E+06  &  8.37&  8.14&  7.42    &  6.64 &  0.5366 &  0.5683       &  0.5684         &  1.946 \\
13&0.00  & 1.84E+06  &  8.41&  8.16&  7.49    &  6.67 &  0.5437 &  0.5741       &  0.5741         &  1.943 \\
14&0.08  & 1.93E+06  &  8.34&  8.15&  7.71    &  6.49 &  0.5509 &  0.5799       &  0.5796         &  1.938 \\
15&0.10  & 2.01E+06  &  8.35&  8.15&  7.71    &  6.45 &  0.5580 &  0.5857       &  0.5852         &  1.932 \\
16&0.12  & 2.10E+06  &  8.33&  8.15&  7.70    &  6.55 &  0.5646 &  0.5911       &  0.5905         &  1.925 \\
17&0.19  & 2.17E+06  &  8.36&  8.15&  7.77    &  6.74 &  0.5711 &  0.5964       &  0.5953         &  1.917 \\
18&0.31  & 2.25E+06  &  8.26&  8.15&  7.73    &  6.49 &  0.5771 &  0.6014       &  0.5996         &  1.907 \\
19&0.41  & 2.32E+06  &  8.35&  8.15&  7.74    &  6.62 &  0.5824 &  0.6060       &  0.6035         &  1.894 \\
20&0.45  & 2.39E+06  &  8.36&  8.14&  7.75    &  6.71 &  0.5873 &  0.6100       &  0.6071         &  1.879 \\
21&0.46  & 2.46E+06  &  8.24&  8.15&  7.75    &  6.77 &  0.5918 &  0.6138       &  0.6107         &  1.860 \\
22&0.46  & 2.52E+06  &  8.44&  8.15&  7.76    &  6.48 &  0.5960 &  0.6172       &  0.6143         &  1.839 \\
23&0.54  & 2.59E+06  &  8.29&  8.16&  7.74    &  6.42 &  0.6002 &  0.6208       &  0.6173         &  1.814 \\
24&0.54  & 2.65E+06  &  8.39&  8.17&  7.77    &  6.65 &  0.6039 &  0.6241       &  0.6204         &  1.783 \\
25&0.51  & 2.70E+06  &  8.35&  8.12&  7.75    &  6.62 &  0.6077 &  0.6272       &  0.6237         &  1.747 \\
26&0.53  & 2.76E+06  &  8.44&  8.15&  7.76    &  6.62 &  0.6116 &  0.6304       &  0.6268         &  1.704 \\
\hline
\multicolumn{11}{c}{M3.z1m2.he07} \\
\hline
1 &0.00  & 0.00E+00  &  8.21&  8.11&  7.71    &  8.03 &  0.6275 &  0.6467       &  0.6468         &  2.971 \\
2 &0.20  & 4.42E+04  &  8.19&  8.11&  7.71    &  7.98 &  0.6304 &  0.6494       &  0.6490         &  2.970 \\
3 &0.40  & 8.76E+04  &  8.18&  8.11&  7.71    &  7.91 &  0.6333 &  0.6526       &  0.6513         &  2.968 \\
4 &0.50  & 1.34E+05  &  8.17&  8.11&  7.71    &  7.26 &  0.6363 &  0.6557       &  0.6535         &  2.965 \\
5 &0.69  & 1.81E+05  &  8.14&  8.11&  7.71    &  6.60 &  0.6397 &  0.6586       &  0.6551         &  2.961 \\
6 &0.78  & 2.31E+05  &  8.12&  8.12&  7.72    &  6.19 &  0.6426 &  0.6610       &  0.6564         &  2.956 \\
7 &0.87  & 2.80E+05  &  8.48&  8.13&  7.71    &  6.76 &  0.6448 &  0.6626       &  0.6572         &  2.949 \\
8 &0.83  & 3.31E+05  &  8.30&  8.15&  7.75    &  6.80 &  0.6465 &  0.6639       &  0.6583         &  2.941 \\
9 &0.94  & 3.78E+05  &  8.47&  8.13&  7.71    &  6.78 &  0.6479 &  0.6647       &  0.6588         &  2.879 \\
10&0.99  & 4.27E+05  &  8.41&  8.13&  7.70    &  7.15 &  0.6490 &  0.6658       &  0.6588         &  2.785 \\
11&1.03  & 4.77E+05  &  8.29&  8.10&  7.74    &  7.81 &  0.6496 &  0.6662       &  0.6586         &  2.666 \\
12&0.86  & 5.26E+05  &  8.28&  8.11&  7.72    &  7.43 &  0.6496 &  0.6661       &  0.6596         &  2.511 \\
13&0.91  & 5.72E+05  &  8.49&  8.13&  7.71    &  6.40 &  0.6508 &  0.6664       &  0.6603         &  2.331 \\
14&0.82  & 6.18E+05  &  8.33&  8.12&  7.75    &  6.25 &  0.6518 &  0.6675       &  0.6615         &  2.051 \\
15&1.05  & 6.61E+05  &  8.15&  8.13&  7.72    &  6.25 &  0.6533 &  0.6684       &  0.6625         &  1.630 \\
\hline
\multicolumn{11}{c}{M3.z2m2.he07} \\
\hline
1 &0.00  & 0.00E+00  &  8.36&  8.20&  7.22    &  6.41 &  0.5681 &  0.5925       &  0.5930         &  2.978 \\
2 &0.00  & 5.60E+04  &  8.35&  8.18&  7.31    &  6.45 &  0.5700 &  0.5944       &  0.5949         &  2.978 \\
3 &0.00  & 1.16E+05  &  8.38&  8.18&  7.31    &  6.46 &  0.5722 &  0.5980       &  0.5982         &  2.977 \\
4 &0.00  & 1.79E+05  &  8.42&  8.17&  7.35    &  6.48 &  0.5764 &  0.6020       &  0.6020         &  2.976 \\
5 &0.09  & 2.43E+05  &  8.41&  8.17&  7.59    &  6.51 &  0.5810 &  0.6064       &  0.6062         &  2.975 \\
6 &0.15  & 3.05E+05  &  8.42&  8.16&  7.67    &  6.52 &  0.5864 &  0.6111       &  0.6105         &  2.973 \\
7 &0.25  & 3.67E+05  &  8.44&  8.16&  7.67    &  6.54 &  0.5920 &  0.6158       &  0.6146         &  2.971 \\
8 &0.40  & 4.27E+05  &  8.42&  8.15&  7.67    &  6.56 &  0.5972 &  0.6201       &  0.6180         &  2.969 \\
9 &0.52  & 4.87E+05  &  8.44&  8.15&  7.72    &  6.58 &  0.6019 &  0.6239       &  0.6209         &  2.965 \\
10&0.54  & 5.46E+05  &  8.46&  8.15&  7.74    &  6.59 &  0.6059 &  0.6271       &  0.6238         &  2.961 \\
11&0.63  & 6.03E+05  &  8.43&  8.15&  7.73    &  6.62 &  0.6096 &  0.6300       &  0.6262         &  2.956 \\
12&0.68  & 6.60E+05  &  8.46&  8.15&  7.71    &  6.63 &  0.6128 &  0.6327       &  0.6283         &  2.951 \\
13&0.66  & 7.15E+05  &  8.47&  8.15&  7.73    &  6.64 &  0.6156 &  0.6350       &  0.6306         &  2.945 \\
14&0.75  & 7.69E+05  &  8.45&  8.14&  7.73    &  6.67 &  0.6185 &  0.6373       &  0.6324         &  2.938 \\
15&0.81  & 8.23E+05  &  8.44&  8.14&  7.71    &  6.68 &  0.6207 &  0.6393       &  0.6337         &  2.930 \\
16&0.85  & 8.77E+05  &  8.47&  8.14&  7.73    &  6.64 &  0.6226 &  0.6409       &  0.6350         &  2.827 \\
17&0.89  & 5.46E+05  &  8.49&  8.14&  7.74    &  6.59 &  0.6253 &  0.6429       &  0.6430         &  2.718 \\
18&0.90  & 6.03E+05  &  8.48&  8.12&  7.73    &  6.62 &  0.6258 &  0.6439       &  0.6439         &  2.585 \\
19&0.92  & 6.60E+05  &  8.26&  8.14&  7.71    &  6.63 &  0.6263 &  0.6444       &  0.6445         &  2.436 \\
20&0.94  & 7.15E+05  &  8.34&  8.15&  7.73    &  6.64 &  0.6285 &  0.6455       &  0.6456         &  2.247 \\
21&0.95  & 7.69E+05  &  8.48&  8.14&  7.73    &  6.67 &  0.6293 &  0.6465       &  0.6466         &  1.993 \\
22&0.94  & 8.23E+05  &  8.26&  8.16&  7.71    &  6.68 &  0.6301 &  0.6480       &  0.6481         &  1.767 \\
23&1.00  & 8.77E+05  &  8.26&  8.04&  7.71    &  6.63 &  0.6309 &  0.6500       &  0.6501         &  0.735 \\
\hline
\multicolumn{11}{l}{TP: TP number.} \\
\multicolumn{11}{l}{ $DUP_\lambda$: DUP Lambda parameter.} \\
\multicolumn{11}{l}{ $t_{TP}$: Time since first TP.} \\
\multicolumn{11}{l}{$T_{FBOT}$: Largest temperature at the bottom of the flash-convective zone.} \\
\multicolumn{11}{l}{$T_{HES}$: Temperature in the He-burning shell during the deepest extend of TDU.} \\
\multicolumn{11}{l}{ $T_{CEB}$: Temperature at the bottom of the convective envelope during the deepest extend of TDU.} \\
\multicolumn{11}{l}{$m_{FBOT}$: Mass coordinate at the bottom of the He-flash convective zone.} \\
\multicolumn{11}{l}{$m_{D,max}$: Mass coordinate of the H-free core at the time of the TP.} \\
\multicolumn{11}{l}{ $M_{\ast}$: Stellar mass at the TP.} 
\label{tab:m3z2m2_TPprop}
\end{longtable}
\end{center}

\clearpage

\begin{landscape}
\begin{table}
\begin{center}
\caption{List of AGB stellar models not included in \tab{tab:model_name} and their relevant parameters: initial mass, initial metallicity, CBM parameterization (see \tab{tab:model_name} for details) and respective modification for the reaction rate reported in the last column, compared to the default nuclear reaction network.
}
\begin{tabular}{lccccccccccc}
\hline
name & mass [M$_{\odot}$] &  metallicity &  CBM & f1 & D2 & f2 & f1* & D2* & f2* & rate test & \\
\hline
\hline
M3.z2m2.zrtest	& 3.0 & 0.02	& df & 0.024 & 10$^{5}$ & 0.14 & 0.014 & 10$^{11}$  & 0.25 & \zrfu\ (n,$\gamma$) \zrse\ / 2 \\ 	
M3.z1m2.zrtest	& 3.0 & 0.01	& df & 0.024 & 10$^{5}$ & 0.14 & 0.014 & 10$^{11}$  & 0.25 & \zrfu\ (n,$\gamma$) \zrse\ / 2 \\ 
M2.z2m2	.zrtest& 2.0 & 0.02	& df & 0.024 & 10$^{5}$ & 0.14 & 0.014 & 10$^{11}$ & 0.25 & \zrfu\ (n,$\gamma$) \zrse\ / 2 \\ 
M2.z1m2.zrtest	& 2.0 & 0.01	& df & 0.024 & 10$^{5}$ & 0.14 & 0.014 & 10$^{11}$  & 0.25 & \zrfu\ (n,$\gamma$) \zrse\ / 2 \\ 
M3.z1m2.hCBM.ntest & 3.0 & 0.01 & df & 0.024 & 10$^{5}$ & 0.14 & 0.014 & 10$^{12}$  & 0.27 & \nvi\ (n,p) \cvi\ x 2 \\ 
M2.z2m2.hCBM.ntest & 2.0 & 0.02 & df & 0.024 & 10$^{5}$ & 0.14 & 0.014 & 10$^{12}$  & 0.27 & \nvi\ (n,p) \cvi\ x 2 \\ 	
Pi13.newnet & 3.0  & 0.02 	& sf & 0.008 & -  & - & 0.126 & -  & - & Pi13 model with updated network\\ 		

\noalign{\smallskip}
\hline
\end{tabular}
\label{tab:model_name_network_test}
\end{center}
\end{table}
\end{landscape}

\clearpage

\begin{landscape}
\begin{table}
\begin{center}
\caption{Final isotopic ratio values of Zr and Ba isotopes calculated in the He-intershell region (final values on the surface are shown in brackets for comparison). 
}
\resizebox{\columnwidth}{!}{%
\begin{tabular}{lccccccccc}
\hline
name & $\delta$($^{90}$Zr/$^{94}$Zr) &  $\delta$($^{91}$Zr/$^{94}$Zr) & $\delta$($^{92}$Zr/$^{94}$Zr) & $\delta$($^{96}$Zr/$^{94}$Zr) & $\delta$($^{134}$Ba/$^{136}$Ba) & $\delta$($^{135}$Ba/$^{136}$Ba) & $\delta$($^{137}$Ba/$^{136}$Ba) & $\delta$($^{138}$Ba/$^{136}$Ba) \\
\hline
\hline
M3.z2m2	           & -393.51 & -181.45 & -154.83 & -426.57 &  83.87 & -878.02 & -387.23 & -52.89 \\ 
                   &(-335.08)&(-161.41)&(-135.18)&(-392.59)& (46.89)&(-789.77)& (-349.64) &(-72.07) \\
M3.z1m2	           & -463.92 & -219.91 & -125.72 &  650.29 &  81.64 & -875.29 & -184.04 &  597.95 \\ 
                   &(-347.54)&(-165.14)&(-112.07)& (122.05)& (32.13)&(-760.49)& (-231.80) &(483.26) \\
M2.z2m2	           & -382.17 & -188.69 & -147.67 & -598.55 &  87.22 & -865.73 & -415.13 & -129.00 \\ 
                   &(-216.94)&(-107.80)&(-100.41)&(-456.40)& (68.68)&(-647.20)& (-323.87) & (-161.59)\\
M2.z1m2	           & -396.44 & -229.26 & -195.42 & -580.93 &  21.78 & -874.24 & -412.71 &  264.87 \\
                   &(-276.76)&(-146.97)&(-126.68)&(-474.66)& (37.41)&(-744.11)& (-350.00) &(146.90)\\
\hline
M3.z1m2.hCBM       & -488.00 & -245.69 & -157.98 &  631.22 &  76.56 & -875.28 & -187.77 &  280.46 \\
                   &(-395.95)&(-190.77)&(-129.68)& (162.06)& (48.45)&(-805.50)& (-239.85) &(217.94)\\
M2.z2m2.hCBM       & -349.64 & -191.38 & -166.33 & -741.10 & 116.10 & -859.01 & -448.36 & -342.55 \\
                   &(-220.55)& (-95.05)&(-103.31)&(-584.30)&(111.88)&(-680.29)& (-359.08) &(-332.49)\\
\hline
M3.z1m2.hCBM.ntest & -474.68 & -224.84 & -142.25 &  639.19 &  91.50 & -872.71 & -194.13 & 102.04 \\ 	
                   &(-387.64)&(-182.62)&(-123.04)& (159.53)& (60.64)&(-792.65)& (-239.59) & (60.02)\\
M2.z2m2.hCBM.ntest & -291.93 & -133.05 & -127.27 & -751.27 & 172.91 & -851.96 & -463.02 & -470.00 \\
 	           &(-161.23)& (-44.19)& (-66.49)&(-533.73)&(124.98)&(-567.77)& -(310.92) &(-352.22)\\
\hline
M3.z2m2.zrtest     & -410.72 & -182.78 &  245.82 & -592.46 &  75.29 & -872.68 & -357.73 & -75.47 \\
                   &(-337.39)&(-161.47)&(-133.07)&(-581.47)& (46.49)&(-789.63)& (-344.24) & (-57.02)\\
M3.z1m2.zrtest     & -465.18 & -215.45 & -129.23 &    6.88 &  70.13 & -874.00 & -158.26 & 585.02 \\
 	           &(-350.91)&(-167.77)&(-111.21)&(-230.81)& (33.72)&(-758.98)& (-216.71) &(494.77)\\
M2.z2m2.zrtest     & -379.70 & -187.06 & -159.57 & -802.90 &  94.57 & -864.86 & -417.91 & -130.13 \\
                   &(-210.93)&(-105.88)&(-100.20)&(-526.57)& (69.21)&(-636.15)& (-319.34) & (-159.15)\\
M2.z1m2.zrtest     & -411.85 & -227.21 & -172.50 & -745.62 &  49.90 & -871.84 & -397.75 &  275.06 \\
                   &(-273.40)&(-142.87)&(-122.68)&(-588.58)& (40.31)&(-738.79)&(-347.05)& (122.01)\\
\hline
M3.z2m2.he07     & -400.18 & -219.03 & -193.04 & -357.52 &  49.65 & -876.58 & -393.20 & -119.29 \\
                   &(-343.19)&(-166.66)&(-135.04)&(-235.48)& (24.19)&(-788.31)& (-324.14) & (-7.66)\\
M3.z1m2.he07     & -414.00 & -158.33 & -88.07 & 832.95 &  55.05 & -877.64 & -127.75 & 574.93 \\
 	           &(-322.07)&(-141.34)&(-92.01)&(261.93)& (21.73)&(-719.77)& (-193.47) &(377.06)\\
M2.z2m2.he07     & -398.38 & -203.37 & -162.19 & -538.58 &  48.18 & -872.15 & -407.46 & -43.56 \\
                   &(-188.99)&(-95.98)&(-87.74)&(-362.46)& (38.41)&(-582.18)& (-285.89) & (-91.69)\\
M2.z1m2.he07     & -412.08 & -223.41 & -166.87 & -479.08 &  34.92 & -873.79 & -395.16 &  367.92 \\
                   &(-237.36)&(-126.75)&(-110.76)&(-402.85)& (22.64)&(-684.83)&(-321.61)& (163.60)\\
\noalign{\smallskip}
\hline
\end{tabular}
}
\label{tab:model_deltaZrBa}
\end{center}
\end{table}
\end{landscape}

\clearpage

\end{document}